\documentclass[nonblind]{informs3} % current default for manuscript submission
\pdfoutput=1
\OneAndAHalfSpacedXI
\usepackage{thmtools}

\usepackage{natbib}
 \bibpunct[, ]{(}{)}{,}{a}{}{,}%
 \def\bibfont{\small}%
\TheoremsNumberedThrough     % Preferred (Theorem 1, Lemma 1, Theorem 2)
\ECRepeatTheorems

\EquationsNumberedThrough    % Default: (1), (2), ...
\MANUSCRIPTNO{ }

\usepackage{amsmath}
\usepackage{amsfonts}
\usepackage{multirow}
\usepackage{tikz}
 \usetikzlibrary{calc} 
\usepackage{mathtools}

\usepackage{booktabs}

\usepackage{xcolor}
\usepackage{bbm}
\usepackage{algorithm,algorithmicx}
\usepackage[noend]{algpseudocode}

\usepackage[normalem]{ulem}

\usepackage{hyperref}
\usepackage{url}
\newcounter{eqname}

\usepackage{comment}

\usepackage{standalone} % for including standalone .tex files
\usepackage{tikz}
\usetikzlibrary{arrows.meta}
\usepackage{wrapfig}
\usepackage{graphicx}

\definecolor{mygreen}{RGB}{0,128,0}

\definecolor{mysilver}{RGB}{220,220,220}

\def\RR{{\mathbb R}}

\def\ZZ{{\mathbb Z}}

\def\P{{\mathcal P}}

\def\C{{\mathcal C}}

\def\S{{\mathcal S}}
\def\E{{\mathcal E}}

\def\argmin{\text{argmin\,}}
\def\C{{\mathcal C}}

\makeatletter
\def\munderbar#1{\underline{\sbox\tw@{$#1$}\dp\tw@\z@\box\tw@}}
\makeatother

\makeatletter
\newcommand{\crossout}[1]{%
  \begingroup
  \settowidth{\dimen@}{#1}%
  \setlength{\unitlength}{0.05\dimen@}%
  \settoheight{\dimen@}{#1}%
  \count@=\dimen@
  \divide\count@ by \unitlength
  \count0=20 \count4=\count@
  \loop
  \count2=\count0 % keep a copy
  \divide\count2\count4 \multiply\count2\count4
  \ifnum\count2<\count0
    \advance\count0 -\count2 % the remainder
    \count2=\count0
    \count0=\count4
    \count4=\count2
  \repeat
  \count0=20 \divide\count0\count4
  \count2=\count@ \divide\count2\count4
  \begin{picture}(0,0)
  \put(0,0){\line(\count0,\count2){20}}
  \put(0,\count@){\line(\count0,-\count2){20}}
  \end{picture}%
  #1%
  \endgroup
}
\makeatother

\newcommand{\Greedy}{Greedy}
\newcommand{\LPheur}{LPH}
\newcommand{\ie}{\emph{i.e.}}
\newcommand{\eg}{\emph{e.g.}}

\newcommand{\MaxHRT}{\textsc{MaxHrt-cap}}
\newcommand{\MinCut}{\textsc{MinCut-cap}}
\newcommand{\AggLin}{\textsc{Agg-Lin}}

\newcommand{\Quad}{\textsc{Quad}}
\newcommand{\CPM}{\textsc{Cpm}}

\usepackage{enumerate}

\usepackage{xr}
\usepackage{subcaption}
\usepackage{mathtools}

\newcommand{\lrp}[1]{\left(#1\right)}

\newcommand{\lrc}[1]{\left[#1\right]}
\newcommand{\lrl}[1]{\left\{#1\right\}}
\newcommand{\lra}[1]{\left|#1\right|}

\algrenewcommand\algorithmicrequire{\textbf{Input:}}
\algrenewcommand\algorithmicensure{\textbf{Output:}}

\allowdisplaybreaks

\renewcommand{\theARTICLETOP}{}

\begin{document}
%%%%%%%%%%%%%%%%

%\setcounter{page}{1}
%\DOI{10.1287/xxxx.0000.0000}%

\RUNAUTHOR{Bobbio et al.}

\RUNTITLE{Capacity Planning in Stable Matching}

% \TITLE{Bundling Information Goods of Decreasing Value}
\TITLE{Capacity Planning in Stable Matching }

\ARTICLEAUTHORS{%
\AUTHOR{Federico Bobbio, Margarida Carvalho}
\AFF{CIRRELT and DIRO, Université de Montr\'eal, \EMAIL{federico.bobbio@umontreal.ca, carvalho@iro.umontreal.ca}} %, \URL{}}
% \AUTHOR{Margarida Carvalho}
% \AFF{CIRRELT and DIRO, Université de Montr\'eal, \EMAIL{carvalho@iro.umontreal.ca}}
\AUTHOR{Andrea Lodi}
\AFF{Jacobs Technion-Cornell Institute, Cornell Tech, \EMAIL{andrea.lodi@cornell.edu}}
\AUTHOR{Ignacio Rios}
\AFF{School of Management, The University of Texas at Dallas, \EMAIL{ignacio.riosuribe@utdallas.edu}}
\AUTHOR{Alfredo Torrico}
\AFF{CDSES, Cornell University, \EMAIL{alfredo.torrico@cornell.edu}}
} % end of the block

\ABSTRACT{%
% We introduce the problem of jointly increasing school capacities and finding a student-optimal assignment in the expanded market.
% Due to the impossibility of efficiently solving the problem with classical methods, we generalize existent mathematical programming formulations of stability constraints to our setting, most of which result in integer quadratically-constrained programs. In addition, we propose a novel mixed-integer linear programming formulation that is exponentially large on the problem size. We show that its stability constraints can be separated by exploiting the objective function, leading to an effective cutting-plane algorithm. We conclude the theoretical analysis of the problem by discussing some mechanism properties. 
% On the computational side, we evaluate the performance of our approaches in a detailed study, and we find that our cutting-plane method outperforms our generalization of existing mixed-integer approaches. We also propose two heuristics that are effective for large instances of the problem. Finally, we use the Chilean school choice system data to demonstrate the impact of capacity planning under stability conditions. Our results show that each additional seat can benefit multiple students and that we can effectively target the assignment of previously unassigned students or improve the assignment of several students through improvement chains. These insights empower the decision-maker in tuning the matching algorithm to provide a fair application-oriented solution.
Motivated by the shortage of seats that the Chilean school choice system is facing, we introduce the problem of jointly increasing school capacities and finding a student-optimal assignment in the expanded market. Due to the theoretical and practical complexity of the problem, we provide a comprehensive set of tools to solve the problem, including different mathematical programming formulations, a cutting plane algorithm, and two heuristics that allow obtaining near-optimal solutions quickly.
On the theoretical side, we show the correctness of our formulations, different properties of the objective and feasible region that facilitate computation, and also several properties of the underlying mechanism to find a student-optimal matching under capacity expansions.
On the computational side, we use data from the Chilean school choice system to demonstrate the impact of our framework and derive insights that could help alleviate the problem. Our results show that each additional seat can benefit multiple students and that we can effectively target the assignment of previously unassigned students or improve the assignment of several students through improvement chains. Nevertheless, our results show that the marginal effect of each additional seat is decreasing and that simply adding seats is insufficient to ensure every student gets assigned to some school. Finally, we discuss several extensions of our framework, showcasing its flexibility to accommodate different needs.
}

 %\HISTORY{
%This paper was first submitted on September 12, 2022 and has been with the authors for 83 years for 65 revisions. }

\maketitle
%%%%%%%%%%%%%%%%%%%%%%%%%%%%%%%%%%%%%%%%%%%%%%%%%%%%%%%%%%%%%%%%%%%%%%

% Samples of sectioning (and labeling) in MNSC
% NOTE: (1) \section and \subsection do NOT end with a period
%       (2) \subsubsection and lower need end punctuation
%       (3) capitalization is as shown (title style).
%
%\section{Introduction.}\label{intro} %%1.
%\subsection{Duality and the Classical EOQ Problem.}\label{class-EOQ} %% 1.1.
%\subsection{Outline.}\label{outline1} %% 1.2.
%\subsubsection{Cyclic Schedules for the General Deterministic SMDP.}
%  \label{cyclic-schedules} %% 1.2.1
%\section{Problem Description.}\label{problemdescription} %% 2.

% Text of your paper here

%!TEX root = ./0_main.tex
\section{Introduction}\label{sec:introduction}
Centralized mechanisms are becoming the standard approach to solve several assignment problems, including the allocation of students to schools, high-school graduates to colleges, residents to hospitals, and refugees to cities. In most of these  markets, a desirable property of the assignment is \emph{stability}, which guarantees that no pair of agents has incentive to circumvent the matching. As discussed in~\citep{RothSotomayor1990} and~\citep{roth02}, finding a stable matching is crucial for the clearinghouse's success, long-term sustainability and also ensures some notion of fairness as it eliminates the so-called \emph{justified-envy}.\footnote{See \cite{romm20} for a discussion of the differences between \emph{stability} and \emph{no justified-envy}.}

A common assumption in these markets is that capacities are fixed and known. However, capacities are only a proxy of how many agents can be accommodated, and there might be some flexibility to modify them in many settings. For instance, in some college admissions systems, colleges may increase their capacities to admit all tied students competing for the last seat~\citep{rios2021}. Moreover, in several colleges/universities, the number of seats offered in a given course or program is adjusted based on their popularity among students.\footnote{Examples include the School of Engineering at the University of Chile, where all students who want to study any of its programs must take a shared set of courses in the first two years and then must apply to a specific program (\eg, Civil Engineering, Industrial Engineering, etc.) based on their GPA, without knowing the number of seats available in each of them. Similarly, in many schools that use course allocation systems such as Course Match~\citet{budish17}, over-subscribed courses often increase their capacities, while under-subscribed ones are merged or canceled.}
In school choice, school districts may experience overcrowding, where some schools serve more students than their designed capacity.\footnote{According to the results of a nationwide survey~\citet{schoolsamerica99}, 22\% of schools in the US experienced some degree of overcrowding, and 8\% had enrollments that exceeded their capacity by more than 25\%.} In response, school districts often explore alternative strategies to accommodate the excess demand, such as utilizing portable classrooms, implementing multi-track or staggered schedules, or adopting other temporary measures, and use students' preferences as input to make these decisions.

An actual example of the latter is the crisis that the Chilean school choice system---known as ``Sistema de Admisión Escolar'' (SAE)---is currently facing, where the strain on the public school system has reached critical levels.
Specifically, in many regions of the country, the number of students applying to the system largely exceeds the number of seats offered. Consequently, more than 3,000 students were not assigned and were forced out of the public system in the last admissions process (2023-2024), with many of them resorting to home-schooling as they cannot afford private schools. One of the most notable examples of the crisis is at the Pre-K level in the Antofagasta region, where there are roughly 0.9 seats per student applying to the system, and 4 out of 10 students do not get assigned in any of their preferred schools after the main round of the admissions process (see Table~\ref{tab:general summary stats Antofagasta Pre-K}).
As a result, there has been public outrage, with parents, teacher associations, members of Congress, and children advocates calling for an immediate solution.

\begin{table}[htp!]
\caption{Summary Statistics: Pre-K in Antofagasta}\label{tab:general summary stats Antofagasta Pre-K}
\centerline{
\scalebox{0.95}{\begin{tabular}[t]{ccccccccc}
\toprule
\multicolumn{5}{c}{ } & \multicolumn{2}{c}{Assignment} & \multicolumn{2}{c}{Enrollment} \\
\cmidrule(l{3pt}r{3pt}){6-7} \cmidrule(l{3pt}r{3pt}){8-9}
Year & Students & Schools & Seats & Seats per student & Total & Rate & Total & Rate\\
\midrule
2018 & 4371 & 75 & 3551 & 0.812 & 2509 & 0.574 & 2422 & 0.554\\
2019 & 4455 & 74 & 3605 & 0.809 & 2435 & 0.547 & 2372 & 0.532\\
2020 & 3918 & 75 & 3675 & 0.938 & 2647 & 0.676 & 2560 & 0.653\\
2021 & 3629 & 76 & 3628 & 1.000 & 2368 & 0.653 & 2312 & 0.637\\
2022 & 3952 & 76 & 3573 & 0.904 & 2417 & 0.612 & 2332 & 0.590\\
2023 & 3795 & 71 & 3558 & 0.938 & 2412 & 0.636 & N/A & N/A\\
\bottomrule
\end{tabular}}
\vspace{0.2cm}
}\centerline{\begin{minipage}[t]{12.5cm}
\footnotesize{\emph{Note.} Summary statistics on applicants, schools, seats, and assignments correspond to the main round of the assignment process.}
\end{minipage}}
\end{table}

In response to the crisis, Chile's Ministry of Education (MINEDUC) implemented the ``Plan de Fortalecimiento de Matricula''~\citep{pfm24}. Since 2022, this plan has increased capacity by more than 13,000 seats, with nearly 8,700 added during the latest admission process. Additionally, MINEDUC launched a platform that enables parents to search for schools with available capacity and request a seat if their child was left unassigned, streamlining the waitlists processing during the after-market period.
Despite these efforts the problem persists, partially because the additional seats often end up allocated sub-optimally, limiting the plan's effectiveness.

Finding a stable matching and making capacity decisions jointly poses several challenges. First, standard approaches like the Deferred Acceptance algorithm (DA)~\citep{gale1962college} require fixed and known capacities, making them unsuitable. 
Second, the clearinghouse may have multiple objectives beyond finding a student-optimal assignment, and these objectives can introduce trade-offs in allocating additional seats. For instance, one possible goal is to maximize \emph{access}, \ie, to allocate extra seats to achieve a stable matching that maximizes the total number of assigned students.
Like most school choice systems, MINEDUC is legally required to guarantee that every student who wishes to attend a public school can do so, making \emph{access} one of their primary goal. However, the clearinghouse may also aim to prioritize \emph{improvement}, \ie, enhancing the assignment of high-priority students. This approach is common in merit-based settings such as college admissions and entry level job-markets, where the optimal placement of candidates is crucial.\footnote{Note that this trade-off between \emph{access} and \emph{improvement} does not arise in the standard version of the problem, as there is a unique student-optimal stable matching under fixed capacities.}
Finally, even if the objective is clear, the computation of a student-optimal matching under capacity planning is theoretically hard~\citep{bobbio2022capacityvariation}, in sharp contrast with the standard case (with no capacity decisions) that can be solved in polynomial time.

In this paper, we study the problem of jointly deciding how to allocate a budget of additional seats and finding a student-optimal assignment that takes advantage of the expanded capacities. The goal of our paper is threefold.
First, to present a novel framework for finding a student-optimal assignment while allocating extra capacities targeting different objectives, such as access, improvements, or any combination.
Second, to develop an optimization toolbox composed of exact algorithms and heuristics that allows us to efficiently obtain optimal or near-optimal solutions. This is of utmost importance, as a flexible and effective set of methods would facilitate policymakers in making capacity decisions. Our last goal is to obtain operational insights that could help MINEDUC navigate the current educational crisis in Chile.

\subsection{Contributions}
Our work makes several contributions that we now describe in detail.

\paragraph{Model and mechanism analysis.} To capture the problem described above, we introduce a stylized model of a many-to-one matching market in which the clearinghouse can make capacity planning decisions while simultaneously finding a student-optimal stable matching, generalizing the standard model by~\cite{gale1962college}. 
We show that the clearinghouse can prioritize different goals by changing the penalty values of unassigned students. Namely, it can obtain the minimum or the maximum cardinality student-optimal stable matchings and, thus, prioritize improvements and access, respectively. In addition, we study other properties of interest, including agents' incentives
and the monotonicity of the mechanism.

\paragraph{Exact solution methods.}
We first formulate our problem as an integer quadratically constrained program by extending existing approaches. On the one hand, to overcome the challenges imposed by the non-linearity of some constraints, we provide a linear reformulation exploiting the concept of stability. Using simulations, we show that this formulation---which we refer to as \emph{compact formulation with L-constraints}---outperforms other linearizations obtained using standard techniques. On the other hand, to overcome the challenges imposed by the integrality of the decision variables, we introduce a second formulation---which we refer to as \emph{generalized comb formulation}---that relaxes the integrality of the matching variables and recovers it by adding an exponential number of comb constraints. To overcome the problem's dimensionality, we propose an efficient cutting plane algorithm that relies on a novel separation algorithm that identifies the most violated constraint for each school in each iteration. Furthermore, we show that we can speed up the search for violated comb constraints by exploiting (i) new structural results characterizing properties of the objective function and the feasible region and (ii) two related matchings to limit the search for new violated combs.
The sum of all these technical enhancements ensures that our cutting-plane method outperforms the benchmarks solved by state-of-the-art mixed-integer programming solvers, both in terms of solving time and solution quality (optimality gap) guarantees. 

\paragraph{Heuristic solution methods.} As shown in~\citep{bobbio2022capacityvariation}, the problem is NP-hard and cannot be approximated within a $\mathcal{O}(n^{(\frac{1}{6}-\varepsilon)})$ factor, where $n$ is the number of students. Moreover, many real-life instances could not be solved in a reasonable time (even using our cutting plane algorithm), making our exact methods unsuitable when time is sensitive (\eg, to perform policy evaluations and counterfactual analyses that require many simulations). These limitations motivated us to devise two efficient heuristics that quickly provide near-optimal solutions. Our first heuristic adapts the standard \emph{Greedy} algorithm for set functions, sequentially adding in each iteration one extra seat to the school leading to the greatest marginal improvement in the objective function. Our second heuristic, called \emph{LPH}, proceeds in two steps: (i) solve the problem without stability constraints to find the allocation of extra seats, and (ii) find the student-optimal stable matching conditional on the capacities defined in the first step. Our computational experiments show that both heuristics significantly reduce the time to find a near-optimal solution, with LPH being the fastest and unaffected by the budget size. Hence, LPH could be a good approach to quickly obtain \emph{high-quality} solutions to large-scale instances of the problem in time sensitive settings.

\paragraph{Practical insights and societal impact.} 
To illustrate the benefits of embedding capacity decisions towards addressing the current crisis in Chile, we adapt our framework to solve the problem at the Pre-K level in the Antofagasta region, including all the specific features described by~\cite{Correa_2022}. First, we show that each additional seat can benefit multiple students, but the marginal improvement is decreasing. Moreover, we observe that having a rate of one seat per student is not enough to solve the problem due to the high correlation in preferences and the short length of preference lists.
Second, in line with our theoretical results, we find that \emph{access} and \emph{improvement} can be prioritized depending on how unassigned students are penalized in the objective. However, for high budget values, the differences in entry are relatively small and, thus, a low penalty that benefits the maximum number of students may be preferable. Third, we show that we can still obtain near-optimal solutions if schools limit the additional seats they can receive to ensure practical feasibility, avoiding solutions that overcrowd some over-demanded schools.

These positive results have led to a fruitful collaboration with MINEDUC and the NGO responsible for the allocation to test our framework in the field and devise potential solutions to the crisis. Our model can be easily adapted to incorporate all the elements of the Chilean system (\eg, static and dynamic priorities, quotas, secured enrollment, etc.) and address other policy-relevant questions. For instance, it can be used to optimally decrease capacities in school districts experiencing significant drops in enrollment~\citep{SFchronicle2022}. Additionally, our model could be employed to optimally allocate tuition waivers under budget constraints, as seen in Hungary's college admissions system.
Our methodology is also applicable in other markets, such as refugee resettlement~{\citep{delacretaz16,Andersson20,ahani2021placement}}, where local authorities determine how many refugees they are willing to receive but could increase their capacity given proper incentives. Similarly, in healthcare rationing~\citep{Pathak20,Aziz21}, policymakers could make additional investments to expand available resources. These examples highlight the importance of jointly optimizing stable assignments and capacity decisions, demonstrating the model's potential to answer crucial questions across various settings.

\subsection{Organization} The remainder of the paper is organized as follows. In Section~\ref{sec:related_work}, we provide a literature review.
In Section~\ref{sec:preliminaries}, we formalize the stable matching problem with capacity decisions and present our exact methods.
In Section~\ref{subsec:heuristics}, we present our heuristics.
In Section~\ref{sec: implementation in chile}, we evaluate our framework using Chilean school choice data. 
In Section~\ref{subsec: properties of mechanism}, we discuss several properties of the mechanism to find a student-optimal stable matching with capacity decisions.
Finally, in Section~\ref{sec: conclusions}, we draw some concluding remarks. All the proofs, examples, extensions and additional discussions can be found in the Appendix.

%!TEX root = ./0_main.tex
\section{Related Work}\label{sec:related_work}

%\textcolor{magenta}{AL: I have not looked at this one again but the comments I shared one week ago stand: (1) please update the verbal form to the past whenever necessary; (2) clarify our TCS paper, and (3) remove the sentence ``We contribute to \ldots" to each of the parts unless we want to change according to the comments in the annotated pdf.}
Our paper lies at the intersection of three streams of literature. First, it relates to the mathematical programming formulations of the stable matching problem and its extensions. Second, it contributes to the emerging field that integrates matching and capacity decisions. Finally, it connects to the literature on school choice.

Our contribution to these areas is threefold: (i) we extend the standard stable matching problem by incorporating capacity decisions; (ii) we provide a comprehensive set of tools to solve the problem, including both exact methods and heuristics; and (iii) we demonstrate the potential impact of our framework in a school choice setting. Below, we provide a detailed review of these three streams of literature and refer interested readers to~\citep{manlove2013algorithmics} for a broader overview.

% Since then, the literature on stable matchings has extensively grown and has focused on multiple variants of the problem. For this reason, we focus on the most closely related work, and we refer the interested reader to~\citep{manlove2013algorithmics} for a broader literature review. 

\paragraph{Mathematical programming formulations.}
The first mathematical programming formulations of the stable matching problem were studied in~\citep{gusfield1989stable, vate1989linear,rothblum1992characterization} and~\citep{roth1993stable}. 
Closer to our work,~\citet{baiou2000stable} provided an exponential-size linear programming formulation describing the convex hull of the set of feasible stable matchings and introduced a polynomial-time separation algorithm. 
Thereafter, this strand of the literature has devised mathematical programming formulations for different extensions of the standard model, including ties in preferences (\ie, when agents are indifferent between two or more options)~\citep{kwanashie2014integer,agoston_etal2021}, regional constraints~\citep{kojima2018designing}, upper and lower quotas~\citep{agoston2016integer}, incomplete lists~\citep{delorme2019mathematical}, and dynamic priorities~\citep{rios24}. 
% \citet{kwanashie2014integer} presented an integer formulation of the problem when there are ties in the preference lists (\ie, when agents are indifferent between two or more options). 
% % \citet{kojima2018designing} introduced a way to represent preferences and constraints to guarantee strategy-proofness. 
% \citet{agoston2016integer} proposed an integer model that incorporates upper and lower quotas.  
% \citet{delorme2019mathematical} devised new mixed-integer programming formulations and pre-processing procedures. 
% More recently,~\citet{agoston_etal2021} proposed similar mathematical programs and used them to compare different policies to deal with ties. 

\paragraph{Capacity expansion.}
Following the preliminary version of our work, several subsequent papers have explored a similar setting.
\citet{bobbio2022capacityvariation} investigated the complexity of the capacity planning problem and its variations, demonstrating that the decision version of the problem is NP-complete and the optimization version inapproximable within an $\mathcal{O}(n^{(\frac{1}{6}-\varepsilon)})$ factor, where $n$ is the number of students. 
\citet{abe2022anytime} introduced a heuristic method to solve the capacity planning problem that relies on the Upper Confidence Tree, a Monte Carlo based search method covering the space of capacity expansions.
\cite{afacan2024capacity}, in an independent work, also analyzed the problem of allocating additional seats across schools in response to students' preferences. The authors introduced an algorithm that characterizes the set of efficient matchings among those who respect preferences and priorities and analyzed its incentives' properties. Their work is complementary to ours in several ways. First, they discussed different applications where capacity decisions are made in response to students' preferences. These include some school districts in French-speaking Belgium, where close to 1\%  of seats are consistently reported after students submit their preferences, and certain college admissions systems, such as in India, where the Ministry of Education plans to increase capacities by up to 50\%.
Second, their proposed approach can recover any Pareto efficient allocation, including the ones returned by our mechanism. However, they cannot target a specific outcome. Our methodology is flexible enough to enable policymakers to target a particular goal when allocating the extra seats, including access, improvement, or any other objective beyond student optimality. Finally, ~\cite{afacan2024capacity} showed that their mechanism is strategy-proof when schools share the same preferences, but it is not in the general case. In this work, we also discuss incentive properties and expand the analysis to study other relevant properties of the assignment mechanism, such as strategy-proofness in the large and monotonicity. 
 Finally, \citep{kumano2022quota} studied and implemented the reallocation of capacities among programs within a restructuring process at the University of Tsukuba. Their capacity allocation constraints could be readily added to our model while maintaining the validity of our methodology.

\paragraph{School choice.} 
Starting with \citep{Abdulkadiroglu2003}, a large body of literature has studied different elements of the school choice problem, including the use of different mechanisms such as DA (introduced in the seminar work by~\citep{gale1962college}), Boston, and Top Trading Cycles~\citep{abdulkadirouglu2005boston,Pathak_2008,Abdulkadiro_lu_2011}; the use of different tie-breaking rules~\citep{Abdulkadiro_lu_2009,Arnosti_2015,Ashlagi_2019}; the handling of multiple and potentially overlapping quotas~\citep{kurata17,Somnez19}; the addition of affirmative action policies~\citep{ehlers2010,Hafalir_2013}; and the implementation in many school districts and countries~\citep{abdulkadirouglu2005boston,calsamiglia2014illusion,Correa_2022,Allman_2022}. Within this literature, the closest papers to ours combine optimizing different objectives with finding a stable assignment. \citet{Caro_2004} introduced an integer programming model to make school redistricting decisions. \citet{shi2016} proposed a convex optimization model to decide the assortment of schools to offer each student to maximize the sum of utilities. \citet{Ashlagi_2016} presented an optimization framework that allowed them to find an assignment pursuing (the combination of) different objectives, such as average and min-max welfare.~\citet{bodon2020} presented an optimization model to find the best stable and incentive-compatible match that maximizes any combination of welfare and diversity and prioritizes the allocation of students to their neighborhood schools. Finally,~\cite{lo2020} introduced a novel mechanism to efficiently reassign vacant seats after an initial round of a centralized assignment and used data from the NYC high school admissions system to showcase its benefits.

\section{Model}\label{sec:preliminaries}

    We start by introducing the standard stable matching problem and extending it to include capacity decisions in Sections~\ref{subsec: stable matching problem} and~\ref{sec:capacity_expansion}, respectively. 
    Then, we present two exact mathematical programming formulations that extend those in~\cite{baiou2000stable} to incorporate capacity decisions: (i) a compact formulation (in Section~\ref{subsec:compact_formulation}), and (ii) a comb formulation (in Section~\ref{subsec:comb_formulation}).
    % Finally, in Section~\ref{subsec:hybrid_formulation}, we introduce a hybrid formulation that leverages the strengths of these approaches to speed up computation. 
    In the remainder of the paper, we use school choice as motivating example.

    \subsection{Stable Matching Problem}\label{subsec: stable matching problem}
        Let $\S=\{s_1,\ldots, s_n\}$ be the set of $n$ students, and let $\C=\{c_1,\ldots, c_m\}$ be the set of $m$ schools. To facilitate the exposition, we assume all students belong to the same grade, \eg, Pre-K. On the one side of the market, each school $c \in \C\cup \lrl{\emptyset}$ ranks the students that applied to it according to a strict order \(\succ_c\). Moreover, we assume that each school \(c\in \C\) has a capacity \(q_c\in\ZZ_+\). 
        When clear from the context, we extend the set of schools to include the outside option of being unassigned, which we denote by $\emptyset$, and we assume that its capacity is sufficiently high (\eg, $q_\emptyset = \lra{\S}$) so every student can potentially be unassigned. Finally, throughout the rest of the paper, we use bold notation to denote vectors/matrices of elements (parameters and variables) and italic notation for one-dimensional ones, \eg, $\mathbf{q} = \lrl{q_c}_{c\in \C\cup \lrl{\emptyset}}$.
        
        On the other side of the market, each student \(s\in \S\) has a strict preference order \(\succ_s\) over the elements in \(\C \cup \lrl{\emptyset}\). Then,
        we use \(c \succ_s c'\) to represent that student $s$ \emph{prefers} school $c$ over school $c'$ and \(\emptyset \succ_s c\) to denote that $s$ prefers to be unassigned over attending school $c$. 
        With a slight abuse of notation, we also refer to $\succ_s$ as student $s$'s preference list, and we assume that it includes all schools that $s$ \emph{weakly} prefers over being unassigned, including the latter option at the bottom of the list. Furthermore, we use $\lra{\succ_s}$ to represent the number of schools that $s$ prefers over being unassigned and $c'\succeq_s c$ to represent that either $c'\succ_s c$ or that $c' =_s c$ (\ie, $s$ is indifferent between $c$ and $c'$).
        Finally, let \(r_{s,c}\) be the position of school \(c\in \C\) in the  preference list of student \(s\in \S\) and \(r_{s,\emptyset}\) be a parameter that represents a penalty for having student \(s\) unassigned.\footnote{For instance, $r_{s,c} = 1$ implies that $c$ is student $s$'s top preference, \ie, $c\succ_s c'$ for all $c'\in C\cup \lrl{\emptyset}\setminus \lrl{c}$. Also, note that the penalty \(r_{s,\emptyset}\) may be different from the position of $\emptyset$ in the preference list of student $s$.}
    
        Let \(\E\subseteq \S\times \big(\C\cup \left\{\emptyset\right\}\big)\) be the set of feasible pairs, with \((s,c) \in \E\) meaning that student \(s\) includes school $c$ in their preference list.
        A \emph{matching} is an assignment \(\mu\subseteq \E\) such that each student is assigned to one school in $\C\cup \left\{\emptyset\right\}$, and each school $c$ receives at most \(q_c\) students. We use \(\mu(s) \in \C\cup \lrl{\emptyset}\) to represent the school of student \(s\) in the assignment \(\mu\), with \(\mu(s) = \emptyset\) representing that \(s\) is unassigned in \(\mu\).
        Similarly, we use \(\mu(c)\subseteq \S\) to represent the set of students assigned to \(c\) in \(\mu\). We say that a school $c \in \C$ is \textit{fully-subscribed} in matching \(\mu\) if $|\mu(c)| = q_c $; otherwise, we say that school \(c\) is \textit{under-subscribed}. A matching \(\mu\) is stable if it has no \emph{blocking pairs}, \ie, no pair \((s,c) \in \E\) would prefer to be assigned to each other compared to their current assignment in \(\mu\).
        Formally, we say that \((s,c)\) is a {blocking pair} if the following two conditions are satisfied: (i) student $s$ prefers school $c$ over $\mu(s) \in \C\cup\lrl{\emptyset}$, and (ii) $|\mu(c)|<q_c$ or there exists \(s'\in \mu(c)\) such that \(s\succ_c s'\), \ie, \(c\) prefers \(s\) over \(s'\).

        An instance of the school choice problem can be fully described as \(\Gamma = \langle \S,\C,\mathbf{\succ}, \mathbf{q} \rangle\), where $\mathbf{\succ} = \lrl{\succ_s}_{s\in \S} \cup \lrl{\succ_c}_{c\in \C\cup \lrl{\emptyset}}$ and $\mathbf{q} = \lrl{q_c}_{c\in \C\cup \lrl{\emptyset}}$.
        \citet{gale1962college} showed that the student-proposing Deferred Acceptance algorithm can find the unique student-optimal stable matching---\ie, the matching that is \emph{weakly} preferred by all students among all stable matchings---in polynomial time for any instance $\Gamma$.\footnote{DA can be adapted to find  the \emph{school-optimal} stable matching, \ie, the unique stable-matching that is weakly preferred by all schools. In Appendix~\ref{app:DA}, we formally describe the DA version that finds the student-optimal stable matching.}

    \subsection{Capacity Expansion}\label{sec:capacity_expansion}
        Let $\Gamma = \langle \S,\C,\mathbf{\succ}, \mathbf{q} \rangle$ be an instance of the school choice problem described in the previous section, and suppose that the clearinghouse aims to allocate a budget $B\in\ZZ_+$ of additional seats. We define the \emph{capacity expansion problem} as finding a vector of additional seats \(\mathbf{t} = \lrl{t_c}_{c\in \C}\in\ZZ_+^{\C}\) that complies with the available budget (\ie, $\sum_{c\in \C} t_c \leq B$) and leads to a student-optimal matching in the instance with expanded capacities $\Gamma_{\mathbf{t}} = \langle \S,\C,\mathbf{\succ}, \mathbf{q} + \mathbf{t} \rangle$.\footnote{Note that \(\Gamma_\mathbf{0}\) corresponds to the original instance $\Gamma$ with no capacity expansion.} We formalize this in Problem~\ref{def: capacity expansion problem}.

        \begin{problem}[Capacity Expansion Problem]\label{def: capacity expansion problem}
            Given an instance $\Gamma = \langle \S,\C,\mathbf{\succ}, \mathbf{q} \rangle$ and a budget $B$, the {capacity expansion problem} is the following:
            \begin{equation}\label{problem_def}
            \min_{\mathbf{t}, \mu} \Bigg\{ \sum_{(s,c)\in \mu} r_{s,c} \; : \; \sum_{c\in \C}t_c\leq B, \; \mu \text{ is a stable matching in instance} \ \Gamma_{\mathbf{t}} \;, \mathbf{t}\in\ZZ_{+}^{\C}
             \Bigg\}.
            \end{equation}            
        \end{problem}

        Both $\mathbf{t}$ and $\mu$ are decision variables in Problem~\ref{def: capacity expansion problem}. The first constraint guarantees that the former satisfies the budget, while the second constraint ensures that the latter is a stable matching in the expanded instance $\Gamma_{\mathbf{t}}$. The third constraint establishes the integrality of $\mathbf{t}$. Finally, the objective is to minimize the sum of the preferences of assignment and the penalty of being unassigned. As we show in Lemma~\ref{lemma: student optimal stable matching with lp}, this objective function ensures that we obtain a student-optimal matching in the expanded instance. We defer the proof to Appendix~\ref{app:missing_proofs_preliminaries}.
        % Note that the choice of the objective function is not arbitrary as it allows us to find an allocation of extra capacities that leads to the best possible student-optimal stable matching among all feasible assignments that result from the ``expanded'' instances. We formalize this in Lemma~\ref{lemma: student optimal stable matching with lp} and we provide its proof in Appendix~\ref{app:missing_proofs_preliminaries}.    
        \begin{lemma}\label{lemma: student optimal stable matching with lp}
          Fix $\mathbf{t}\in\ZZ_+^{\C}$. Then, minimizing $\sum_{(s,c)\in \mu} r_{s,c}$ over the space of stable matchings $\mu$ in $\Gamma_{\mathbf{t}}$ is equivalent to finding the unique student-optimal stable matching in $\Gamma_{\mathbf{t}}$.
        \end{lemma}
        
        Note that an optimal solution of Problem~\ref{def: capacity expansion problem} always exists (\eg, $\mathbf{t} = 0$ and $\mu$ obtained from using student-proposing DA in the instance $\Gamma_0 = \Gamma$ are a feasible solution) but may not be unique, as we show in Appendix~\ref{app:example_multiple_optimal_solutions}. Moreover, the optimal solution may not use the entire budget. For instance, this would hold if every student is assigned to their top choice in the initial instance $\Gamma$.
        
        \begin{remark}
            It is important to highlight that $\mathbf{\succ} = \lrl{\succ_s}_{s\in \S} \cup \lrl{\succ_c}_{c\in \C\cup \lrl{\emptyset}}$ is an input of Problem~\eqref{problem_def}; thus, capacity decisions respond to students' preferences and schools' orderings. This assumption matches Chile's current practice and is suitable in settings involving short time horizons, \eg, adding seats in a course/grade/program, merging neighboring schools, or re-organizing classrooms to accommodate different courses based on their popularity. Our model could also accommodate longer-term policies that go beyond the specific preferences of a cohort of students (\eg, building an entirely new school building) if the number of applicants and their preferences are consistent over time. Finally, note that our framework is flexible enough to accommodate other objectives beyond student optimality, such as minimizing implementation (\eg, adding teachers, portable classrooms, etc.) or transportation costs, students' estimated welfare, or any combination of goals.
        \end{remark} 
    
    \subsection{Compact Formulation}\label{subsec:compact_formulation}
        One natural approach to solve Problem~\ref{def: capacity expansion problem} is to generalize the formulation of the stable matching problem studied in~\citep{gusfield1989stable, vate1989linear, baiou2000stable}, incorporating the decision vector \(\mathbf{t}\). This na\"ive approach results in the following integer quadratically constrained program:
        \begin{subequations}\label{IntMathProgr}
            \begin{alignat}{2} %(\maxhrtcap) \qquad
            \min_{\mathbf{x},\mathbf{t}} \quad & \sum_{(s,c)\in \E} r_{s,c}\cdot x_{s,c}  \label{obj-fun.1a}\\
             s.t. \quad & (t_c +q_c) \cdot \Bigg( 1-\sum_{c'\succeq_s c} x_{s,c'}  \Bigg)  \leq  \sum_{s'\succ_c s} x_{s',c}, \qquad \forall \ (s,c)\in \E, c \succ_s \emptyset \label{constraint.1b}\\
              \quad & (\mathbf{x},\mathbf{t})\in\P_{\ZZ}, \label{constraint.1c}
            \end{alignat}
        \end{subequations}
        where the decision variables $x_{s,c}$ indicate whether student $s$ is assigned to school $c$, and $t_c$ represents the number of additional seats added to school $c$. The family of constraints in~\eqref{constraint.1b} guarantees that the matching $\mathbf{x}$ is stable in $\Gamma_{\mathbf{t}}$ for any fixed $\mathbf{t}$. Moreover,
        $\P_{\ZZ}$ is the set of integer points of 
        \begin{equation}\label{eq: feasible region base model}
            \P = \Bigg\{(\mathbf{x},\mathbf{t})\in[0,1]^{\E}\times [0,B]^{\C} \colon \ \sum_{c: (s,c)\in \E }x_{s,c} =  1  \ \forall \ s\in  \S, \
            \sum_{s: (s,c)\in \E} x_{s,c} \leq q_c+t_c \ \forall \ c\in  \C,  \ \sum_{c\in  \C}t_c\leq B \Bigg\},
        \end{equation}        
        which ensures the feasibility of the assignment under the expanded capacities, \ie, that (i) every student is matched to exactly one school (including $\emptyset$), (ii) every school receives at most $q_c+t_c$ students, and (iii) the budget is not exceeded. 
        
        Formulation~\eqref{IntMathProgr} may seem attractive due to the polynomial number of variables and constraints. However, directly solving it with an off-the-shelf solver remains challenging because of (i) the non-convexity of the quadratic constraints in~\eqref{constraint.1b} and (ii) the integrality of the decision variables. 
        To address the former, we propose a linear reformulation, obtained by replacing~\eqref{constraint.1b} with
        \begin{equation}\label{eq:L_constraints}
            t_c + q_c -(B +q_c) \cdot \sum_{c'\succeq_s c} x_{s,c'}    \leq  \sum_{s'\succ_c s} x_{s',c}, \qquad \forall \ (s,c)\in \E, c \succ_s \emptyset. \tag{2d}
        \end{equation}
        As we show in Lemma~\ref{lemma:validity_L_constraints}, this reformulation is a valid formulation of Problem~\ref{def: capacity expansion problem}. In Appendix~\ref{app: compact formulation}, we provide the proof and discuss other more natural linear reformulations obtained via McCormick envelopes. However, preliminary computational results show that the family of constraints in~\eqref{eq:L_constraints}---henceforth referred to as L-constraints---performs better than those alternatives, so we focus on the implementation of the former in the remainder of the paper.
        \begin{lemma}\label{lemma:validity_L_constraints} The compact formulation with L-constraints, obtained by replacing~\eqref{constraint.1b} with~\eqref{eq:L_constraints} in Formulation~\eqref{IntMathProgr}, is a valid formulation of Problem~\eqref{def: capacity expansion problem}.
            %Formulation~\eqref{IntMathProgr} remains valid when we replace~\eqref{constraint.1b} by~\eqref{eq:L_constraints}. 
        \end{lemma}

        To deal with the challenges imposed by the integrality of the decision variables, one natural approach is to allow variables $\mathbf{x}\in[0,1]^{\S\times\C}$ in~\eqref{IntMathProgr}. However, as observed by~\cite{baiou2000stable} in the base case (\ie, with $\mathbf{t} = 0$), the resulting formulation may not be valid, \ie, the optimal solution may be fractional. In the next section, we focus on devising a valid formulation for Problem~\ref{def: capacity expansion problem} that relaxes the integrality of the matching variables $\mathbf{x}$, while keeping that of capacity decisions (\ie,~$\mathbf{t}$).

    \subsection{Comb Formulation}\label{subsec:comb_formulation}
        To overcome the challenges involved with the integrality of $\mathbf{x}$, \cite{baiou2000stable} introduced a valid linear programming formulation that relies on an exponential number of inequalities, called \emph{comb constraints}, and also provided a separation algorithm that allows to efficiently solve the base problem (with $\mathbf{t}=0$) using a cutting plane algorithm. Since these are the building blocks for our exact proposed methods, we provide more background on the comb formulation in Section~\ref{subsec: background combs}. Then, in Section~\ref{subsec:cutting_plane}, we generalize it to incorporate capacity decisions. Finally, in Section~\ref{subsec: cutting plane algorithm for generalized comb formulation}, we discuss our cutting plane method to solve Problem~\ref{def: capacity expansion problem}.

        \subsubsection{Background.}\label{subsec: background combs}
            \citet{baiou2000stable} analyze the stable matching problem using a graph representation of $\Gamma$. In their \emph{admissions graph}, each node corresponds to a pair $(s,c)\in\E$, and there are two types of (directed) arcs: (i) horizontal, which go from a node $(s,c)$ to a node $(s',c)$ whenever $s'\succ_c s$; and (ii) vertical, which go from a node $(s,c)$ to a node $(s,c')$ whenever $c'\succ_s c$.\footnote{Arcs that are implied by transitivity are omitted. For example, if $c''\succ_s c'\succ c$, then the admissions graph only includes the arcs $((s,c),(s,c'))$ and $((s,c'),(s,c''))$.}
            
            Let $\E_c^+$ be the set of pairs $(s,c) \in \E$ such that $c$ prefers at least $q_c-1$ students over $s$ for $c \in \C$. For $(s,c) \in \E^+_c$, a \emph{shaft} with base $(s,c)$ is denoted by $S(s,c)$ and consists of $(s,c)$ and all pairs $(s',c)$ such that $s' \succ_c s$. A \emph{tooth} $T(s,c)$ with \emph{base} $(s,c) \in \E$ consists of $(s,c)$ and all pairs $(s,c')$ such that $c' \succ_s c$. Based on these elements, we can now define a comb. 
            \begin{definition}[Comb]\label{def: combs} 
                     For $(s,c) \in \E^+_c$, a comb with base $(s,c)$, denoted by $C(s,c)$, consists of the union of $S(s,c)$ and exactly $q_c$ teeth $T(s',c)$ with $(s',c) \in S(s,c)$, including $T(s,c)$. Furthermore, let $\mathbf{C}_c$ be the family of combs for school $c \in \C$. 
            \end{definition}  
            Note that combs can only be defined for schools that are over-subscribed, \ie, that have at least as many applicants as seats available. Additionally, there may be multiple combs for each base $(s,c)$ since the shaft can have more than $q_c$ elements and, therefore, we can select different subsets of $q_c-1$ teeth. 

            \begin{example}
                 In Figure~\ref{fig:comb_constraint}, we present a subset of the admissions graph of a given instance that includes all students who applied to $c$, where $q_c = 3$. Then, the set of feasible pairs that can be the base of a comb is $\E_c^+ = \{ (s_3',c), (s_4',c), (s,c), (s_5',c)\}$. If we pick $(s,c)$ as the base of the comb, then we have shaft $S(s,c) = \{ (s,c), (s_1',c), (s_2',c), (s_3',c), (s_4',c)\}$. Finally, each comb based on $(s,c)$ must include $q_c$ teeth based in the shaft, one of which must be $T(s,c)$. In this case, there are $\binom{4}{2}$ possible teeth choices (and so is the number of combs based on $(s,c)$); Figure~\ref{fig:comb_constraint}
                 shows in blue the comb that results from selecting $T(s_2',c)$ and $T(s_3',c)$ as the two additional teeth. 

                 In Figure~\ref{fig:L_constraint}, we clarify the inspiration for the naming of the L-constraints. The terms on the left-hand side of~\eqref{eq:L_constraints} are highlighted in yellow, while the terms on the right-hand side are in red.
                 \hfill $\square$
                 %\input{OR/2nd_round/plots/comb_constraint}
                 % \documentclass{article}
% \usepackage{tikz}
% \usepackage{caption}
% \usepackage{subcaption}
% \usetikzlibrary{calc}
% \usepackage{xcolor}

% \begin{document}

\definecolor{royalblue}{rgb}{0.255, 0.412, 0.882}
\definecolor{canaryyellow}{rgb}{1, 0.937, 0.0}
\definecolor{darkyellow}{rgb}{0.8, 0.8, 0.0}
\begin{figure}[h!]
    \centering
 \begin{subfigure}[b]{0.5\textwidth}
        \centering
        \begin{tikzpicture}[node distance=1.5cm, auto]

        % Define the style for the nodes
        \tikzstyle{state}=[rectangle, draw, fill=gray!5, 
            text centered, minimum height=1em, minimum width=1em,font=\fontsize{7}{8}\selectfont]

        % Define the nodes
        \node[state] (v1) {$(s'_5,c)$};
        \node[state] (v2) [right of=v1] { $(s,c)$};
        \node[state] (v3) [right of=v2] {$(s'_4,c)$};
        \node[state] (v4) [right of=v3] {$(s'_3,c)$};
        \node[state] (v5) [right of=v4] {$(s'_2,c)$};
        \node[state] (v6) [right of=v5] {$(s'_1,c)$};

        \node[state] (v7) [above of=v2] {$(s,c'_2)$};
        \node[state] (v8) [above of=v4] {$(s'_3,c'_2)$};
        \node[state] (v9) [above of=v5] {$(s'_2,c'_2)$};
        \node[state] (v10) [above of=v8] {$(s'_3,c'_1)$};
        \node[state] (v11) [above of=v7] {$(s,c'_1)$};
        \node[state] (v12) [right of=v11] {$(s'_4,c'_2)$};

        % \node (comb1) [left of=v7] {\hspace{-2cm}\tiny ${\color{royalblue}\displaystyle \sum_{(s,c) \in C}x_{s,c}}
        % \geq \displaystyle q_c+ \sum_{k=1}^3 y_c^k$};

        % Draw the edges
        \draw[->] (v1) -- (v2);
        \draw[->] (v2) -- (v3);
        \draw[->] (v3) -- (v4);
        \draw[->] (v4) -- (v5);
        \draw[->] (v5) -- (v6);
        \draw[->] (v2) -- (v7);
        \draw[->] (v7) -- (v11);
        \draw[->, bend left] (v4) to (v10);
        \draw[->] (v10) -- (v8);
        \draw[->] (v5) -- (v9);
        \draw[->] (v3) -- (v12);

        % Draw the dashed ellipse around v2, v7, and v10
        \fill[royalblue, rounded corners, opacity=0.1] 
            ($(v2) + (-0.5, -0.5)$) -- ($(v7) + (-0.5, 0.5)$) -- ($(v11) + (-0.5, 0.5)$) -- ($(v11) + (0.5, 0.5)$) -- ($(v2) + (0.5, 0.5)$) -- ($(v4) + (-0.5, 0.5)$) -- ($(v10) + (-0.5, 0.5)$) -- ($(v10) + (0.5, 0.5)$) -- ($(v4) + (0.5, 0.5)$) -- ($(v5) + (-0.5, 0.5)$) -- ($(v9) + (-0.5, 0.5)$) -- ($(v9) + (0.5, 0.5)$) -- ($(v5) + (0.5, 0.5)$) -- ($(v6) + (-0.5, 0.5)$) -- ($(v6) + (0.5, 0.5)$) -- ($(v6) + (0.5, -0.5)$) --cycle;
        \draw[dashed, line width=1.5pt, royalblue, rounded corners] 
            ($(v2) + (-0.5, -0.5)$) -- ($(v7) + (-0.5, 0.5)$) -- ($(v11) + (-0.5, 0.5)$) -- ($(v11) + (0.5, 0.5)$) -- ($(v2) + (0.5, 0.5)$) -- ($(v4) + (-0.5, 0.5)$) -- ($(v10) + (-0.5, 0.5)$) -- ($(v10) + (0.5, 0.5)$) -- ($(v4) + (0.5, 0.5)$) -- ($(v5) + (-0.5, 0.5)$) -- ($(v9) + (-0.5, 0.5)$) -- ($(v9) + (0.5, 0.5)$) -- ($(v5) + (0.5, 0.5)$) -- ($(v6) + (-0.5, 0.5)$) -- ($(v6) + (0.5, 0.5)$) -- ($(v6) + (0.5, -0.5)$) --cycle;

        \end{tikzpicture}
        \caption{\tiny Student-school pairs in a comb with base $(s,c)$.} \label{fig:comb_constraint}
    \end{subfigure}
 % \vspace{1cm} % Space between the subfigures
    \begin{subfigure}[b]{0.5\textwidth}
        \centering
     \begin{tikzpicture}[node distance=1.5cm, auto]

        % Define the style for the nodes
        \tikzstyle{state}=[rectangle, draw, fill=gray!5, 
            text centered, minimum height=1em, minimum width=1em,font=\fontsize{7}{8}\selectfont]

        % Define the nodes
        \node[state] (v1) {$(s'_5,c)$};
        \node[state] (v2) [right of=v1] { $(s,c)$};
        \node[state] (v3) [right of=v2] {$(s'_4,c)$};
        \node[state] (v4) [right of=v3] {$(s'_3,c)$};
        \node[state] (v5) [right of=v4] {$(s'_2,c)$};
        \node[state] (v6) [right of=v5] {$(s'_1,c)$};

        \node[state] (v7) [above of=v2] {$(s,c'_2)$};
        \node[state] (v8) [above of=v4] {$(s'_3,c'_2)$};
        \node[state] (v9) [above of=v5] {$(s'_2,c'_2)$};
        \node[state] (v10) [above of=v8] {$(s'_3,c'_1)$};
        \node[state] (v11) [above of=v7] {$(s,c'_1)$};
        \node[state] (v12) [right of=v11] {$(s'_4,c'_2)$};

        % \node (l1) [left of=v7] {\hspace{-3cm}\tiny$q_c+\displaystyle\sum_{k=1}^By_c^k- (q_c+B){\color{darkyellow}\displaystyle \sum_{c' \succeq c}x_{s,c'}}\leq \color{red}\displaystyle \sum_{s' \succ s}x_{s',c}$};

         % Draw the edges
        \draw[->] (v1) -- (v2);
        \draw[->] (v2) -- (v3);
        \draw[->] (v3) -- (v4);
        \draw[->] (v4) -- (v5);
        \draw[->] (v5) -- (v6);
        \draw[->] (v2) -- (v7);
        \draw[->] (v7) -- (v11);
        \draw[->, bend left] (v4) to (v10);
        \draw[->] (v10) -- (v8);
        \draw[->] (v5) -- (v9);
        \draw[->] (v3) -- (v12);

        % Draw the dashed ellipse around v2, v7, and v10
        \fill[canaryyellow, rounded corners, opacity=0.1]
        ($(v2) + (-0.5, -0.5)$) -- ($(v7) + (-0.5, 0.5)$) -- ($(v11) + (-0.5, 0.5)$) -- ($(v11) + (0.5, 0.5)$) -- ($(v2) + (0.5, -0.5)$) -- cycle;
        \draw[dashed, line width=1.5pt, canaryyellow, rounded corners]
        ($(v2) + (-0.5, -0.5)$) -- ($(v7) + (-0.5, 0.5)$) -- ($(v11) + (-0.5, 0.5)$) -- ($(v11) + (0.5, 0.5)$) -- ($(v2) + (0.5, -0.5)$) -- cycle;
        \fill[red, rounded corners, opacity=0.1]
        ($(v3) + (-0.5, -0.5)$) -- ($(v6) + (0.5, -0.5)$) -- ($(v6) + (0.5, 0.5)$) -- ($(v3) + (-0.5, 0.5)$) -- cycle;
        \draw[dashed, line width=1.5pt, red, rounded corners]
        ($(v3) + (-0.5, -0.5)$) -- ($(v6) + (0.5, -0.5)$) -- ($(v6) + (0.5, 0.5)$) -- ($(v3) + (-0.5, 0.5)$) -- cycle;

        \end{tikzpicture}
        \caption{\tiny Student-school pairs appearing in the L-constraint for pair $(s,c)$. }\label{fig:L_constraint}
    \end{subfigure}  
    \caption{Example of a subset of the admissions graph when school $c$ has capacity 3.}
\label{fig:L_comb_constraint}

\end{figure}

% \end{document}

            \end{example}
            \citet{baiou2000stable} showed that, if $\mathbf{t} = 0$, the family of constraints in~\eqref{constraint.1b} can be replaced by 
            \begin{equation}\label{eq:BB_comb_constraint}
                \sum_{(s',c')\in C} x_{(s,c)}\geq q_c, \hspace{5em} \forall\; c\in\C, \ C\in \mathbf{C}_c. 
            \end{equation}
            To see why these two families of constraints are equivalent, let $(s,c)\in \E$ be such that $x_{s,c} = 0$. The stability constraints in~\eqref{constraint.1b} guarantee that either (i) $s$ is assigned to something more preferred, \ie, $\sum_{c'\succ_s c} x_{s,c'} = 1$, or (ii) there are at least $q_c$ students with higher priority than $s$ currently assigned to $c$, \ie, $\sum_{s'\succ_c s} x_{s',c} \geq q_c$. If $\sum_{(s',c)\in S(s,c)} x_{s',c} \geq q_c$, then it is direct that (ii) holds by definition of the shaft. Conversely, if $\sum_{(s',c)\in S(s,c)} x_{s',c} < q_c$,~\eqref{eq:BB_comb_constraint} and the definition of the combs imply that each pair $(s',c)\in S(s,c)$ for which $x_{s',c} = 0$ is such that $\sum_{c'\succ_{s'} c} x_{s',c'} = 1$ and, thus, (i) holds for $(s,c)$.
            
            The key result in~\citet{baiou2000stable} is that the polytope $\left\{\mathbf{x}: \ (\mathbf{x},\mathbf{0})\in\P, \ \mathbf{x} \text{ satisfies } \eqref{eq:BB_comb_constraint} \right\}$ that allows for fractional solutions coincides with the convex hull of all stable matchings and, thus, the polytope is integral. Moreover, the authors showed that the constraints in~\eqref{eq:BB_comb_constraint} can be separated in polynomial time, which aligns with the result on the equivalence between optimization and separation~\citep{gls81}.
            
        \subsubsection{Generalized Comb Formulation.}\label{subsec:cutting_plane}
            For $k \in \ZZ_+$ and $c \in \C$, let $\E_c^+(k)$ be the set of pairs $(s,c) \in \E$ such that $c$ prefers at least $q_c+k-1$ students over $s$. In Definition~\ref{def:generalized_combs}, we extend Definition~\ref{def: combs} to allow for extra capacities. 
            % \alfredo{I deleted this example, because it is not clear -- to be discussed.}
            % \delete{One can easily construct an example in which the family of vanilla comb constraints~\eqref{eq:BB_comb_constraint} are not valid for Problem~\eqref{problem_def}; we refer to Appendix~\ref{appendix:counterexample_comb_formulation} for an example. Therefore, we must appropriately adapt the definition of a comb to our setting and define the family of constraints by using additional decision variables.} 
            % The comb constraints~\eqref{eq:BB_comb_constraint} do not directly generalize to the capacity planning setting, since the comb structure depend on the capacity of each school, which is a variable in our problem. An example displaying such a critical issue can be found in Appendix~\ref{appendix:counterexample_comb_formulation}.
            % To address this problem, 
                %To accomplish this, we start by generalizing the definition of comb.
                % \begin{definition}
            \begin{definition}[Generalized Combs]\label{def:generalized_combs}
                     For $(s,c) \in \E^+_c(k)$ and $k \in \ZZ_+$, a \emph{generalized comb} with base $(s,c)$, denoted by $C_k(s,c)$, consists of the union of the shaft $S(s,c)$ and exactly $q_c+k$ teeth $T(s',c)$ with $(s',c) \in S(s,c)$, including $T(s,c)$. 
                     Furthermore, let $\mathbf{C}_c(k)$ be the family of combs for school $c \in \C$ with $k$ extra seats.
            \end{definition}
            
            Given Definition~\ref{def:generalized_combs}, note that one could na\"ively attempt to design a formulation with generalized comb constraints in the form of~\eqref{eq:BB_comb_constraint} fixing the right-hand side to $q_c+k$ for each school $c \in \C$, $k\in \ZZ_+$ and combs in $\mathbf{C}_c(k)$. However, this may lead to infeasible or suboptimal solutions because the student-optimal stable matchings of two expanded instances $\Gamma_{\mathbf{t}}$ and $\Gamma_{\mathbf{t}'}$ for $\mathbf{t}\neq\mathbf{t}'$ may not coincide. 
            We address this by considering the capacity expansion variables in the right-hand side of the constraints (instead of fixing their value) and appropriately connecting them with the families of generalized combs constraints $\mathbf{C}_c(\cdot)$. This approach aligns with our goal of jointly deciding the extra capacity and the assignment of students to schools.
            Thus, we model the capacity expansion problem using the following mixed-integer programming formulation:
            \begin{subequations}\label{Baiou_Balinski-binary-expansion}
                \begin{alignat}{2}
                \min_{\mathbf{x},\mathbf{y}} \quad & \sum_{(s,c)\in \E} r_{s,c}\cdot x_{s,c}  \\
                 s.t. \quad & \sum\limits_{ (s',c') \in C }  x_{s',c'} \geq q_c + \sum_{j=1}^k y^j_c, \qquad  \forall \ c\in \C\,, \, \forall \  k=0,\ldots, B\,\,,  \forall \ C\in \mathbf{C}_c(k),   \label{constraint.binarization.stability}\\
                  \quad & (\mathbf{x},\mathbf{y})\in \P^{\text{ext}}, \label{constraint.binarization.polytope}
                \end{alignat}
            \end{subequations}
            where $\P^{\text{ext}}$ is the set 
            \begin{equation}\label{eq: feasible region generalized comb formulation}
                \begin{split}
                    \P^{\text{ext}} = \bigg\{(\mathbf{x},\mathbf{y})\in[0,1]^{\E} \times \{0,1\}^{\C\times \{1,\ldots,B\}} \colon 
                    & \sum\limits_{c: (s,c)\in \E}x_{s,c}=  1, \hspace{2cm} \forall \ s\in  \S, \\
                    & \sum\limits_{s:(s,c)\in \E} x_{s,c} \leq q_c+\sum_{k=1}^B y_c^k, \quad \forall \ c\in  \C,  \\ & \sum\limits_{c\in \C}\sum\limits_{k=1}^B  y_c^k\leq B, \\ 
                    & y_c^k \geq y_c^{k+1}, \hspace{2.8cm} \forall \ c \in \C, \; k=1,\ldots, B-1 \bigg\}.
                \end{split}
            \end{equation}
            The decision variable $y_{c}^k$ indicates whether school $c$ receives at least $k$ extra seats. 
            As a result, the vector $\mathbf{y}$ corresponds to the pseudo-polynomial description (or unary expansion) of an allocation vector $\mathbf{t}\in\ZZ_+^{\C}$ since $t_c = \sum_{k=1}^B y_c^k$ and, consequently, there is a one-to-one correspondence between the elements of $\P^{\text{ext}}$ and $\P$ for integral points $\mathbf{t}$. In line with~\eqref{eq: feasible region base model}, the first three families of constraints in~\eqref{eq: feasible region generalized comb formulation} ensure that each student is assigned to one school (including $\emptyset$), each school receives at most their expanded capacity, and the budget is not exceeded, respectively. The last family of constraints ensures the consistency of $\mathbf{y}$. In Theorem~\ref{thm: correctness of comb formulation}, we show the correctness of our generalized comb formulation; we defer the proof to Appendix~\ref{app:proof_of_thm_correctness_comb_formulation}.
            \begin{theorem}\label{thm: correctness of comb formulation} The generalized comb formulation,
                Formulation~\eqref{Baiou_Balinski-binary-expansion},
                is a valid formulation of Problem~\eqref{def: capacity expansion problem}.
            \end{theorem}

           \begin{remark}
                For each school $c \in \C$, only the generalized comb constraints for $k^\star = \max \lrl{k \in \ZZ_+: y_c^{k\star} = 1}$ are required to obtain a stable solution; see Lemmas~\ref{lemma: combs positive expansion} and~\ref{lemma: combs negative expansion} in Appendix~\ref{app: results_generalized_combs}. For instance, if school $c \in \C$ has capacity $q_c+k^\star$ in the optimal solution, we only need the constraints for the combs in $\mathbf{C}_c(k^\star)$. Note that this reduces considerably the number of comb constraints needed to ensure stability, although it requires knowledge about the optimal expansion $\mathbf{y}^\star$.    
           \end{remark}
           
        \subsubsection{Cutting-plane Algorithm.}\label{subsec: cutting plane algorithm for generalized comb formulation}
            Formulation~\eqref{Baiou_Balinski-binary-expansion} has exponentially many generalized comb constraints, so it cannot be solved directly with an off-the-shelf solver.\footnote{Same holds for the comb formulation in~\citet{baiou2000stable}.} For this reason, in Algorithm~\ref{alg: cutting-plane}, we present a cutting plane approach that relies on a separation algorithm to iteratively identify the most violated constraints (if any).\footnote{Note that in Formulation~\eqref{Baiou_Balinski-binary-expansion}, the decision vector $\mathbf{y}$ is binary. Therefore, our separation method works for $\mathbf{x}$ fractional and $\mathbf{y}$ binary. Thus, the fact that the separation runs in polynomial time does not guarantee that the cutting-plane method runs in polynomial time.} 
            Specifically, we start by solving a relaxation of~\eqref{Baiou_Balinski-binary-expansion} involving only a subset $\mathcal{J}$ of the comb constraints (in Step~\ref{step: solve MP}); note that $\mathcal{J}$ may be empty. 
            Then, for each school $c\in \C$, we run our separation algorithm that identifies the most violated constraint (if any) and adds it to the set of violated constraints $\mathcal{J}'$ (Steps~\ref{step: definition of C star} to~\ref{step:add new constraint}).
            If the solution $\mathbf{x}^\star$ is not stable, then $\mathcal{J}' \neq \emptyset$ and, thus, we add these constraints to the main program (Step~\ref{step:addConstr}) and resolve it, repeating the process. Conversely, if $\mathbf{x}^\star$ is stable and no additional comb constraints are added to the main problem (as $\mathcal{J}' = \emptyset$), the solution is stable and optimal (due to Theorem~\ref{thm: correctness of comb formulation}). Note that Algorithm~\ref{alg: cutting-plane} terminates in a finite number of steps since the set of comb constraints is finite. %Moreover, in each iteration (Steps~\ref{step: solve MP} to~\ref{step:addConstr}), a feasible solution is obtained, $\mathbf{y}^\star$ in Step~\ref{step: solve MP} and its corresponding stable matching in Step~\ref{step: definition of C star} (as explained in Appendix~\ref{app:separation_algorithm}), ensuring that stopping the algorithm before it proves optimality still provides a valid solution.

            \begin{algorithm}[htp!]
                \caption{Cutting-plane method}\label{alg: cutting-plane}
                \begin{algorithmic}[1]
                \Require  An instance $\Gamma = \langle \S,\C,\succ, \mathbf{q} \rangle$ and a budget $B$.
                \Ensure The student-optimal stable matching $\mathbf{x}^\star$ and the optimal vector of additional seats $\mathbf{t}^\star$.
                \State $\mathcal{J} \leftarrow$ subset of initial comb  constraints~\eqref{constraint.binarization.stability}   \label{step:J}
                \State $\lrp{\mathbf{x}^\star,\mathbf{y}^\star} \leftarrow \argmin  \lrl{\sum\limits_{(s,c)\in \E} r_{s,c}\cdot x_{s,c} \;:\; (\mathbf{x},\mathbf{y})\in \P^{\text{ext}}\cap \mathcal{J}}  $ \label{step: solve MP} %\Comment{Relaxation of Formulation~\eqref{Baiou_Balinski-binary-expansion}}
                \State $\mathcal{J}' \leftarrow \emptyset$
                \For{$c\in \C$}
                    \State $C^\star \in \argmin \biggl\{ \sum_{(s',c') \in C} x^\star_{s',c'}: \; C \in \mathbf{C}_c(t_c^\star) \ \text{where} \ t_c^\star= {\sum_{j=1}^{B}  y_c^{j\star}}  \biggr\}$\label{step: definition of C star}
                    \If{$\sum_{(s',c') \in C^\star} x^\star_{s',c'} < q_{c} + t^\star_c$}
                        \State $\mathcal{J}' \leftarrow \mathcal{J}' \cup \lrl{C^\star}$\label{step:add new constraint}
                    \EndIf
                \EndFor
                \If{$\mathcal{J}' \neq \emptyset$}
                \State $\mathcal{J} \leftarrow \mathcal{J} \cup \mathcal{J}'$ and go to Step~\ref{step: solve MP}. \label{step:addConstr}
                \Else
                \State Terminate and return $\mathbf{x}^\star$ and $\mathbf{t}^\star$ where $t_c^\star= {\sum_{j=1}^{B}  y_c^{j\star}}$ for every $c\in \C$.
                \EndIf
                % \State \Return 
                \end{algorithmic}
            \end{algorithm}

            We can exploit some structural properties of the objective function and the incumbent solution (discussed in Appendix~\ref{app:separation_algorithm}) to speed up the search for violated comb constraints. Specifically, for a given incumbent solution $(\mathbf{x}^\star,\mathbf{y}^\star)$, we can guide our search using two matchings: (i) the (potentially fractional and non-stable) matching given by $\mathbf{x}^\star$, and (ii) the feasible (but potentially suboptimal) stable matching $\mu^\star$ obtained using the DA algorithm over the instance $\Gamma_{\mathbf{t}^\star}$, where $t^\star_c=\sum_{j=1}^B y_c^{\star j}$ for all $c\in\C$. 
            As we show in Appendix~\ref{app:separation_algorithm}, we can restrict the search for violated comb constraints to schools $c\in \C$ such  that $\sum_{s:(s,c)\in\E}x_{s,c}^\star = |\mu^\star(c)|$ and $\lrl{s: x^\star_{s,c} > 0} \setminus \mu^\star(c) \neq \emptyset$. 
            The first condition guarantees that the school is fully subscribed, while the second ensures that there are students matched (at least fractionally) to school $c$ but for whom their assignment is not stable. 
            Considering schools that satisfy these two conditions allows us to focus on schools that certainly have a blocking pair, effectively reducing the search space and resulting in a efficient method. We formalize our method in Algorithm~\ref{alg: separation method} in Appendix~\ref{app:separation_algorithm}.

    \begin{remark}
        As we will show in Section~\ref{sec: computational time and heuristics}, we can speed up computation by combining the compact and comb formulations introduced in Sections~\ref{subsec:compact_formulation} and~\ref{subsec:comb_formulation}, respectively. %Specifically, we propose a \emph{hybrid} approach that starts with a main problem including combs for those pairs $(s,c)$ whose students have secured admission in the school (as they would get assigned with no additional budget) and L-constraints for all the other pairs. 
        % \margarida{I do not think this was what was implemented (the option initializing with combs was false in the code run in the server). Is this intended to be revised (I did not read Appendix~\ref{app: hybrid details})?  It should be: 
        Specifically, we propose a \emph{hybrid} approach with input parameter $\gamma$ that is exactly Algorithm~\ref{alg: cutting-plane}, except that the main program of Step~\ref{step: solve MP} has $\mathcal{J}$ replaced by L-constraints for all $(s,c)\in \E$ such that school $c$ ranks student $s$ as $q_c+\lfloor \gamma \cdot B +0.5 \rfloor$ or higher. Then, it separates infeasible solutions through the separation algorithm (Algorithm~\ref{alg: separation method}), \ie, it cuts unstable matchings by adding comb constraints. Appendix~\ref{app: hybrid details} thoroughly discusses this approach.
        % \nacho{I just moved what was explained in the simulation sections here without going over the details. I will check it.}
    \end{remark}

\section{Heuristics}\label{subsec:heuristics}
    Our toolkit for solving Problem~\ref{def: capacity expansion problem} includes not only exact formulations but also efficient heuristics that provide near-optimal solutions. In this section, we introduce two natural methods: (i) a greedy approach (\Greedy), and (ii) a linear programming-based heuristic (\LPheur). Both heuristics rely on the computation of a student-optimal stable matching, which can be achieved in polynomial time using the DA algorithm.
    A description of the DA {algorithm can be found} in Appendix~\ref{app:DA}. With a slight abuse of notation, we use $\mu_{\mathbf{t}}$ and $f(\mathbf{t})$ to denote the student-optimal matching and the corresponding value of the objective function that result from running DA in the instance $\Gamma_{\mathbf{t}}$ , \ie, 
    \begin{equation}\label{eq: function f to evaluate objective given t}
        f(\mathbf{t}) := \min_{\mu\subseteq \E}\left\{\sum_{(s,c)\in \mu} r_{s,c} : \ \mu \ \text{is a stable matching in} \ \Gamma_{\mathbf{t}} \right\}    .
    \end{equation}
    
    \paragraph{Greedy Approach.} Under this approach, we exploit the fact that the objective function is component-wise decreasing, \ie, for $\mathbf{t}\leq\mathbf{t}'$ component-wise, we have $f(\mathbf{t}')\leq f(\mathbf{t})$. {\Greedy} iteratively assigns an extra seat to the school with the best marginal improvement. Specifically, {\Greedy} performs at most $B$ sequential iterations---until exhausting the budget, each involving the allocation of one extra seat through two steps: (i) evaluating the objective function for each possible allocation of one extra seat using DA, and (ii) allocating the extra seat to the school leading to the lowest objective value. This is equivalent to look for the highest marginal $f(\mathbf{t})-f(\mathbf{t}+\mathbf{1}_c)$ where $\mathbf{1}_c\in\{0,1\}^{\C}$ denotes the indicator vector whose value is 1 in component $c\in\C$ and 0 otherwise.
    We formalize this heuristic in Algorithm~\ref{alg: greedy}. 

    \begin{algorithm}[htp!]
        \caption{\Greedy}\label{alg: greedy}
        \begin{algorithmic}[1]
            \Require An instance $\Gamma = \langle \S,\C,\succ, \mathbf{q} \rangle$ and a budget $B$.
            \Ensure A feasible allocation $\mathbf{t}$ and a stable matching $\mu_{\mathbf{t}}$ in the expanded instance $\Gamma_{\mathbf{t}}$.
            \State Initialize $\mathbf{t} \leftarrow \mathbf{0}$
            \While{$\sum_{c\in\C} t_c < B$}
                \State $c^\star \in \argmin\left\{f(\mathbf{{t}}+\mathbf{1}_c): \ c\in \C\right\}$, where $f$ is defined as in~\eqref{eq: function f to evaluate objective given t}.
        	\State $\mathbf{{t}}\leftarrow \mathbf{{t}}+\mathbf{1}_{c^\star}$
            \EndWhile
            \State Compute $\mu_{\mathbf{t}}$ using the DA algorithm over $\Gamma_{\mathbf{t}}$.
        \end{algorithmic}
    \end{algorithm}

    \paragraph{LP-based Heuristic.}
    Under this approach, we exploit the fact that  the optimal solution of $\min\Big\{\sum_{(s,c)\in\E} r_{s,c}\cdot x_{s,c}: \; (\mathbf{x},\mathbf{t})\in\P \Big\}$, when projected in $\mathbf{t}$, is integral since $\P$ has integral vertices in terms of $\mathbf{t}$.\footnote{This can be concluded by noting that $\P$ can be reformulated with flow conservation constraints.}
    Hence, {\LPheur} starts by solving this linear program. As a result, we obtain an allocation of extra seats $\mathbf{t}$ and an assignment $\mathbf{x}$ that may be fractional and non-stable. Nevertheless, since $\mathbf{t}$ is integral, {\LPheur} obtains a student-optimal stable matching using the DA algorithm in the instance with expanded capacities $\Gamma_{\mathbf{t}}$. We formalize this heuristic in Algorithm~\ref{alg:LPH}. Note that the solution obtained by {\LPheur} is equivalent to the solution generated from solving the problem in Step~\ref{step: solve MP} in Algorithm~\ref{alg: cutting-plane} considering $\mathcal{J} = \emptyset$.
    
    \begin{algorithm}[htp!]
        \caption{\LPheur}\label{alg:LPH}
        \begin{algorithmic}[1]
        \Require An instance $\Gamma = \langle \S,\C,\succ, \mathbf{q} \rangle$ and a budget $B$.
        \Ensure A feasible allocation $\mathbf{{t}}$ and a stable matching $\mu_{\mathbf{t}}$ in the expanded instance $\Gamma_{\mathbf{{t}}}$.
        \State Obtain $(\mathbf{{x}},\mathbf{{t}}) \in \argmin\left\{\sum_{(s,c)\in\E} r_{s,c}\cdot x'_{s,c}: \ (\mathbf{x}',\mathbf{t}')\in\P\right\}$.
        \State Compute the student-optimal matching $\mu_{\mathbf{{t}}}$ in $\Gamma_{\mathbf{{t}}}$ using the DA algorithm.
        \end{algorithmic}
    \end{algorithm}

%!TEX root = ./0_main.tex
\section{Application to School Choice in Chile}\label{sec: implementation in chile}

    As discussed in Section~\ref{sec:introduction}, the third goal of our work is to illustrate the potential benefits of our proposed methodology and devise some insights that could help guide a solution to the crisis currently affecting the Chilean school choice system. In Section~\ref{sec: background Chile}, we provide additional background on the Chilean school choice system and discuss why (i) it is a good application for our framework and (ii) a general enough setting to showcase the applicability of our approach. Then, in Section~\ref{sec: data and simulation setting}, we discuss the data and simulation setup in detail. Finally, in Section~\ref{sec:  simulation results}, we present our simulation results and discuss some key insights.  

    \subsection{Background}\label{sec: background Chile}
        Starting with the 2015 School Inclusion Law, every school that receives public funding (\eg, public and voucher schools) must use the centralized assignment system at the core of the ``Sistema de Admisión Escolar'' (SAE). After a scaling process, the new system reached its full implementation in 2020 and currently serves every year more than half a million students and eight thousand schools nationwide across all levels, from Pre-K to 12th grade.
        % The admissions process consists of two rounds: (i) the Main round, where every student that seeks admission to a non-private school (close to 90\% of the total enrollment) participates; and (ii) the complementary round, where all students who did not get assigned in the former can apply and seek admission in the schools that report seats left. In each of these rounds, students submit a strict preference order (no limit on the number of schools they can include), and schools sort their applicants based on a series of priorities and quotas, breaking ties among students belonging to the same group using a multiple tie-breaking rule at the family level, \ie, each family gets a random tie-breaker in each school they apply to. Then, using students' preferences and the schools' orderings, SAE runs a variant of the Deferred Acceptance algorithm. \delete{whereby grade levels are solved sequentially in decreasing order, \ie, starting from the highest level (12th grade),  SAE (i) solves the allocation for the current grade level, (ii) updates preferences and priorities, and (iii) moves on to the next grade level until solving the allocation for the lowest level (Pre-K).} 
        We refer to~\citep{Correa_2022} for a detailed description of the Chilean school choice system and the algorithm used to perform the allocation.

        The Chilean school choice system is a good application for our methodology for multiple reasons. First, the system's goal is to find a student-optimal stable assignment and, thus, uses a variant of the student-proposing Deferred Acceptance algorithm. Second, we can easily adapt our framework to incorporate all the features of the Chilean system, including the block application, the dynamic siblings' priority, etc. 
        Third, the Ministry of Education manages all schools participating in the system and, thus, can ask them to modify their vacancies within a reasonable range of variation.
        Finally, as discussed earlier, the system is under scrutiny due to the insufficient seats offered in many regions of the country. As a result, MINEDUC introduced a plan in 2022 aiming to increase school capacities and, in most cases, these capacity decisions are made after students submit their preferences and before performing the allocation of the main round, which is the exact timing assumed in our framework. Given the direct applicability of our work, we contacted MINEDUC and the NGO in charge of the implementation of the system, and we have been collaborating since then to devise solutions to mitigate the current crisis.

    \subsection{Data}\label{sec: data and simulation setting}
        We consider data from the 2023-2024 admissions process and, specifically, from students applying to Pre-K in the Antofagasta region during the main round.\footnote{All the data is publicly available and can be downloaded from this \href{https://datosabiertos.mineduc.cl/sistema-de-admision-escolar-sae/}{datosabiertos.mineduc.cl}.} As discussed in Section~\ref{sec:introduction}, we focus on the main round of the admissions process because it involves the highest share of students and seats---only 15\% of students in our sample participated in the complementary round, and the number of seats offered was close to 16\% of those available in the main round. Also, we restrict ourselves to Antofagasta because it is the region with the highest need of additional seats. We focus on a single level to speed up the computation, and we choose Pre-K because it is one of the largest entry levels with the lowest number of seats per applicant and assignment rate, as shown in Table~\ref{tab:general summary stats entry levels Antofagasta}.

        \begin{table}
            \caption{Summary Statistics: Entry Levels in Antofagasta region, Admissions Process 2023-2024} \label{tab:general summary stats entry levels Antofagasta}
            \centerline{\scalebox{0.85}{\begin{tabular}[t]{lcccccc}
            \toprule
            \multicolumn{5}{c}{ } & \multicolumn{2}{c}{Assignment} \\
            \cmidrule(l{3pt}r{3pt}){6-7}
            Level & Students & Schools & Seats & Seats per student & Total & Rate\\
            \midrule
            Pre-K & 3795 & 71 & 3558 & 0.938 & 2412 & 0.636\\
            K & 1443 & 85 & 1472 & 1.020 & 734 & 0.509\\
            1st & 3818 & 121 & 4829 & 1.265 & 2924 & 0.766\\
            7th & 1454 & 111 & 1938 & 1.333 & 1009 & 0.694\\
            9th & 5440 & 67 & 4961 & 0.912 & 3788 & 0.696\\
            \bottomrule
            \end{tabular}}}
        \end{table}

        For all simulations conducted, we use the actual preferences reported by students during the main round. 
        Additionally, our simulations consider the seat availability reported by each school during the main round, the prioritization established by the priority groups, and the actual lotteries drawn. To simplify the exposition,  we omit the siblings' priority scheme used in the actual admission system (for more details, see~\citep{Correa_2022, rios24, bobbio2024dynamic}), and we also do not consider the quotas for students with special needs and those for disadvantaged and high-achieving students. Finally, we perform all our simulations for two possible values of the penalty for unassigned students, namely, $r_{s,\emptyset} \in \lrl{\lra{\succ_s} + 1, \lra{\C}+1}$ for all $s\in \mathcal{S}$. As will be detailed in Section~\ref{subsec: properties of mechanism}, these penalty values allow us to measure the trade-off between \emph{access}, \ie, prioritizing unassigned students by using a hefty penalty ($r_{s, \emptyset} = \lra{\C}+1$), and \emph{improvement}, \ie, using each additional seat to reach the longest chain of improvements and benefit the largest share of students by using a minor penalty (\(r_{s,\emptyset} = \lra{\succ_s} + 1\)). Unless otherwise stated, all these simulations were solved considering a MipGap tolerance of $1e^{-4}$ and using the hybrid approach discussed at the end of Section~\ref{subsec:comb_formulation} with a time limit of ten hours. 
        % \margarida{I would always say ``hybrid approach'' and not ``hybrid formulation''. If we use the term ``hybrid formulation'', we need to prove why the formulation is correct and what is the approach to solve it. If we use the term ``hybrid approach'', where we are simply enriching the initial main program of the cutting-plane method with L-constraints, then the cutting-plane is trivially correct and there is nothing to prove. I think the latter simplifies the exposition.}

    \subsection{Results}\label{sec:  simulation results}
        In Section~\ref{sec: effect of budget and penalty}, we study the impact on students' outcomes when varying the budget of the extra capacities, including access and improvement with respect to their initial assignment. Then, in Section~\ref{sec: effect of bounds on expansion}, we analyze the effect of imposing bounds on the number of additional seats added to each school. In Section~\ref{sec: effect of preferences}, we provide a sensitivity analysis of the results to changes in students' preferences. Finally, in Section~\ref{sec: computational time and heuristics}, we compare the computing time of the proposed hybrid approach with other relevant benchmarks, and we analyze these results with respect to those obtained from the two heuristics discussed in Section~\ref{subsec:heuristics}.

        \subsubsection{Effect of Budget.}\label{sec: effect of budget and penalty}
\begin{figure}[htp!]
                \begin{subfigure}{0.49\textwidth}
                    \caption{\(r_{s,\emptyset} = \lra{\succ_s}+1\)}\label{fig: effect budget of students low penalty}
                    \includegraphics[width=\textwidth]{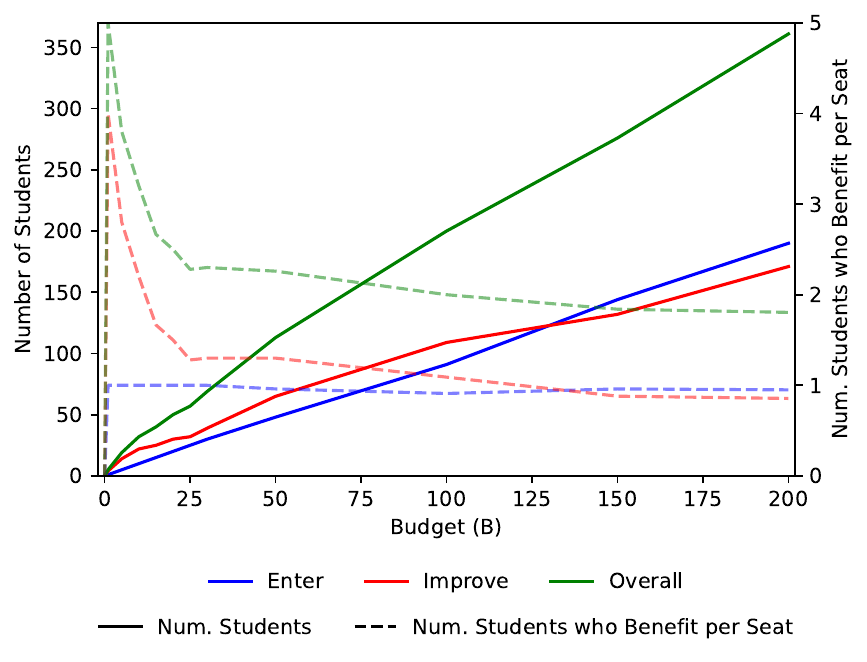}
                \end{subfigure}
                \begin{subfigure}{0.49\textwidth}
                    \caption{\(r_{s,\emptyset} = \lra{\C}+1\)}\label{fig: effect budget of students high penalty}
                    \includegraphics[width=\textwidth]{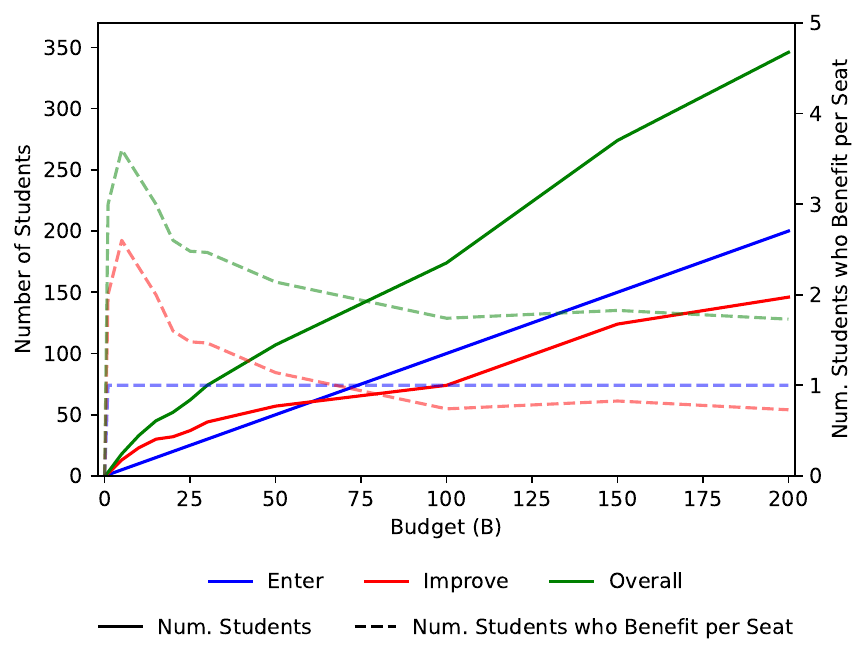}
                \end{subfigure}                
                \caption{Effect of the Budget on Students' Outcome}\label{fig: effect budget for students}
            \end{figure}

            In Figure~\ref{fig: effect budget for students}, we report simulation results showing the impact of the budget ($B \in \lrl{1, 2, 5, 10, 15, 20, 25, 30, 50, 100, 150, 200}$) on the number of students who (i) \emph{enter} the assignment, \ie, students who are initially unassigned (with \(B=0\)) and get assigned to one of their preferences under the expanded capacities; and who (ii) \emph{improve} their assignment, \ie, students who are initially assigned and get assigned to a more preferred school (according to their reported preferences) under the expanded capacities.  
            We also include in the analysis the \emph{overall} number of students who benefit from the extra capacity relative to the baseline (with $B=0$), which is equal to the sum of the number of students who \emph{enter} and \emph{improve}. 
            In each case, we report the total number of students who benefit (solid lines), and we also include the number of students who benefit per additional seat (dashed lines).
            Note that  \emph{enter}  captures the reduction in unassigned students. 
            
            First, we observe that the total number of students who benefit (in green) is greater than the number of additional seats and, thus, the overall number of students who benefit per additional seat is above one. However, we also observe that this rate decreases as the budget increases. For instance, when the penalty is low (Figure~\ref{fig: effect budget of students low penalty}), five students benefit per additional seat when the budget is low, while this decreases to approximately two when the budget is high. This result is intuitive, as the chains of improvement targeted with a low budget will likely be triggered when the budget is increased and, thus, the additional chains reach fewer students.
            
            Second, comparing Figures~\ref{fig: effect budget of students low penalty} and~\ref{fig: effect budget of students high penalty}, we observe that the total number of students who benefit per additional seat is greater under a low penalty if the budget is low. As we formally show in Section~\ref{subsec: properties of mechanism}, we know that each chain of improvement ends either in a under-demanded school (after benefiting some students) or in an unassigned student, with the latter chains being the primary target when the penalty is high. 
            % \margarida{We only know these from the next section. So we could indicate that this is formally shown in the next section.} 
            Then, since the former are more likely to cover a larger set of students, it is expected that a low penalty results in a greater number of students who benefit per seat. Nevertheless, we observe that the difference becomes negligible as we increase the budget, with both penalty values leading to an improvement of roughly two students per additional seat. 
            
            Finally, in line with the previous result, we observe that the number of students who enter the assignment per additional seat (blue dashed line) is equal to one when the penalty is high (Figure~\ref{fig: effect budget of students high penalty}), while it is slightly below one when the penalty is low. Since the budget provides an upper bound on the maximum number of students who can enter the assignment, this result suggests that considering a high penalty maximizes entry, consistent with our theoretical results in Section~\ref{subsec: properties of mechanism}.
            
            Overall, the results in this section suggest that each additional seat can improve the allocation of multiple students and that a high penalty for unassigned students can minimize the number of unassigned students. However, these results also show that ensuring a minimum of one seat per student in the system (which could be achieved with a budget $B=200$) does not guarantee that every student will get assigned, as there were initially 1,383 (=3,795-2,412 from Table~\ref{tab:general summary stats entry levels Antofagasta}) unassigned students. The most likely explanation lies around the high correlation between students' preferences and the small number of schools that students report in their preference list. Hence, MINEDUC should combine capacity planning with other information policies to expand students' consideration sets, encouraging them to apply to a more diverse set of schools.

        \subsubsection{Effect of Bounds on Expansion.}\label{sec: effect of bounds on expansion}
\begin{figure}[htp!]
                \begin{subfigure}{0.49\textwidth}
                    \caption{\(r_{s,\emptyset} = \lra{\succ_s}+1\)}\label{fig: effect of students low penalty}
                    \includegraphics[width=\textwidth]{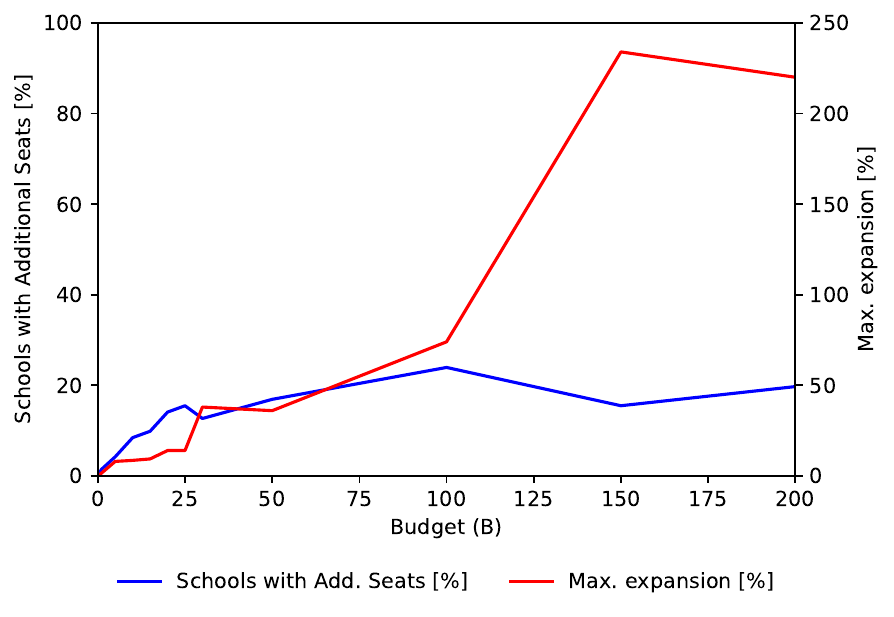}
                \end{subfigure}            
                \begin{subfigure}{0.49\textwidth}
                    \caption{\(r_{s,\emptyset} = \lra{\C}+1\)}\label{fig: effect of students high penalty}
                    \includegraphics[width=\textwidth]{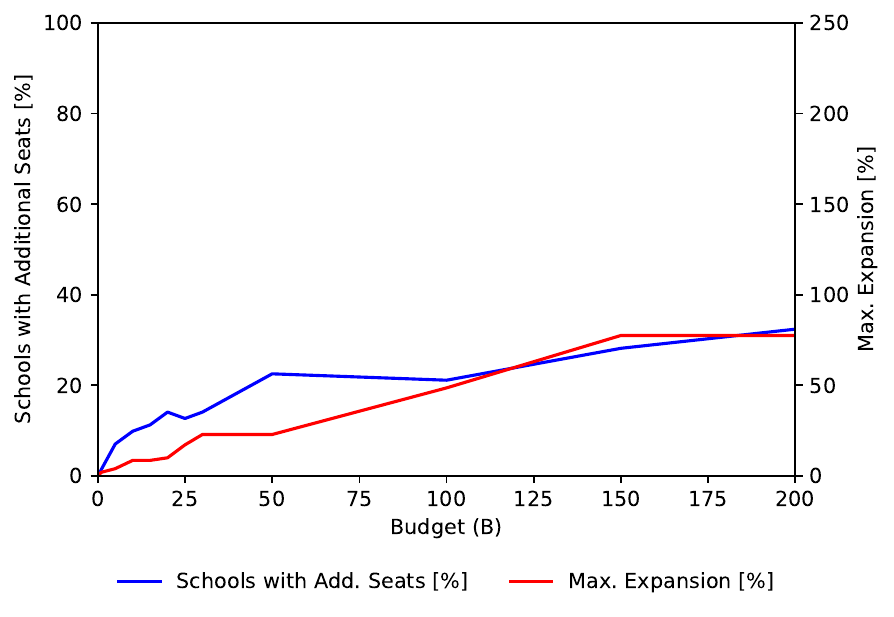}
                \end{subfigure}
                \caption{Effect for Schools}\label{fig: effect for schools}
            \end{figure}

            A valid concern raised during our discussions with MINEDUC and the NGO in charge of the allocation is that our approach would assign most extra seats to a few over-demanded schools if preferences are highly correlated. In fact, from  Figure~\ref{fig: effect for schools}, we observe that the maximum expansion relative to the original capacity (in red) is quite high for certain budgets (\eg, doubling the capacity of a school when the penalty is low, as shown in Figure~\ref{fig: effect of students low penalty} for $B=150$), while the fraction of schools that receive at least one additional seat (in blue) is relatively small (less than 20\%, as shown in Figure~\ref{fig: effect of students low penalty} for $B=150$). 
            The results for a high penalty (in Figure~\ref{fig: effect of students high penalty}) are along the same lines, although the maximum relative expansion is more attenuated; this may occur because the optimal solution aims to maximize the number of students who enter by adding at least one seat to more schools (roughly 32\% when $B=150$ in blue). These results confirm MINEDUC's concern. Additionally, MINEDUC has highlighted that some schools have limited flexibility to accommodate more students, such as those already operating in two shifts—serving different groups of students in the morning and afternoon—or those with space constraints that prevent the installation of modular classrooms.

\begin{figure}[htp!]
                \begin{subfigure}{0.49\textwidth}
                    \caption{Objective}\label{fig: effect bound on gap in obj}
                    \includegraphics[width=\textwidth]{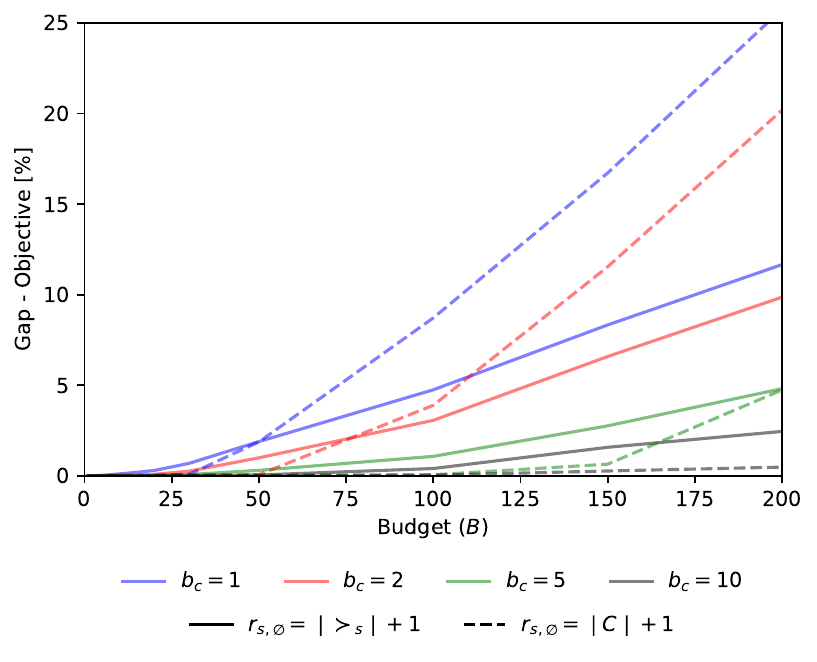}
                \end{subfigure}
                \begin{subfigure}{0.49\textwidth}
                    \caption{Entry}\label{fig: effect bound on gap in ent}
                    \includegraphics[width=\textwidth]{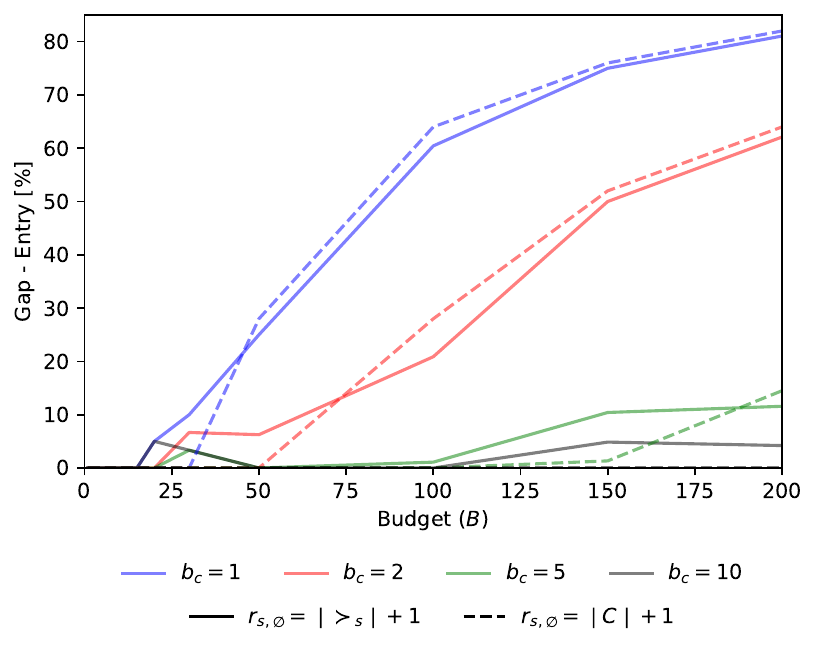}
                \end{subfigure}
                \caption{Effect of Bounds on Expansion}\label{fig: bounds on expansion}
            \end{figure}

            To overcome this practical limitation, we can easily adapt our framework by incorporating the new set of constraints
            \(
            t_c \leq b_c,
            \) for all \(c\in \C,\)
            where \(b_c\) is the maximum number of additional seats that can be allocated to school \(c\).\footnote{In Appendix~\ref{app: model extensions}, we discuss some model extensions to incorporate other relevant aspects from a practical standpoint.} 
            Let $\overline{\mathbf{t}}$ and $\mathbf{t}^\star$ be the optimal allocations of extra seats with and without these bounds, respectively.
            In Figure~\ref{fig: bounds on expansion}, we compare the optimal (unconstrained) solution with that obtained for different bounds (\(b_c \in \lrl{1, 2,5,10}\) for all \(c\in \C\)) focusing on (i) the gap in the objective function (Figure~\ref{fig: effect bound on gap in obj}), \ie, $(f(\overline{\mathbf{t}})-f(\mathbf{t}^\star))/f(\overline{\mathbf{t}})$; and (ii) the gap in the number of students that \emph{enter} the assignment (Figure~\ref{fig: effect bound on gap in ent}), \ie, $\lrp{\lra{\mu_{\mathbf{t}^\star}} - \lra{\mu_{\overline{\mathbf{t}}}}}/\lra{\mu_{\mathbf{t}^\star}}$, where $f(\mathbf{t})$ and $\mu_{\mathbf{t}}$ are as defined in Section~\ref{subsec:heuristics}.
            We include the gap in entry because MINEDUC's primary concern is reducing the number of unassigned students; thus, maximizing entry is one of the most relevant outcomes. 
            In both cases, the solid (resp. dashed) lines report the results obtained considering a low (resp. high) penalty for unassigned students, \ie, $r_{s,\emptyset} = \lra{\succ_s} + 1$ (resp. $\lra{\mathcal{C}}+1$) for all $s\in \mathcal{S}$.

            On the one hand, from Figure~\ref{fig: effect bound on gap in obj}, we observe that the gap in the objective function increases as we increase the budget. However, we also observe that this gap is relatively small when the upper bound is sufficiently high ($b_c \in \lrl{5,10}$) and when we prioritize \emph{access} over \emph{improvement} (\ie, $r_{s,\emptyset} = \lra{\mathcal{C}}+1$). The small gap in the latter case suggests that the problem has many near-optimal solutions and, thus, we can select one that does not over-expand some schools. 
            On the other hand, from Figure~\ref{fig: effect bound on gap in ent}, we observe that imposing tight limits (\ie, $b_c \in \lrl{1,2}$) can significantly reduce the number of students who enter the assignment, especially when the budget is sizable. Nevertheless, we also observe that considering $b_c = 10$ leads to no reduction in entry when coupled with a high penalty, while the gap in entry is minor (less than 5\%) with a low penalty.
            These results suggest that limiting expansions to ten seats per school results in a high-quality solution without overcrowding schools.

        \subsubsection{Effect of Preferences.}\label{sec: effect of preferences}
            Our methodology relies on students' reported preferences to optimally decide the allocation of extra seats to maximize the desired objective (which also depends on the penalty values $r_{s, \emptyset}$'s). The timing of capacity decisions matches the current practice in Chile, as MINEDUC has previously expanded capacities (i) after students apply to the main round but before performing the assignment or (ii) after the main round but before students apply to the complementary round. However, in other school districts, policy-makers may want to make capacity decisions before students submit their preferences.

\begin{figure}[htp!]
                \begin{subfigure}{0.49\textwidth}
                    \caption{Objective}\label{fig: effect prev pref on gap in obj}
                    \includegraphics[width=\textwidth]{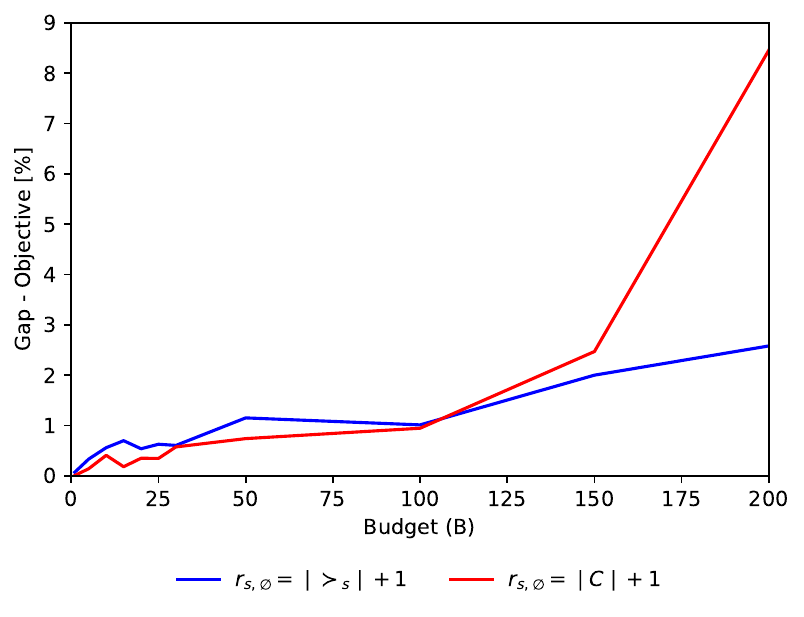}
                \end{subfigure}
                \begin{subfigure}{0.49\textwidth}
                    \caption{Entry}\label{fig: effect prev pref on gap in ent}
                    \includegraphics[width=\textwidth]{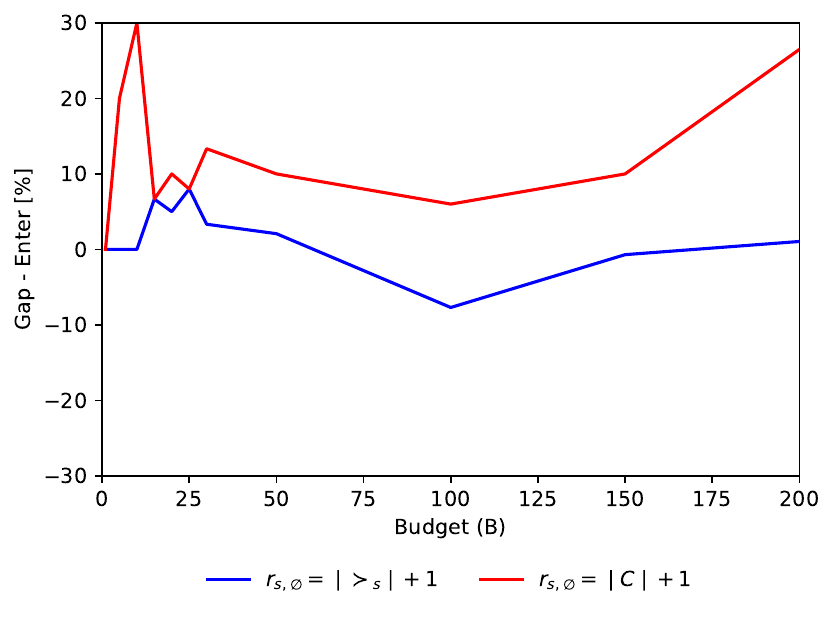}
                \end{subfigure}
                \caption{Effect of Using Previous Preferences}\label{fig: effect prev pref}
            \end{figure}

            If preferences across years are highly correlated and the number of applicants per year is stable (as it is the case in Chile), one approach would be to use the applicants and their realized preferences from the previous admissions process, and combine this information with the current set of schools and their capacities to decide the allocation of additional seats.
            In Figure~\ref{fig: effect prev pref}, we report the results evaluating this alternative approach. Specifically, for each budget $B$, we proceed in three steps that lead us to two solutions, $(\tilde{\mathbf{x}},\tilde{\mathbf{t}})$ and $(\mathbf{x}^\star,\mathbf{t}^\star)$. First, we obtain the optimal allocation of extra seats $\tilde{\mathbf{t}}$ considering (i) the students, their preferences, and the schools' orderings of the 2022-2023 admission process, and (ii) the school capacities of the 2023-2024 admission process. Second, we compute the student-optimal stable assignment $\tilde{\mathbf{x}}$ considering the instance of the 2023-2024 admission process (with its corresponding sets of students, schools, preferences and orderings) but assuming that capacities are $q_c + \tilde{t}_c$ for each school $c \in \mathcal{C}$. Finally, we directly obtain the optimal allocation of extra seats and assignment $(\mathbf{x}^\star,\mathbf{t}^\star)$ with the instance of 2023-2024.
            By comparing the outcomes obtained from  $(\tilde{\mathbf{x}},\tilde{\mathbf{t}})$ and $(\mathbf{x}^\star,\mathbf{t}^\star)$, we can assess the value of knowing students' actual preferences when making capacity decisions.

            In Figure~\ref{fig: effect prev pref}, we report the results comparing (i) the gap in the objective function (Figure~\ref{fig: effect prev pref on gap in obj}) and (ii) the gap in entry (Figure~\ref{fig: effect prev pref on gap in ent}) achieved with $(\tilde{\mathbf{x}},\tilde{\mathbf{t}})$ relative to $(\mathbf{x}^\star,\mathbf{t}^\star)$. On the one hand, from Figure~\ref{fig: effect prev pref on gap in obj}, we observe that using students' preferences from the previous admissions process to decide the allocation of extra seats leads to a gap in the objective that is (i) increasing in the budget, (ii) higher for high values of the penalty, and (iii) lower than 10\% for all budgets and penalties. These results suggest that the distribution of preference of assignment (which mainly affects the objective) is not significantly affected by whether we use the current or previous preferences to decide the allocation of extra seats.

            On the other hand, from Figure~\ref{fig: effect prev pref on gap in ent}, we observe that the gap in entry is small when the penalty is low, taking either positive or negative values depending on the budget. In contrast, when the penalty is high, we observe that the gap in entry is always positive and can be relatively high (up to 30\%) for some budget values. One possible explanation is that when the penalty is high, the optimal solution is tailored to ensure that students with particular circumstances (\eg, applying to rural schools, with short preference lists, etc.) get assigned, so any differences in these specific cases (from one year to the other) largely affect the effectiveness in entry. This is not the case when the penalty is low since the distribution of preferences, which is more stable across years compared to the particular cases mentioned above, is the primary driver of the allocation of additional seats. 

        \subsubsection{Methods' Computational Performance.}\label{sec: computational time and heuristics}
            One of the primary goals of the exact and heuristic methods introduced in Sections~\ref{sec:preliminaries} and~\ref{subsec:heuristics} is to speed up computation relative to the quadratic formulation described in~\eqref{IntMathProgr}, enabling us to handle realistically-sized instances. To evaluate their effectiveness and understand the drivers of improvement, we compare the quality of the solutions (gap) and the computational time required by each approach to solve the Antofagasta instance described previously for different budget values. Specifically, we compare these two elements when solving (i) the quadratic formulation (\emph{Quad}) and (ii) the compact formulation with L-constraints (\emph{Lin}) with an off-the-shelf solver, when running (iii)  Algorithm~\ref{alg: cutting-plane} for the generalized comb formulation (\emph{Comb}), (iv) the hybrid approach (\emph{Hybrid}), and (v) our two heuristics, \ie, \emph{Greedy} and \emph{LPH}. For (iii) and (iv), we use the initialization and acceleration tools for Algorithm~\ref{alg: cutting-plane} discussed in Appendix~\ref{app: computational details of cutting-plane method}.
            In Table~\ref{tab: comp formulations}, we report the running time and the gap achieved (in brackets) by each method. These results were obtained on an Intel(R) Xeon(R) Gold 6226 CPU clocked at 2.70GHz, using Gurobi~11.0.0 to solve optimization problems with code written in Julia~1.10.2. For each instance, the run of each method was restricted to a single thread and, for the exact methods (i) to (iv), a time limit of one hour and a MipGap tolerance of $10^{-4}$. 
            % \margarida{Maybe we should \textbf{not} write the mipgap tolerance in \% because the table does not provide it in \% which can be confusing for the reader}. 
            To facilitate the comparison across methods, we compute the gap with respect to the best solution obtained with \emph{Hybrid} within a time limit of ten hours and MipGap tolerance of $10^{-4}$ (\ie, the solutions used in the previous sections).  
            % \margarida{Double check previous sentence which was written based on slack discussion on the gap definition (I am not sure of the mipgap).} 
            
            \begin{table}[htp!]
                \caption{Methods Comparison}\label{tab: comp formulations}
                \centerline{\scalebox{0.85}{
                \begin{tabular}{ccccccc|cccccc}
                \toprule
                \multicolumn{1}{c}{} & \multicolumn{6}{c}{$r_{s,\emptyset} = \mid \succ_s \mid + 1$} & \multicolumn{6}{c}{$r_{s,\emptyset} = \mid \mathcal{C} \mid + 1$} \\
                \cmidrule(lr){2-7} \cmidrule(lr){8-13}
                $B$ & Quad & Lin & Comb & Hybrid & Greedy & LPH & Quad & Lin & Comb & Hybrid & Greedy & LPH  \\
                \midrule
                    0 & 24.916 & 12.294 & 6.478 & 5.741 & 0.008 & 0.730 & 25.362 & 10.256 & 4.368 & 5.158 & 0.007 & 0.482\\
                     & [0.000] & [0.000] & [0.000] & [0.000] & [0.000] & [0.000] & [0.000] & [0.000] & [0.000] & [0.000] & [0.000] & [0.000]\\
                    1 & 56.837 & 18.885 & 63.156 & 10.705 & 31.000 & 0.428 & 62.163 & 16.747 & 40.624 & 9.126 & 33.774 & 0.441\\
                     & [0.000] & [0.000] & [0.000] & [0.000] & [0.000] & [0.000] & [0.000] & [0.000] & [0.000] & [0.000] & [0.000] & [0.000]\\
                    2 & 38.474 & 21.602 & 94.173 & 21.774 & 63.451 & 1.447 & 40.415 & 16.880 & 80.941 & 16.788 & 78.872 & 0.466\\
                     & [0.000] & [0.000] & [0.000] & [0.000] & [0.000] & [0.001] & [0.000] & [0.000] & [0.000] & [0.000] & [0.000] & [0.000]\\
                    5 & 346.913 & 39.580 & 1499.293 & 23.489 & 124.519 & 0.459 & 206.549 & 37.673 & 916.264 & 17.103 & 134.592 & 0.446\\
                     & [0.000] & [0.000] & [0.000] & [0.000] & [0.000] & [0.002] & [0.000] & [0.000] & [0.000] & [0.000] & [0.000] & [0.000]\\
                    10 & 2549.262 & 121.820 & tl & 65.105 & 233.805 & 0.457 & tl & 61.807 & tl & 31.775 & 222.407 & 0.498\\
                     & [0.000] & [0.000] & [0.000] & [0.000] & [0.001] & [0.003] & - & [0.000] & [0.000] & [0.000] & [0.000] & [0.001]\\
                    25 & tl & tl & tl & 2118.371 & 611.750 & 0.994 & tl & 226.337 & tl & 201.855 & 592.394 & 1.750\\
                     & - & [0.000] & [0.002] & [0.000] & [0.002] & [0.003] & - & [0.000] & [0.000] & [0.000] & [0.000] & [0.003]\\
                    50 & tl & tl & tl & tl & 1114.221 & 0.718 & tl & tl & tl & 862.415 & 1404.633 & 0.624\\
                     & - & - & [0.005] & [0.000] & [0.008] & [0.007] & - & - & [0.001] & [0.000] & [0.001] & [0.001]\\
                    100 & tl & tl & tl & tl & tl & 3.357 & tl & tl & tl & tl & tl & 2.984\\
                     & - & - & [0.006] & [0.001] & [0.016] & [0.009] & - & - & [0.003] & [0.000] & [0.001] & [0.004]\\
                    150 & tl & tl & tl & tl & tl & 2.895 & tl & tl & tl & tl & tl & 2.101\\
                     & - & - & [0.012] & [0.009] & [0.023] & [0.023] & - & - & [0.009] & [0.000] & [0.002] & [0.010]\\
                    200 & tl & tl & tl & tl & 2906.369 & 3.373 & tl & tl & tl & tl & 2642.118 & 3.720\\
                     & - & - & [0.014] & [0.006] & [0.029] & [0.026] & - & - & [0.010] & [0.000] & [0.004] & [0.016]\\
                \bottomrule
                \end{tabular}             
                }}
                
                \vspace{0.2cm}
                \centerline{\begin{minipage}[t]{14.5cm}
                \footnotesize{\emph{Note.} For each method and budget value, we report the time needed to solve the problem (within the tolerance for exact methods) and the gap to the best objective value found in brackets. We use ``tl'' to denote instances not solved within the time limit and ``-'' to represent instances where a method did not find a feasible solution. 
                %\margarida{I checked the output file for Lin, $r_{s,\emptyset} = \mid \succ_s \mid + 1$ and $B=25$, and there is a MipGap of 0.006. However, the table presents no gap. I guess we would need to run this instance again for Lin to recover the best objective found by gurobi, command objective\_value(model), or we use the fact that in the output files we have the mipgap and Gurobi's lower bound to get the upper bound (best objective). For $B\geq 50$, the mipgap for Lin seems to be always 100\%, so no action to take. I verified manually that all other entries with gap reported as ``-'' were fine,\ie, the mipgap was 100\% :-)}
                }
                \end{minipage}}
            \end{table}

            First, note that we cannot solve the problem using \emph{Quad} for high budget values. Second, we observe that \emph{Lin} helps speed up computation relative to \emph{Quad} but fails to solve the instance for greater budgets and even to output a feasible solution on those cases. Third, we observe that the proposed methods involving our cutting plane method (\ie, \emph{Comb} and \emph{Hybrid}) find a feasible solution within the time limit for all the budgets considered. This is not surprising, since when the separation algorithm is called, a student-optimal stable matching for the current seat allocation is computed. Fourth, comparing \emph{Comb} and \emph{Hybrid}, we find that combining the two formulations helps reduce computing times and achieve lower gaps. Finally, as expected, we observe that \emph{LPH} is substantially faster than all the other methods considered, and its running time is unaffected by the budget, unlike all the other methods.
            These results suggest that our proposed methods are flexible enough to accommodate different needs. On the one hand, if time is not a concern and the main focus is to obtain a low optimality gap---which is the case when solving the allocation as the timeline of the process is such that there is almost a month between the deadline to submit preferences and the publication of the assignment results---, then \emph{Hybrid} may be a good approach. On the other hand, if speed is required---for instance, to perform simulations on different counterfactual policies---then \emph{LPH} becomes handy. It should be noted that, unlike the \emph{Comb} or \emph{Hybrid}, which can output optimality gaps (\ie, a measure of how far the method is from proving optimality), heuristics offer no guarantee of solution performance.

\begin{figure}[htp!]
                \begin{subfigure}{0.49\textwidth}
                    \caption{Gap - Objective}\label{fig: heuristics gap}
                    \includegraphics[width=\textwidth]{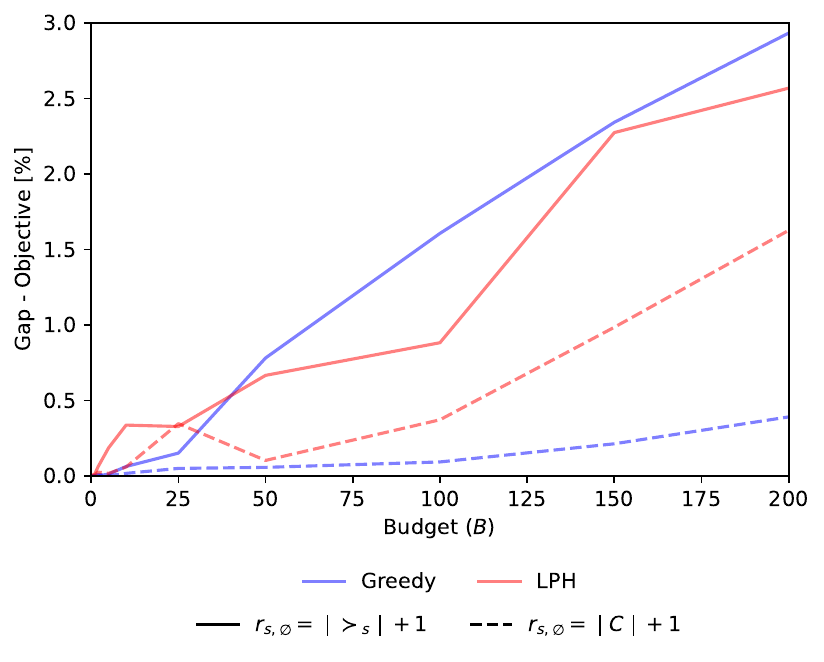}
                \end{subfigure}
                \begin{subfigure}{0.49\textwidth}
                    \caption{Gap - Entry}\label{fig: heuristics entry}
                    \includegraphics[width=\textwidth]{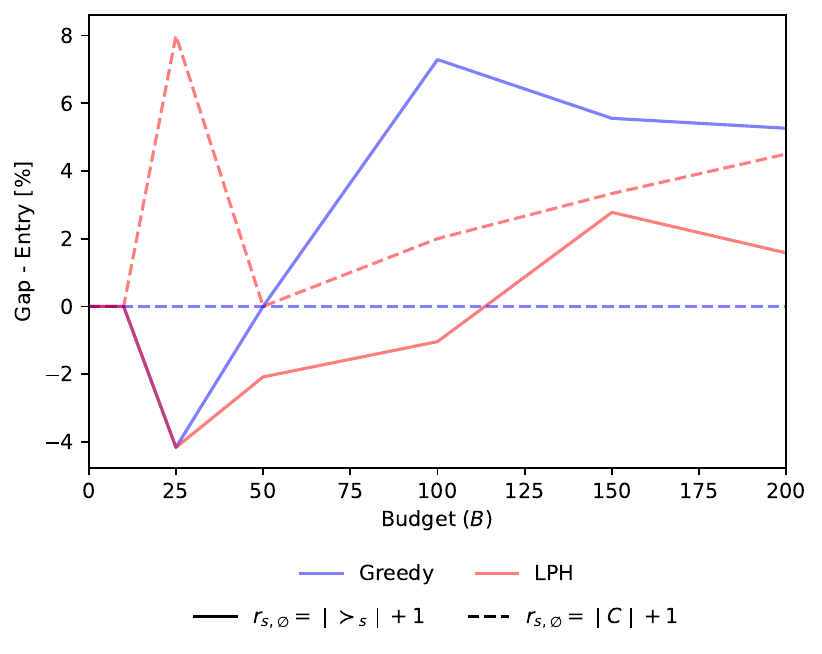}
                \end{subfigure}
                \caption{Heuristics}\label{fig: heuristics}
            \end{figure}

            To further compare our heuristics, we report (i) the gap in the objective (in Figure~\ref{fig: heuristics gap}) and (ii) the gap in entry (in Figure~\ref{fig: heuristics entry}) obtained relative to the best solution found.
            % \margarida{From where does this optimal solution come? From previous section? Note that in the previous sections that mipgap tolerance was higher.} 
            On the one hand, we observe that the gap in the objective is relatively small, remaining below 3\% for both heuristics and all budget values and penalty levels. Moreover, we observe that \emph{Greedy} outperforms \emph{LPH} when (i) the penalty is high and (ii) for low budget values when the penalty is low. 
            On the other hand, we observe that both methods achieve small gaps in entry, with \emph{LPH} leading to the best results when the penalty is low. These results, combined with the fact that \emph{LPH} swiftly returns a near-optimal solution, suggest that this heuristic can be a practical approach for large instances (involving high budgets) and time-sensitive tasks, especially when the goal is to minimize the number of unassigned students. Furthermore, as pointed out by MINEDUC, \emph{LPH} has the advantage that it does not rely on students' random tie-breakers, so it can be used to make capacity decisions in cases where it may require time to contact schools to request additional seats and, thus, delaying capacity decisions may not be feasible.
            
%!TEX root = ./0_main.tex

\section{Properties of the Mechanism}\label{subsec: properties of mechanism}

Now that we have devised a framework to solve Problem~\eqref{problem_def}, we briefly discuss some properties of the mechanism\footnote{Mechanism in this case stands for the optimal method that solves Problem~\eqref{problem_def}.} and the underlying optimal solutions. We provide a thorough discussion of these properties in Appendix~\ref{app: properties mechanism}.

    \paragraph{Cardinality.} In the standard setting with no capacity decisions, we know from the Rural Hospital theorem~\citep{roth1986allocation} that the set of students assigned in any stable matching is the same. This result no longer holds when we add capacity decisions, as the cardinality of the matching largely depends on the penalty values \(r_{s,\emptyset}\) of unassigned students. In Appendix~\ref{subsec:accessVsImpro}, we show that if these values are sufficiently small, then there exists an optimal solution of Problem~\ref{problem_def} whose allocation of capacities yields a student-optimal stable matching of minimum cardinality among all the possible student-optimal stable matchings (Theorem~\ref{thm: minimum cardinality}).\footnote{Interestingly, the minimum cardinality student-optimal stable matching is not necessarily the most preferred by the set of students initially assigned when \(B=0\) (see Example~\ref{app: min cardinality is not optimal for initially assigned students} in Appendix~\ref{app: properties mechanism}).}
    In contrast, if the penalty values are sufficiently high, we show that the optimal solution of the problem corresponds to a student-optimal stable matching of maximum cardinality (Theorem~\ref{thm:card_maximized}).

     %I deleted the examples of penalties; if the reader wants they can go to the appendix, and we reduce the saliency of the negative penalty that may be problematic.

    Note that this result is not surprising in hindsight, as the objective function in Problem~\eqref{problem_def} is the weighted sum of the students' preference of assignment and the value of unassigned students. Nevertheless, Theorems~\ref{thm: minimum cardinality} and~\ref{thm:card_maximized} are valuable from a policy standpoint, as they provide policymakers a tool to obtain an entire spectrum of stable assignments controlled by capacity planning where two extreme solutions stand out: (i) the solution that maximizes the number of assigned students (access), and (ii) the solution that allocates the extra seats to benefit the preferences of the students in the initial assignment (improvement). The former is a common goal in school choice settings, where the clearinghouse must guarantee a spot to each student that applies to the system, while the latter is common in college admissions, where merit plays a more critical role. Independent of the goal (or any intermediate point), our framework allows policymakers to achieve it by simply modifying the models' parameters resulting in a solution approach that is flexible and easy to communicate and interpret. %\fede{, indeed the penalty $r_{s,\emptyset}$ can be viewed as the social cost of having  student $s$ unassigned.} }

    \paragraph{Incentives.} A property that is commonly sought after in any mechanism is strategy-proofness, \ie, that students have no incentive to misreport their preferences in order to improve their allocation. \citet{roth1982economics} and \citet{dubins1981machiavelli} show that the student-proposing version of DA is strategy-proof for students in the case with no capacity decisions. Unfortunately, this is not the case when students know that there exists a budget of extra seats to be allocated, as we formally show in Appendix~\ref{subsec: incentives} (see Proposition~\ref{proposition_strategyproof_student}). Nevertheless, we also show that our mechanism is strategy-proof in the large~\citep{azevedo18}, which guarantees that it is approximately optimal for students to report their true preferences for any i.i.d. distribution of students' reports. As \citet{azevedo18} argue, this is a more appropriate notion of manipulability in large markets, as students are generally unaware of other students' realized preferences and priorities. Thus, the lack of strategy-proofness is not a major concern in our setting.\footnote{Note that \cite{afacan2024capacity} independently show that their mechanism is also manipulable, and in the special case of schools having the same preference list, they provide a mechanism that is efficient and strategy-proof.}

    \paragraph{Monotonicity.} Another commonly desired property in any mechanism is student-monotonicity, which guarantees that any improvement in students' priorities (in school choice) or scores (in college admissions) cannot harm their assignment.
    ~\cite{BalinskiSonmez1999} and~\cite{baiou2004student} show that this property holds for the student-proposing DA algorithm in the standard setting. However, in Appendix~\ref{app: monotonicity}, we show that students can be harmed when adding extra seats if their rank improves in a given school. Nevertheless, we can adapt our model to incorporate additional constraints that would rule out non-monotone allocations if this is a concern for policymakers. 
    
    % \fede{Providing a higher lottery to some groups of students is a standard affirmative action policy. Interestingly, our couterexample shows that using extra capacities for affirmative action is not a complementary policy to improving the lottery of students. This insight opens the question of leveraging extra capacities along with improvements in lottery scores to provide an efficient affirmative action policy.}
    % \alfredo{I am not familiar with monotonicity, so why is it harmful? Maybe we could add a small comment stating the implications of this negative result, basically how the students can unilateral improve their scores as these are centrally assigned?} \fede{Perhaps the question that I wrote before can be deleted so that we address it in the next paper about affirmative action in Chile. }
    
%    \nacho{I do not think we should add the complexity result that we previously had, because here we focus on properties of the mechanism. Maybe we can add it before in the manuscript.}\alfredo{agree} \fede{I think we can simply include the complexity results in the Appendix. I think it is important not to remove them.}\margarida{Maybe at the end of Section 3.1, we could say something as: Motivated by the theoretical complexity of our problem and the lack of potential to design approximation algorithms (See Appendix~\ref{subsec: complexity}), next, we concentrate on mathematical programming approaches.}
%    \alfredo{I would add that comment but not the submodular result in the appendix.}

%!TEX root = ./0_main.tex
\section{Conclusions}\label{sec: conclusions}
Motivated by the urgent needs of the Chilean school choice system, where insufficient capacity has led to a significant numbers of students being unassigned, we study the joint problem of capacity planning and stable matching in many-to-one matching markets. 
Specifically, we extend the standard stable matching model by integrating capacity decisions, allowing policymakers to prioritize different objectives, such as maximizing access or enhancing assignments for high-priority students, by modifying the penalty incurred for having unassigned students. We show that the problem can be modeled using a quadratically-constrained integer program. To overcome the challenges imposed by the non-linearity of constraints and by the intergrality of the matching variables, we provide two alternative formulations that exploit the properties of the problem, enabling us to design an efficient cutting plane algorithm. This approach relies on a novel separation algorithm aimed at enhancing the one introduced by~\citet{baiou2000stable}. Furthermore, we develop two efficient heuristics that enable us to find near-optimal solutions, with \emph{LPH} being the most suitable for for time-sensitive tasks and when information about lotteries may not be available.

Finally, we apply our framework to real-world data from the Pre-K level in the Antofagasta region of Chile and identify several key insights. First, we find that each additional seat benefits multiple students, but the marginal improvement diminishes as we increase the number of seats added to the system. Second, we observe that the penalty for unassigned students impacts the allocation, so policymakers can use it to prioritize different goals. However, we observe that the differences in access depending on the penalty decrease for greater budgets. Finally, we show that our framework is flexible enough to accommodate different needs, including school-specific budget constraints, the lack of information about school lotteries or students' current preferences, and other extensions.
Building on our positive results, we are collaborating with Chile's Ministry of Education (MINEDUC) and the NGO in charge of performing the allocation to test our framework in the field and develop solutions to the ongoing crisis. 

Overall, our results showcase how to extend the classic stable matching problem to incorporate capacity decisions and how to solve the problem effectively. Our methodology is flexible to accommodate different settings, such as capacity reductions, allocations of tuition waivers, quotas, secured enrollment, and arbitrary constraints on the extra seats per school (see Appendix~\ref{app: model extensions} for more details). Moreover, we can adapt our framework to other stable matching settings, such as the allocation of budgets to accommodate refugees, the assignment of scholarships or tuition waivers in college admissions, and the rationing of scarce medical resources. All of these are exciting new areas of research in which our results can be used.

% Appendix here
% Options are (1) APPENDIX (with or without general title) or
%             (2) APPENDICES (if it has more than one unrelated sections)
% Outcomment the appropriate case if necessary
%
% \begin{APPENDIX}{<Title of the Appendix>}
% \end{APPENDIX}
%
%   or
%

% Acknowledgments here
\ACKNOWLEDGMENT{This work was funded by FRQ-IVADO Research Chair in Data Science for Combinatorial Game Theory, and the NSERC grant 2019-04557.}

% References here (outcomment the appropriate case)

% CASE 1: BiBTeX used to constantly update the references
%   (while the paper is being written).

\bibliographystyle{informs2014} % outcomment this and next line in Case 1
\bibliography{bibliography} % if more than one, comma separated

% CASE 2: BiBTeX used to generate mypaper.bbl (to be further fine tuned)
%\input{mypaper.bbl} % outcomment this line in Case 2

%If you don't use BiBTex, you can manually itemize references as shown below.
\renewcommand{\theHchapter}{A\arabic{chapter}}
\begin{APPENDICES}
\section{Appendix to Section~\ref{sec:capacity_expansion}}

\subsection{Proof of Lemma~\ref{lemma: student optimal stable matching with lp}}\label{app:missing_proofs_preliminaries}

    We prove the theorem assuming that Problem~\ref{problem_def} has no extra capacities, \ie, $B=0$. Indeed, given $\textbf{t}$, the expanded instance $\Gamma_\textbf{t}$ can be seen as an instance of the school choice problem with $B=0$. Note that, if $B=0$, an optimal solution of Problem~\ref{problem_def} is of the form $(\textbf{0},\mu)$.
    
   % We begin by observing that  \citet{baiou2000stable} show that the feasible region of {\maxhrt} corresponds to the set of stable matchings. Therefore, in the following proof, we only need to focus on proving the equivalence between the student-optimal matching and a matching minimizing the sum of the students' rank over the set of stable matchings.

    Next, we prove that a student-optimal stable matching is a stable matching that minimizes the average rank of the students, and vice-versa.

    \textbf{If}: In a student-optimal stable matching, each student is assigned to the best school they could achieve in any stable matching \citep{gale1962college}. Thus, each unassigned student is unassigned in every stable matching. Moreover, by the Rural Hospital Theorem \citep{roth1982economics,roth1984evolution,GaleSotomayor1985,roth1986allocation}, the same students are assigned in all stable matchings. Suppose that $\mu$ is the student-optimal stable matching but it is not optimal to Problem~\ref{problem_def}.
    %$x$ be an optimal solution to the Integer Program~\eqref{eq: optimization to find stable matching for fixed capacities} and let $\mu=\{(s,c) \in \E: x_{sc}=1 \}$ be the corresponding student-optimal stable matching. Suppose that $\mu$ does not minimize the cost.
     Let $\mu'$ be an optimal solution to Problem~\ref{problem_def} (recall that $\mathbf{t}$ is fixed). Hence, the following inequality holds:
     $\sum_{(s,c) \in \mu} r_{s,c}> \sum_{(s,c) \in \mu'} r_{s,c}$. This means that there is at least one student $s'$ who prefers the stable matching $\mu'$ to the stable matching $\mu$, which is a contradiction.

    \noindent
    \textbf{Only if}: Let $\mu$ be a (stable) matching corresponding to an optimal solution of Problem~\ref{problem_def} (with $\mathbf{t}$ is fixed). As before, by the Rural Hospital Theorem, we observe that the set of students unassigned in $\mu$ is the same set of students unassigned in every stable matching. Hence, the set of assigned students in $\mu$, is the same set for every stable matching. Let us suppose, again by contradiction, that  $\mu$ is not a student-optimal stable matching. Let $\mu'$ be a student-optimal stable matching. Denote by $S'$ the set of students whose assignment to schools differs in the two matchings. By construction, the objective value of Problem~\ref{problem_def} for  $\S\setminus S'$ is the same in both $\mu$ and $\mu'$:
    $$\sum_{(s,c) \in \mu: \ s \in \S\setminus S'} r_{s,c}=\sum_{(s,c) \in \mu': \ s \in \S\setminus S'} r_{s,c}.$$
    Furthermore, $S'$ is the disjoint union of the following two sets of students: $S_1'$, the set of assigned students who prefer their school in $\mu'$, and $S_2'$, the set of assigned students who prefer their school in $\mu$. By \citet{gale1962college}, $\mu'$ is a stable matching in which all students are assigned the best school they could achieve in any stable matching. Therefore, the set $S_2'$ is empty. Hence, $S'=S_1'$, and by hypothesis \[\sum_{(s,c) \in \mu: \ s \in S_1'} r_{s,c}< \sum_{(s,c) \in \mu': \ s \in S_1'} r_{s,c}.\]
    However, from the definition of $S_1'$, $r_{s,\mu'(s)}\leq r_{s,\mu(s)}$ for every $s\in S_1'$, which leads to a contradiction.
   \Halmos 

\subsection{Multiple Optimal Solutions}\label{app:example_multiple_optimal_solutions}
        Consider an instance with three schools \(\C = \lrl{c_1, c_2, c_3}\), four students \(\S= \lrl{s_1, s_2, s_3, s_4}\), and capacities \(q_{c_1} = q_{c_2} = 1, \; q_{c_3} = 2\). In addition, consider preferences given by:
        \[
        \begin{aligned}
            c\;\, &: s_1\succ s_2 \succ s_3 \succ s_4, \; \forall c\in \C \\
            s_1&: c_1 \succ c_2 \succ c_3 \\
            s_2&: c_2 \succ c_1 \succ c_3 \\
            s_3&: c_1 \succ c_3 \succ c_2 \\
            s_4&: c_2 \succ c_3 \succ c_1. \\
        \end{aligned}
        \]
        Notice that, with no capacity expansion, the student-optimal stable matching is
        \[\mu^* = \lrl{(s_1, c_1), (s_2, c_2), (s_3, c_3), (s_4, c_3)},\]
        which leads to a value of 6. On the other hand, if we have a budget \(B=1\), note that we can allocate it to either \(c_1\) and obtain the matching \(\mu' = \lrl{(s_1, c_1), (s_2, c_2), (s_3, c_1), (s_4, c_3)}\), or to school \(c_2\) and obtain the matching \(\mu'' = \lrl{(s_1, c_1), (s_2, c_2), (s_3, c_3), (s_4, c_2)}\). In both cases, one student moves from their second choice to their top choice, and thus in both cases the sum of preferences of assignment is 5. Hence, we conclude that this problem has more than one optimal solution.
%!TEX root = ./0_main.tex
\section{Appendix to Section~\ref{subsec:compact_formulation}}\label{app: compact formulation}

\subsection{Proof of Lemma~\ref{lemma:validity_L_constraints}}
The lemma follows by noting that since $\mathbf{x} \in \P_{\ZZ}$, then $\sum_{c'\succeq_s c} x_{s,c'}$ is either 0 or 1. When $\sum_{c'\succeq_s c} x_{s,c'}=0$, the L-constraint becomes $\sum_{s'\succ_c s} x_{s',c}\geq q_c+t_c$, which coincides with~\eqref{constraint.1b}. When $\sum_{c'\succeq_s c} x_{s,c'}=1$, the L-constraint is $\sum_{s'\succ_c s} x_{s',c}\geq t_c-B$. Given that $t_c-B\leq 0$, this constraint is redundant because $\sum_{s'\succ_c s} x_{s',c}$ is always non-negative. Constraint~\eqref{constraint.1b} is also redundant under this case. \Halmos

    % \begin{example}\label{ex: multiple optimal solutions}

    % \end{example}
    % \vspace{0.3cm}

% \subsection{McCormick Background and  Linearizations}\label{sec:mccormick_formulations}

\subsection{Background on M{c}Cormick Envelopes}\label{app:McCormick}

        In this section we describe the McCormick convex envelope used to obtain a linear relaxation for bi-linear terms~\citep{McCormick1976}; if one of the terms is binary, the linearization provides an equivalent formulation.
        Consider a bi-linear term of the form $x_i\cdot x_j$ with the following bounds for the variables $x_i$ and $x_j$: $l_i\leq x_i\leq u_i$ and $l_j\leq x_j\leq u_j$. Let us define $y= x_i\cdot x_j$, $m_i = (x_i - l_i)$, $m_j = (x_j - l_j)$,  $n_i = (u_i - x_i)$ and $n_j = (u_j - x_j)$.
        Note that $m_i \cdot m_j \geq 0$, from which we derive the under-estimator $y\geq  x_i \cdot l_j  + x_j \cdot l_i  - l_i \cdot l_j $.
        Similarly, it holds that $n_i \cdot n_j \geq 0$, from which we derive the under-estimator $y\geq  x_i \cdot u_j  + x_j \cdot u_i  - u_i \cdot u_j $.
        Analogously, over-estimators of $y$ can be defined. Make $o_i = (u_i - x_i)$, $o_j = ( x_j - l_j )$, $p_i = ( x_i - l_i )$ and $p_j = (u_j - x_j)$.
        From $o_i \cdot o_j \geq 0$ we obtain the over-estimator $y\leq  x_i \cdot l_j  + x_j \cdot u_i  - u_i \cdot l_j $, and from $p_i \cdot p_j \geq 0$ we obtain the over-estimator $y\leq  x_j \cdot l_i  + x_i \cdot u_j  - u_j \cdot l_i $.
        The four inequalities provided by the over and under estimators of $y$, define the McCormick convex (relaxation) envelope of $x_i \cdot x_j$.

\subsection{McCormick Linearizations of Formulation~\eqref{IntMathProgr}}
    Note that the quadratic term $t_c \cdot \sum_{c'\succeq_s c} x_{s,c'}$ in constraint~\eqref{constraint.1b} can be linearized in at least two ways. Specifically, we call
    \begin{itemize}

        \item \emph{Aggregated Linearization}, when for each $(s,c) \in \E$, we define $ \alpha_{s,c} := t_c \cdot \sum_{c'\succeq_s c} x_{s,c'}$;

        \item \emph{Non-Aggregated Linearization}, when for each $(s,c) \in \E$ and $c'\succeq_s c$, we define $ \beta_{s,c,c'} := t_c \cdot x_{s,c'}$.

    \end{itemize}
    The mixed-integer programming formulation of the McCormick envelope for the aggregated linearization reads as
    \begin{subequations}
    \label{MathProgr_agg_appendix}
    \begin{alignat}{3}
    %(\AggLin) \qquad
    \min_{\mathbf{x},\mathbf{t},\mathbf{\alpha}} \quad & \sum_{(s,c)\in \E} r_{s,c}\cdot x_{s,c}   \\
    s.t. \quad & t_c - \alpha_{s,c} + q_c \cdot \Bigg( 1-\sum_{c'\succeq_s c} x_{s,c'}  \Bigg) \leq  \sum_{s'\succ_c s} x_{s',c},  \hspace{3em} & \forall \ (s,c)\in \E, c \succ_s \emptyset \\
    & -\alpha_{s,c} + t_c +B\cdot \sum\limits_{c'\succeq_s c} x_{s,c'}\leq  B, &\forall \ (s,c)\in \E, c \succ_s \emptyset \\
    & \alpha_{s,c} \leq  t_c, &\forall \ (s,c)\in \E, c \succ_s \emptyset \\
    & \alpha_{s,c} \leq B\cdot \sum_{c'\succeq_s c} x_{s,c'}, & \forall \ (s,c)\in \E, c \succ_s \emptyset   \\
    & (\mathbf{x},\mathbf{t})\in\P_{\ZZ},  \mathbf{\alpha} \geq 0.  
    \end{alignat}
    \end{subequations}
    
    Next, we prove that the set of feasible solutions for the original problem coincides with the set of feasible solutions obtained from the linearization.
    \begin{corollary}\label{cor:aggregated}
    The projection of the feasible region in Formulation~\eqref{MathProgr_agg_appendix} in the variables $\mathbf{t}$ and $\mathbf{x}$ coincides with the region given by  Formulation~\eqref{IntMathProgr}.
    \end{corollary}

    We now discuss the mixed-integer programming formulation of the McCormick envelope for the non-aggregated linearization. Formally, we have the following
    \begin{subequations}
    \label{MathProgr_nonaggregated}
    \begin{alignat}{3}
     \min_{\mathbf{x},\mathbf{t},\mathbf{\beta}} \quad & \sum_{(s,c)\in \E} r_{s,c}\cdot x_{sc}   \label{obj-fun.ext}\\
    s.t. \quad & t_c - \sum_{c'\succeq_s c}\beta_{s,c,c'} + q_c \cdot \Bigg( 1-\sum_{c'\succeq_s c} x_{s,c'}  \Bigg)  \leq  \sum_{s'\succ_c s} x_{s',c}, &\forall \ (s,c)\in \E, c \succ_s \emptyset \label{constraint.ext.b}\\
    & -\beta_{s,c,c'} + t_c +B\cdot x_{s,c'}\leq  B,  &\forall \ (s,c)\in \E, \forall \ c'\succeq_s c \succ_s \emptyset, \label{constraint.ext.c}\\
    & \beta_{s,c,c'} \leq  t_c, &\forall \ (s,c)\in \E,\forall \ c'\succeq_s c \succ_s \emptyset,   \label{constraint.ext.d}\\
    & \beta_{s,c,c'} \leq B\cdot x_{s,c'}, &\forall \ (s,c)\in \E,\forall \ c'\succeq_s c \succ_s \emptyset,   \label{constraint.ext.e}\\
    & (\mathbf{x},\mathbf{t})\in\P_{\ZZ}, \mathbf{\beta} \geq 0.  \label{constraint.ext.f}
    \end{alignat}
    \end{subequations}
    Constraints~\eqref{constraint.ext.c},  \eqref{constraint.ext.d}, \eqref{constraint.ext.e} and the non-negativity constraints for $\beta_{s,c,c'}$ form the McCormick envelope. Similar to the case of Corollary \ref{cor:aggregated}, we know that this is an exact formulation since
    $x_{s,c'} \in \{0,1\}$.
    \begin{corollary}\label{cor:nonaggregated}
    The projection of the feasible region given by Formulation~\eqref{MathProgr_nonaggregated} in the variables $\mathbf{t}$ and $\mathbf{x}$ coincides with the region given by  Formulation~\eqref{IntMathProgr}.
    \end{corollary}
    Therefore, Formulation~\eqref{MathProgr_agg_appendix} and~\eqref{MathProgr_nonaggregated} yield the same set of feasible solutions.
    Interestingly, the feasible region of the relaxed aggregated linearization (\ie, $(\mathbf{x},\mathbf{t})\in\P_\ZZ$ in Formulation~\eqref{MathProgr_agg_appendix} is changed to $(\mathbf{x},\mathbf{t})\in\P$) is  contained in the feasible region of the relaxed non-aggregated linearization.
    \begin{theorem}\label{relaxed_theo}
    The feasible region of the relaxed aggregated linearization model is contained in the feasible region of the relaxed non-aggregated linearization model.
    \end{theorem}
\begin{proof}{\it Proof.}
    The constraints that do not involve the linearization terms $\alpha_{s,c}$ or $\beta_{s,c,c'}$ are trivially satisfied by a feasible solution in both formulations. Therefore, we will restrict our analysis to the remaining constraints. Let $(\mathbf{x},\mathbf{t},$\mbox{\boldmath$\alpha$}) be a feasible solution of the relaxed aggregated linearized program. It is easy to verify that by defining $\bar{\beta}_{s,c,c'}= \alpha_{s,c}\cdot x_{s,c'}$ for every $s\in  \S, \, c \in \C$ and $c' \succeq_s c$, the constraints of the relaxed non-aggregated linearization are all met. \Halmos
        \end{proof}
    Theorem \ref{relaxed_theo} implies that the optimal value of the relaxed aggregated linearized model is greater than or equal to the optimal value of the relaxed non-aggregated linearized model.
    %\delete{In Appendix~\ref{app:missing_example_inclusion}, we provide an example that shows that the inclusion in Theorem \ref{relaxed_theo} is strict.
    % Since solution approaches to mixed-integer programming formulations are based on the quality of their continuous relaxation, we conclude that the aggregated linearization dominates the non-aggregated one, and thus we expect it to perform better in practice.}
    % \alfredo{Remove example of strict inclusion?}
%!TEX root = ./0_main.tex
\section{Appendix to Section~\ref{subsec:comb_formulation}}\label{app:appendix_section_combs}

\subsection{Results on the Generalized Comb Constraints}\label{app: results_generalized_combs}
 In the following lemma, we show that if a capacity variable is set to 1, then all the comb constraints determined by supersets are satisfied.

        \begin{lemma}\label{lemma: combs positive expansion}
        Let $\bar{\mathbf{x}} \in \{0,1\}^{\E}$ be a stable matching for some vector of extra seats $\bar{\mathbf{y}} \in \{0,1\}^{\C\times \{1,\ldots,B\}}$. If $\bar y_c^k=1$ then all constraints~\eqref{constraint.binarization.stability} for the combs in $\mathbf{C}_c(k+\alpha)$ with $\alpha>0$ are satisfied.
        \end{lemma}
        \proof{\it Proof.} When $\bar y_c^k=1$, the right-hand-side of constraints~\eqref{constraint.binarization.stability} for combs in $\mathbf{C}_c(k+\alpha)$ is $q_c +k$ since $y_c^k=1$ implies that $\bar y_c^\ell=1$ for all $\ell=1,\ldots,k-1$. Thus, we need to show that the left-hand-side of these constraints is guaranteed to be greater than or equal to $q_c+k$. By the definition of comb, for all   $C \in \mathbf{C}_c(k+\alpha)$ there is $\hat C \in  \mathbf{C}_c(k)$ such that $\hat C \subseteq C$. Hence, $$\sum_{(s',c') \in C} \bar x_{s',c'}\geq \sum_{(s',c') \in \hat C} \bar x_{s',c'}\geq q_c +k , $$
        where the second inequality follows from the hypothesis that $\bar{\mathbf{x}}$ is stable for $\bar{\mathbf{y}}$.
         \Halmos
        \endproof

        We now prove the analogous of the previous lemma, but for comb constraints defined by subsets.

        \begin{lemma}\label{lemma: combs negative expansion}
        Let $\bar{\mathbf{x}} \in \{0,1\}^{\E}$ be a stable matching for some vector of extra seats $\bar{\mathbf{y}} \in \{0,1\}^{\C\times \{1,\ldots,B\}}$.  If $\bar y_c^k=1$, then  all constraints~\eqref{constraint.binarization.stability} for combs in $\mathbf{C}_c(k-\alpha)$ with $\alpha<k$ are satisfied. % if there is a $ \bar C \in C_c(k)$ such that $C \subseteq \bar C$.
        \end{lemma}
        \proof{\it Proof.}
        When $\bar y_c^k=1$, the right-hand-side of constraint~\eqref{constraint.binarization.stability} for each comb $C \in C_c(k-\alpha)$ is $q_c+k-\alpha$ because $\bar{y}_{c}^k=1$ implies that $\bar{y}_c^\ell = 1$ for all $\ell=1,\ldots,k-1$. Thus, we need to show that the left-hand-side of this constraint is guaranteed to be greater than or equal to $q_c+k-\alpha$. Let $(s,c)$ be the base of the comb $C$.

        First, we show the result when there is $ \hat C \in \mathbf{C}_c(k)$ such that $C \subseteq \hat C$. By the definition of generalized comb (see Subsection~\ref{subsec:cutting_plane}), $C$ contains the shaft $S(s,c)$. Consequently, since $\hat C$ contains $C$, if $(s',c)$ is the base of $\hat C$, then $S(s,c) \subseteq S(s',c)$. %and $s'$ is less preferred than $s$ by $c$.

        Denote by $T_{s_r}$ for $r=1,\ldots, q_c+k$ be the teeth of comb $\hat C$ which (obviously) contains the teeth of comb $C$. Therefore,
        $$\hat C = \bigcup_{r=1}^{q_c+k} T_{s_r} \cup  S(s',c).$$
        Let $I \subseteq \{1,\ldots, q_c+k\}$ be the set of indices corresponding to the teeth of $C$; thence, $|I|=q_c+k-\alpha$. There are two cases:
        \begin{itemize}
            \item If there is $r \in I$ such that $$\sum_{(s_r,j) \in T_{s_r}} \bar x_{{s_r},j} = 0, $$ then student $s_r$ is matched with a school that they prefer less than school $c$. Under the stable matching $\bar{\mathbf{x}}$, this is only possible  if school $c$ is matched with exactly $q_c+k$ students that school $c$ prefers more than student $s_r$: $$\sum_{(i,c) \in S(s,c)} \bar x_{i,c} = q_c+k. $$
        Since $S(s,c) \subseteq C$, then
         $$\sum_{(i,j) \in C} \bar x_{i,j} \geq q_c+k\geq q_c .$$
            \item If $\sum_{(s_r,j) \in T_{s_r}} \bar x_{{s_r},j} = 1 $ for all $r \in I$, then
         $$ \sum_{r \in I}\sum_{(s_r,j) \in T_{s_r}} \bar x_{s_r,j} =|I|= q_c+k-\alpha\geq q_c.$$
         Since $\bigcup_{r\in I} T_{s_r} \subseteq C$ the results holds.
         \end{itemize}
        Second, we prove the lemma when no comb in $\mathbf{C}_c(k)$ contains $C$. Then, such situation can only occur if it is not possible to find a shaft $S(s',c)$ where $s'=s$ or $s \succ_c s'$. This means that the number of seats $q_c+k$ is greater than the number of students that the school ranks. Consequently, the set $\mathbf{C}_c(k)$ is empty and there is nothing to prove.
        \Halmos
        \endproof

        % We now prove our main result in this section.

\subsection{Proof of Theorem~\ref{thm: correctness of comb formulation}}\label{app:proof_of_thm_correctness_comb_formulation}

        %Preliminaries: Lemmata~\ref{lemma: combs positive expansion} and~\ref{lemma: combs negative expansion}.}
        % The preceding lemmas lead us to conclude the statement when the integrality of $\mathbf{x}$ is imposed: For each fixed $\mathbf{y}$, only the comb constraints~\eqref{constraint.binarization.stability} associated to the overall capacity of each school are enforced; the remaining are redundant. When the integrality of $\mathbf{x}$ is relaxed, 
        The statement follows directly from~\citet{baiou2000stable}:
        When $\mathbf{y}$ is fixed, the comb inequalities for the associated capacities provide the convex hull of stable matchings for $\Gamma_{\mathbf{t}}$ where $t_c =\sum_{k=1}^B y_c^{k}$ for all $c \in \C$. Thus, as long as the number of seats for each school is integer,  constraints~\eqref{constraint.binarization.stability} contain the comb constraints from~\citet{baiou2000stable} for $\Gamma_{\mathbf{t}}$.
        \Halmos
        % \endproof

\subsection{Our Separation Algorithm and Structural Results}\label{app:separation_algorithm}

% A separation algorithm is a method that, given a point and a polyhedron, produces a valid inequality that is violated (if any) by that point. This is our goal in Step~\ref{step:determineviolated} of Algorithm~\ref{alg: cutting-plane}. In fact, given an infeasible $(\mathbf{x}^\star, \mathbf{y}^\star)$ to {\baioub}, for each school we aim to find the comb constraint~\ref{constraint.binarization.stability} that is the most violated by $(\mathbf{x}^\star, \mathbf{y}^\star)$. 
        {Before introducing our separation method, formalized in Algorithm~\ref{alg: separation method}, we present key structural results that reduce significantly the running time. 
        %can be further improved  In this case, it is possible to reduce the number of students and schools involved in Algorithm~\ref{alg: separation method}. 
        }
%\fede{As shown in Example~\ref{example:combs_counterexample},} since capacities may change, we cannot use the cutting plane procedure based on the formulation by~\citet{baiou2000stable}. In order to address this issue, we propose Algorithm~\ref{alg: cutting-plane}. \delete{However, we could use their separation algorithm, since in this step the capacities are fixed (recall Problem~\eqref{subproblem:separation}).\footnote{The separation by~\citet{baiou2000stable} runs in $\mathcal{O}(m\cdot n^2)$, where $m$ is the number of schools and $n$ is the number of students, and is valid to separate fractional solutions when capacities are fixed (\ie, $\mathbf{y}^\star$ is binary).}  Nevertheless} 
        Given an optimal solution $(\mathbf{x}^\star, \mathbf{y}^\star)$ obtained from Step~\ref{step: solve MP} in Algorithm~\ref{alg: cutting-plane}, let $\mathbf{t}^\star$ be the projection of $\mathbf{y}^\star$ in the original problem, \ie, $t^\star_c=\sum_{j=1}^B  y^{j\star}_c$ for all $c \in C$.
        In addition, let $\mu^\star_{\mathbf{t}^\star}$ be the student-optimal stable matching for instance $\Gamma_{\mathbf{t}^\star}$.\footnote{We can obtain this by applying DA on the expanded instance $\Gamma_{\mathbf{t}^\star}$.} To simplify, we assume that $\mathbf{y}^\star$ (and thus $\mathbf{t}^\star$) is fixed, and we use $\mu^\star$ to represent $\mu^\star_{\mathbf{t}^\star}$. Throughout this section, the role of $\mu^\star$ will be to guide the search of violated comb constraints. %We say that a school $c \in \C$ is \textit{fully-subscribed} in $\mathbf{x}^\star$ if $\sum_{s: (s,c)\in \E} x_{s,c}^\star = q_c + t_c^\star$; otherwise, we say that school \(c\) is \textit{under-subscribed}. %\footnote{Example~\ref{example:counter-example} in Appendix~\ref{app:missing_proofs} shows that the set of fully-subscribed schools is not necessarily the same for $\mathbf{x}^\star$ and $\mu^\star$.}
        %\alfredo{I suggest to use $\C^f(\mathbf{x}^\star)$ as this is a subset of $\C$, $\mathcal{FS}$ is not intuitive} We denote by $\mathcal{FS}(\mathbf{x}^\star)$  the set of schools full subscribed in matching $\mathbf{x}^\star$.
        
        Given a school $c$, let \(\text{exceeding}(c)= \{ s \in \S: x_{s,c}^\star>0 \textrm{ and }   \mu^\star_{s,c}=0\}\) be the set of exceeding students, \ie, the set of students assigned (possibly fractionally) to school $c$ in $\mathbf{x}^\star$ that are not assigned to $c$ in $\mu^\star$,
        and let \[\text{block}(\mathbf{x}^\star) = \Big\{c\in \C \colon \sum_{s: (s,c)\in \E }x^\star_{s,c} = \sum_{s: (s,c)\in \E }\mu^\star_{s,c} =   q_c + t_c^\star ,\,\, \text{exceeding}(c) \neq \emptyset\Big\}\] be the set of fully-subscribed schools in both $\mathbf{x}^\star$ and $\mu^\star$ that have a non-empty set of exceeding students.
        Finally, a student-school pair $(s,c)$ is called a fractional blocking pair for $\mathbf{x}^\star$ if the following two conditions hold: (i) there is a school $c'$ such that $x_{s,c'}^\star>0$ and $c \succ_s c'$, and (ii)  $c$ is not fully-subscribed or there is a student $s'$ such that $x_{s',c}^\star>0$ and $s\succ_c s'$.\footnote{\citet{kesten2015theory} introduce the notion of ex-ante justified envy, which, in the case of a fully-subscribed school, is equivalent to our definition of fractional stability. \citet{kesten2015theory} define ex-ante justified envy of student $s$ towards student $s'$ if both $x_{s,c'}^\star, x_{s',c}^\star>0 $ with $c\succ_s c'$ and $s\succ_c s'$.  Note that the definition of fractional blocking pair implies that $x_{s,c}^\star<1$. Moreover, a blocking pair is also a fractional blocking pair. }

        %\paragraph{\fede{Structural results.}} In our first structural result, formalized in Lemma~\ref{lemma: fully subscribed schools}, \fede{ we show that, while performing Algorithm~\ref{alg: cutting-plane}, if a school is part of a blocking pair then it must be fully subscribed.} %we can restrict the search for the most violated combs to the schools that are fully-subscribed, reducing significantly the number of schools that we need to check at every iteration of the separation algorithm. 
        
        Given that the family of comb constraints is exponentially large, it is crucial to  reduce the scope of our search. Our main structural result, formalized in Lemma~\ref{lemma: blocking pair in LPH}, accomplishes this by {further} restricting the search of violated combs only among those schools that are in block($\textbf{x}^\star$). To prove Lemma~\ref{lemma: blocking pair in LPH}, we first show   Lemma~\ref{lemma: fully subscribed schools}, which states that while performing Algorithm~\ref{alg: cutting-plane}, if a school is part of a blocking pair then it must be fully subscribed. %we can restrict the search for the most violated combs to the schools that are fully-subscribed, reducing significantly the number of schools that we need to check at every iteration of the separation algorithm. 
        %All the proofs in this subsection can be found in Appendix~\ref{app:missing_proofs_separation}.
    \begin{lemma}\label{lemma: fully subscribed schools}
          Let $(\mathbf{x}^\star,\mathbf{y}^\star)$ be the optimal solution obtained at Step~\ref{step: solve MP} in some iteration of Algorithm~\ref{alg: cutting-plane}, and suppose that there is a fractional blocking pair  $(s,c)$. Then, $c$ is fully-subscribed in $\mathbf{x}^\star$.
          %\delete{Let $\mathbf{x}^\star$ be a matching with a blocking pair $(s,c)$. Then, $c$ is fully-subscribed in $\mathbf{x}^\star$.}\alfredo{agree} \fede{agree} \margarida{Agree}
        \end{lemma}

        \proof{\it Proof of Lemma~\ref{lemma: fully subscribed schools}.}
        We prove that $c$ is fully-subscribed in $\mathbf{x}^\star$ by contradiction.
        Let $(s,c)$ be a fractional blocking pair for $ \mathbf{x}^\star$, where $c$ is an under-subscribed school in $ \mathbf{x}^\star$. Let $s$ be such that $x_{s,c'}^\star>0$ and $c\succ_s c'$.
        So $s$ prefers $c$ over $c'$. Moreover, $c$ has some available capacity which would allow to increase $x_{s,c}^\star$ by decreasing $x_{s,c'}^\star$.
        Clearly, such modification would decrease the objective value of the main program.
        This contradicts the optimality of $\mathbf{x}^\star$.\Halmos\endproof

    In Appendix~\ref{ex:fully_subscribed_difference}, we present an example that shows that the set of fully-subscribed schools is not necessarily the same for $\mathbf{x}^\star$ and $\mu^\star$, and, thus, illustrates the potential of using $\mu^\star$. In our next, result we provide a sufficient condition for stopping the separation algorithm for a given allocation $\mathbf{t}^\star$. Specifically, we show that we only need to check for violated constraints among the fully-subscribed schools, thus reducing significantly the number of schools that we need to check at every iteration of the separation algorithm. %to also guide the search for violated comb constraints.

        \begin{lemma}\label{lemma: fully subscribed schools DA}
        Let $(\mathbf{x}^\star,\mathbf{y}^\star)$ be the optimal solution obtained at Step~\ref{step: solve MP} in some iteration of Algorithm~\ref{alg: cutting-plane} and $\mu^\star$ the student-optimal stable matching in instance $\Gamma_{\mathbf{t}^\star}$, where $\mathbf{t}^\star$ is the allocation defined by $\mathbf{y}^\star$. If $\mathcal{J}$ contains all the comb constraints of every fully-subscribed school in $\mu^\star$, then both matchings $\mathbf{x}^\star$ and $\mu^\star$ coincide, \ie, $\mathbf{x}^\star$ is the characteristic vector of $\mu^\star$. %  blocking pair $(s,c)$. Then $c$ is fully-subscribed in $\mathbf{x}^\star$. % and in $\mu^\star$.
        \end{lemma}

    \proof{\it Proof of Lemma~\ref{lemma: fully subscribed schools DA}.}
        % Due to Lemma~\ref{lemma: student optimal stable matching with lp}, we know that if $\mathbf{x}^\star$ is a stable matching, then it is student-optimal in $\Gamma_{\mathbf{t}^\star}$ and because of uniqueness it must coincide with $\mu^\star$. Let us prove that $\mathbf$
        Let $\C'$ be the set of fully-subscribed schools in $\mu^\star$ and $\S'$ be the set of students matched to the schools $\C'$ in $\mu^\star$. We denote by $\C''$ and  $\S''$ the complementary sets of $\C'$ and $\S'$ respectively.
        Let us analyze the matches in $\S'$. Some facts: 
        \begin{itemize}
            \item Note that for every student $s\in\S'$ we have $\sum_{c\succeq_s \mu^\star(s)}x^\star_{s,c}=1$, \ie, student $s$ must be matched in $\mathbf{x}^\star$ to some school that she prefers at least as much as $\mu^\star(s)\in C'$. This is because $\mathcal{J}$ contains all the comb constraints of the schools that are fully subscribed in the unique student-optimal stable matching $\mu^\star$. Since $\mathbf{x}^\star$ minimizes the rank of the schools that students are assigned to, then some fully subscribed schools may be part of a blocking pair, which can be addressed by introducing a comb constraint, however, all the comb constraints of the fully subscribed schools are already in $\mathcal{J}$.
            \item Every student $s\in\S'$ must be matched in $\textbf{x}^\star$ to a school that is in $\C'$, \ie, $\sum_{c\in\C'}x^\star_{s,c}=1$. This is because, otherwise, $s$ would create a blocking pair in $\mu^\star$ with a under-subscribed school in $\C''$ due to our first argument. As a consequence, we cannot have $x^\star_{s,c}>0$ for a student $s\in\S''$ and a school $c\in\C'$. This is because we proved above that all students in $\S'$ are matched in $\C'$ and schools in $\C'$ are fully-subscribed.
        \end{itemize}
        
         All the above imply that we can consider $\S'$ and $\C'$ as a sub-instance of the problem and the integrality of $\mathbf{x}^\star$ is given by the result proved by \cite{baiou2000stable}. Because of optimality, $\mu^\star$ and $\mathbf{x}^\star$ must coincide in $\S'$ and $\C'$.
        %\federico{Third, every student $s\in\S'$ must be matched in $\textbf{x}^\star$ to the same school as in $\mu^\star$. Indeed, if that would not be the case, then there would be a blocking pair involving a school in $\C'$, which would violate the assumption that MP has all the comb constraints of the schools in $\C'$.} \alfredo{The third point is not clear because $\mathbf{x}^\star$ could be fractional among schools in $\C'$. Also the paragraph show that $\S'$ is matched to $\C'$, but someone in $\S''$ could be potentially matched in $\C'$ (in $\mathbf{x}^\star$) and I guess that's the purpose of the next paragraph.}
        
        Let us now analyze the assignments of the students in $\S''$. First, note that the integrality of $\mathbf{x}^\star$ in the sub-instance defined by $\S''$ and $\C''$ is given by the integrality of the matching polytope without stability constraints, \ie, $\P$ (restricted to $\S''$ and $\C''$ with $\mathbf{t}^\star$ fixed) has integral vertices (see e.g. \citep{schrijver2003combinatorial}). Now, we will show that the assignments coincide in  $\mu^\star$ and $\textbf{x}^\star$.
        By contradiction, assume that the students in $\S''$ are matched differently in $\textbf{x}^\star$ and in $\mu^\star$. First, if there is a student $s\in \S''$ that prefers a school $c\in\C''$ over $\mu^\star(s)$, then $(s, c)$ would be a blocking pair in $\mu^\star$ because $c$ is under-subscribed in $\mu^\star$, which is not possible. Therefore, every student $s\in \S''$ weakly prefers $\mu^\star(s)$  over their assignment in $\textbf{x}^\star$, and, by the hypothesis in our contradiction argument, there is at least one student $s$ that strictly prefers $\mu^\star(s)$  over their match in $\textbf{x}^\star$ (\ie, the two schools are different). Given that $\S''$ and $\C''$ form a separate sub-instance, we can rearrange the matching $\textbf{x}^\star$ of every student in $\S''$ by matching the student to her matching in $\mu^\star$. This new matching would have a lower objective than the one of $\textbf{x}^\star$, which would violate its optimality. Therefore, both $\mu^\star$ and $\mathbf{x}^\star$ also coincide in $\S''$ and $\C''$.  %Otherwise, student $s\in \S''$ that prefers $\mu(s)$  over $\textbf{x}^\star (s)$. If $\mu(s)$ is not fully-subscribed in $\textbf{x}^\star (s)$, then it would contradict the optimality of $\textbf{x}^\star (s)$, since we can only rearrange the matching of $s$ and obtain a better matching. Therefore, $\mu(s)$ must be fully-subscribed in $\textbf{x}^\star (s)$. Hence, since $\textbf{x}^\star$ has less comb constraints than $\mu$, there must be at least another student $s''$ in $\S''$ who is matched better off with $\mu(s)$ in $\textbf{x}^\star (s)$, rather than her corresponding match in $\mu(s'')$. Then, $(s'', \mu(s))$ would be a blocking pair in $\mu$.
        \Halmos
        \endproof

Our next structural result, formalized in Lemma~\ref{lemma: blocking pair in LPH}, further reduces the search by  restricting the search of violated combs only among those schools that are in block($\textbf{x}^\star$).

        \begin{lemma}\label{lemma: blocking pair in LPH}
            If $\mathbf{x}^\star$ has a fractional blocking pair, then there is at least one student-school pair $(s,c) $, where $c$ is in \emph{block(}$\mathbf{x}^\star$\emph{)} and $s$ is preferred by $c$ over the least preferred student in $\text{exceeding}(c)$, such that the value of the tooth $T(s,c)$ is smaller than 1, \ie, $\sum\limits_{(s,c') \in T(s,c)} x^\star_{s,c'}<1$.
        \end{lemma}

        \proof{\it Proof of Lemma~\ref{lemma: blocking pair in LPH}.}
        Suppose that the statement is false. Then, for every school $c$ in block($\mathbf{x}^\star$), and for every student $s$ more preferred by $c$ to the least preferred student in $\text{exceeding}(c)$, {the pair $(s,c)$} has the tooth  evaluated to 1. By Lemma~\ref{lemma: fully subscribed schools}, we know that the fractional blocking pairs must all involve a school that is fully-subscribed in $\mathbf{x}^\star$. Among these schools, it is sufficient to focus our attention on the schools $c \in C$ such that there exists $s \in S$ with $x^\star_{s,c}>0$ and student $s$ is not matched to $c$ in $\mu^\star$.  Therefore, for these schools the set of exceeding students is non-empty.

        We now observe that, by hypothesis, every school $c$ in block($\mathbf{x}^\star$), has the property that all the teeth are evaluated to 1 until the least preferred exceeding student. This implies that all the combs in $\mathbf{C}_c(\mathbf{t}_c^\star)$ are not violated since such combs are composed by $q_c+t_c^\star$ teeth. Hence, $\mathbf{x}^\star$ is a stable matching with a better objective for the students than the student-optimal stable matching. %\margarida{I do not understand where is the contradiction.} %have violated the hypothesis of non-stability of $\mathbf{x}$.
        \Halmos
        \endproof

       We now describe our separation method, which is illustrated in Algorithm~\ref{alg: separation method}.
        Algorithm~\ref{alg: separation method} begins by initializing the set of violated combs $\mathcal{J}'$ equal to $\emptyset$. 
        Based on Lemma~\ref{lemma: blocking pair in LPH}, we focus only on the schools in block($\mathbf{x}^\star$). 
        At Step~\ref{step:sep_alg:for_schools}, we iterate to find the most violated (\ie, least valued) comb of every school in block($\mathbf{x}^\star$). To do so, we first initialize as empty the list of teeth $\mathcal{T}$ (Step~\ref{step:sep_alg:initialization_teeth}). The list $\mathcal{T}$ will be updated to store the set of students whose teeth are part of the least valued comb. 
        At Step~\ref{step:sep_alg:last_student}, we look for the least preferred exceeding student $\underline{s}$ in school \(c\in\)block($\mathbf{x}^\star$), and, finally, we introduce the preference list of school $c$ that terminates with student $\underline{s}$. 
        As previously mentioned, the key idea is to recursively update \(\mathcal{T}\) so that it contains the students whose teeth form the least valued comb. To accomplish this, we iterate over the set of students following the preference list \(\succ_c\), going up to \(\underline{s}\).\footnote{Note that, since school \(c\) is fully-subscribed, we know that there are \(q_c+t_c^\star\) students \(s'\succeq_c \underline{s}\) assigned to \(c\), and thus the comb constraint is always satisfied.}
        At Step~\ref{step:sep_alg:for_students} we select the student $s'$ and we find the value $v_{s',c}$ of $T^-(s',c):=T(s',c)\setminus \lrl{(s',c)}$ in $\mathbf{x}^\star$ (Step~\ref{step:sep_alg:tooth_value}). 
        If the list \(\mathcal{T}\) does not contain a sufficient number of students to build a comb (\ie, $q_c+t^\star_c$), we introduce $s'$ in the list \(\mathcal{T}\) while respecting a descending ordering of the elements of \(\mathcal{T}\) according to $v_{s,c}$ (Step~\ref{step:sep_alg:not_minimum_number_teeth}). Moreover, if we have obtained a set \(\mathcal{T}\) of cardinality equal to the capacity of school $c$ then,  we create the first comb $C$ at Step~\ref{step:sep_alg:first_comb}. 
        %Then, we create the set $\mathcal{T}$, which is initialized with the $q_c+t^\star_c$ most preferred students of $c$. The set $\mathcal{T}$ will be updated to store the set of students whose teeth are part of the least valued comb.
        Once we have obtained the first comb, at Step~\ref{step:sep_alg:stud_highest_tooth} we define $s^\star$  as the student in $\mathcal{T}$ with the highest value $T^-(s^\star,c)$
        %\margarida{What is $T^-$? This is not defined. Can we just say tooth in $\mathcal{T}$? Note that this undefined notation appears in the pseudocode}. 
        %Then, we build the initial comb $C$ by considering the students in $\mathcal{T}$ (Step~\ref{step:comb-sel:5}).
        At Step~\ref{step:sep_alg:if_new_candidate}, we compare the value of $T^-(s',c)$ with the value of $T^-(s^\star,c)$. If the former value is smaller, then it is worth pursuing the search for a comb valued less than $C$ with a tooth based in $(s',c)$; we build such a comb $C'$ at Step~\ref{step:sep_alg:compute_new_comb}. If the value of $C'$ is smaller than the value of $C$, then we update $\mathcal{T}$ to include $s'$ at the place of $s^\star$ (Step~\ref{step:sep_alg:student_removed} and Step~\ref{step:sep_alg:new_tooth}) and we update $C$ as $C'$  (Step~\ref{step:sep_alg:update_comb}). 
        At the end of the inner \textbf{for} cycle, we obtain the least valued comb $C$ of school $c$. If $C$ has a value in $\mathbf{x}^\star$ smaller than $q_c+t_c$, then the stability condition is violated for school $c$. Hence, at Step~\ref{step:sep_alg:add_comb_constraint}, we add the violated comb $C$ to the set $\mathcal{J}'$. In Appendix~\ref{ex: separation algorithm}, we exemplify the application of Algorithm~\ref{alg: separation method}.
\begin{algorithm}[ht]
            \caption{Separation method}\label{alg: separation method}
            \begin{algorithmic}[1]
            \Require An instance $\Gamma = \langle \S,\C,\succ, \mathbf{q}+\mathbf{t}^\star \rangle$ and a (fractional) matching $\mathbf{x}^\star$. % , a school $c$ in block($\mathbf{x}^\star$) A (fractional) matching $\mathbf{x}^\star$ and a (binary) vector of additional seats $\mathbf{y}^\star$.
            \Ensure A non-empty set $\mathcal{J}'$ of constraints~\eqref{constraint.binarization.stability} violated by $(\mathbf{x}^\star,\mathbf{y}^\star)$, if it exists.
            \State $\mathcal{J}' \leftarrow \emptyset$ {\footnotesize \Comment{(empty set of constraints)}}\label{step:sep_alg:initialization_const}
            \For{$c \in$ block($\mathbf{x}^\star$)} \label{step:sep_alg:for_schools}%{$c \in \mathcal{FS}(\mathbf{x}^\star) \cap \mathcal{FS}(\mu)$}
            \State \(\mathcal{T} \leftarrow \emptyset\) {\footnotesize \Comment{(empty list of teeths)}}\label{step:sep_alg:initialization_teeth}
            \State \(\underline{s} \leftarrow \) least preferred student in \(c\) such that \(x_{s,c}^* > 0\) {\footnotesize \Comment{(last student in excess)}} \label{step:sep_alg:last_student}
            \State \(\succ_c[\underline{s}] \leftarrow \) preference list of $c$ until $\underline{s}$  \label{step:sep_alg:preference_list}
            \For{\(s' \in\,  \succ_c[\underline{s}]\)} {\footnotesize \Comment{(in order following \(\succ_c\))}} \label{step:sep_alg:for_students}
            \State \(v_{s',c} \leftarrow \sum_{c'\succ c} x_{s',c'}^*\) {\footnotesize \Comment{(value of \(T^{-}(s',c)\))}}\label{step:sep_alg:tooth_value}
            \If{$ |\mathcal{T}| < q_c + t^\star_c$} \label{step:sep_alg:not_minimum_number_teeth}
            \State $\mathcal{T} \leftarrow  \{ s'\}$ {\footnotesize\Comment{($s'$ enters $\mathcal{T}$ in descending order according to $v_{s,c}$)}}\label{step:sep_alg:completing_teeth}
            \If{$ |\mathcal{T}| = q_c + t^\star_c$}\label{step:sep_alg:if_minimum_number_teeth}
            \State  \(C \leftarrow \lrl{(s,c): s\in \mathcal{T}} \cup \bigcup_{s\in \mathcal{T}}  T^{-}(s,c)\) {\footnotesize \Comment{(initial comb, made of shaft and selected teeths)}}\label{step:sep_alg:first_comb}
            \EndIf
            \Else 
            \State $s^\star \leftarrow $ first student in $\mathcal{T} $  {\footnotesize \Comment{(student $s\in \mathcal{T}$ such that $T^-(s,c)$ has the highest value)}}\label{step:sep_alg:stud_highest_tooth}
                \If{\(v_{s',c} < v_{s^\star,c}\)}\label{step:sep_alg:if_new_candidate}
                    \State \(C' \leftarrow S^{t_c^\star}(s',c) \cup \bigcup_{s \in \mathcal{T}\setminus \lrl{s^\star} } T^{-}(s,c) \cup T^{-}(s',c)\)\label{step:sep_alg:compute_new_comb} %\margarida{The shaft must depend on the capacity expansion. See definition of comb}
                    \If{\(\sum_{(s,h) \in C'} x_{s,h}^* < \sum_{(s,h) \in C} x_{s,h}^*\) }\label{step:sep_alg:if_new_comb_better} %\margarida{I believe we need to use a different index than $c$ for these sums. Note that they include teeth so we make sums involving schools other than $c$}
                        \State $\mathcal{T} \setminus \{ s^\star\}$ {\footnotesize\Comment{($s^\star$ is removed from  $\mathcal{T}$)}}\label{step:sep_alg:student_removed}
                        \State $\mathcal{T} \leftarrow  \{ s'\}$ {\footnotesize\Comment{($s'$ enters $\mathcal{T}$ in descending order according to $v_{s,c}$)}}\label{step:sep_alg:new_tooth}
                        %\State \(\mathcal{T} =\left( \mathcal{T} \setminus \lrl{s^\star} \right) \cup \lrl{s'}\)\label{step:comb-sel:13}
                        %\State \(s^\star \in \arg \max_{s\in \mathcal{T}}\lrl{v_{s,c}} \)\label{step:comb-sel:14} %\margarida{Again, should we use $\in$ instead of $=$?}
                        \State \(C\leftarrow C'\)\label{step:sep_alg:update_comb}
                    \EndIf
                \EndIf
            \EndIf
            \EndFor
            %\For{\((s,c) \in S^{t_c^\star}(\underline{s}, c)\)} \label{step:innerloop}
            %\State \(v_{s,c} \leftarrow \sum_{c'\succ c} x_{s,c'}^*\) {\footnotesize \Comment{(value of \(T^{-}(s,c)\))}}\label{step:comb-sel:2} %\margarida{Should the index of $x^\star$ in the sum be $x_{s,c'}^\star$? What is a node in a teeth? Note that we did not introduce the graph representation so we cannot use its interpretation.}
            %\EndFor
            %\State \(\mathcal{T} \leftarrow \lrl{s \in \S: s \text{ is in the top } q_c + t_c^\star \text{ most preferred students in } \succ_c}\) {\footnotesize \Comment{(select first teeths)}}\label{step:comb-sel:3}
             %\margarida{What if there is more than one? Should we put instead $s^\star \in \arg\max$?}
            %\State \(\bar{s} \leftarrow \) most preferred student in \(\succ_c\) not included in \(\mathcal{T}\)\label{step:comb-sel:6}
            %\State \(C \leftarrow \lrl{(s,c): s\in \mathcal{T}} \cup \bigcup_{s\in \mathcal{T}}  T^{-}(s,c)\) {\footnotesize \Comment{(initial comb selected, consisting of selected teeths and shaft)}}\label{step:comb-sel:5}
            \If{\( \sum_{(s,h) \in C} x_{s,h}^* < q_c + t_c^\star \) }\label{step:sep_alg:if_comb_value_small} %\margarida{I believe that in the left-hand-side sum the index $c$ must be replaced by something else. Note that we are within a cycle that fixes $c$}
            \State $\mathcal{J}' \leftarrow \mathcal{J}' \cup \{C\} $   \label{step:sep_alg:add_comb_constraint} %\margarida{I prefer $\mathcal{J}' \leftarrow \mathcal{J} \cup \{C\} $}%\Comment{Apply Algorithm~\eqref{alg: implemented comb selection algorithm}}
            \EndIf
            \EndFor
            \State \Return $\mathcal{J}'$
            \end{algorithmic}
        \end{algorithm}
        
        {Note that Algorithm~\ref{alg: separation method} resembles the one introduced in \cite{baiou2000stable}.\footnote{The separation by~\citet{baiou2000stable} runs in $\mathcal{O}(m\cdot n^2)$, where $m$ is the number of schools and $n$ is the number of students, and is valid to separate fractional solutions when capacities are fixed (\ie, $\mathbf{y}^\star$ is binary).}  However, there is one key difference: We begin the search of the most violated comb at the head of the preference list of the school rather than at the tail (as it is done in \cite{baiou2000stable}), which guarantees that our method finds the most violated comb constraint.}\footnote{In Appendix~\ref{ex: baiou separation algorithm counterexample} we show that the separation algorithm by \cite{baiou2000stable} may not find the most violated constraint.}

    \begin{theorem}\label{theorem_separation_algorithm}
        Algorithm~\ref{alg: separation method} finds the combs solving  Step~\ref{step: definition of C star} in Algorithm~\ref{alg: cutting-plane} for every school in $\mathcal{O}(m \cdot n\cdot \Bar{q})$  time, where $n$ is the number of students, $m$ is the number of schools and  $\Bar{q}= \max\limits_{c\in \C } \{ q_c +t^\star_c\}$. 
        Moreover, Algorithm~\ref{alg: separation method} returns a set of combs such that the corresponding constraints~\eqref{constraint.binarization.stability} are violated by $(\mathbf{x}^\star,\mathbf{y}^\star)$, if $\mathbf{x}^\star$ is not stable for $\Gamma_{\mathbf{t}^\star}$.

        \end{theorem}

\begin{remark}
Algorithm~\ref{alg: separation method} can be easily adapted to separate the comb constraints with any linear objective function. Indeed, it is sufficient to iterate over the whole set of schools at Step~\ref{step:sep_alg:for_schools}, remove Steps~\ref{step:sep_alg:last_student} and~\ref{step:sep_alg:preference_list}, and iterate over the whole preference list $\succ_c$ at Step~\ref{step:sep_alg:for_students}. However, this generalization deteriorates the running-time of the separation algorithm.
\end{remark}

\proof{{\it Proof of Theorem}~\ref{theorem_separation_algorithm}.}
        The proof is based on the validity of the following statement, that we prove by induction: For every school $c \in \text{block}(\mathbf{x}^\star)$, Steps~\ref{step:sep_alg:initialization_teeth}-\ref{step:sep_alg:update_comb} of Algorithm~\ref{alg: separation method} provide the comb solving the problem in Step~\ref{step: definition of C star} in Algorithm~\ref{alg: cutting-plane} in  $\mathcal{O}(n\cdot \Bar{q})$.
        First, let us observe that the search for the least valued comb in $\mathbf{C}_c(t_c^\star)$ terminates in a finite number of steps because the preference list $\succ_c$ is finite. 
        % Moreover, it returns the least valued comb found so far, since the school has a non-empty set of exceeding students.
        On the other hand, at the beginning of every cycle in Step~\ref{step:sep_alg:for_students}, we select a student $s'$ and by the end of the $(q_c+t_c^\star)$-th iteration, we claim that the comb $C$ is the smallest valued comb in $\mathbf{C}_c(t_c^\star)$.
        If we prove the main statement presented above integrated with the two previous observations, we prove that the algorithm solves Step~\ref{step: definition of C star} in Algorithm~\ref{alg: cutting-plane} for $c$. Note that at the first $q_c+t_c^\star -1$ steps, a comb cannot be defined; indeed, a comb needs $q_c+t_c -1$ teeth and a base with its tooth. The proof follows an inductive argument:

        \textbf{Base:} At the $(q_c+t_c^\star)$-th iteration, we create comb $C$ at Step~\ref{step:sep_alg:first_comb}. 
        The base of comb $C$ is the student $s'$ that is ranked $q_c+t_c^\star$ by $c$, and the bases of the other $q_c+t_c^\star-1$ teeth are the students preferred more than $s'$. Since this comb is the only possible that we can create with a base preferred more or equal than $s'$, it is also the least valued comb possible. 

        \textbf{Inductive step:} The inductive hypothesis states that at the end of iteration $\ell\geq q_c+t_c^\star$, after selecting student $s'$, the selected comb $C$ is the comb with the lowest value among all combs with a base student preferred equal to or more than $s'$. We want to prove that at the end of step $\ell+1$, the algorithm selects the comb with lowest value among all combs with a base student ranked less or equal than $(\ell+1)$-th one. 
        Let $s'$ be the student ranked $\ell+1$ and let $s^\star$ be the student in $\mathcal{T}$ with the highest valued $T^-$. 
        We need to verify if adding $T^-(s',c)$ to $\mathcal{T}$ may produce a comb of smaller value than the one recorded. If $T^-(s',c)\geq T^-(s^\star,c)$, then adding $s'$ in $\mathcal{T}$ would produce a comb with equal or higher value. Otherwise, $T^-(s',c)< T^-(s^\star,c)$, and adding $s'$ in $\mathcal{T}$ may provide a better comb. If $T^-(s',c)< T^-(s^\star,c)$, then we create a new comb $C'$ with base in $(s',c)$ and teeth in $\mathcal{T}\setminus \{ s^\star \}$.
        Any other choice of teeth is sub-optimal for the base $(s',c)$ since they are the least valued $q_c+t_c^\star$ teeth among those in the first $\ell+1$ positions. If the new comb $C'$ has value smaller than the comb $C$, then we have found the best comb among those with base ranked at most $\ell+1$.

        {To establish the correctness of Algorithm~\ref{alg: separation method}, it remains to show that it finds the violated combs (if any) that cut-off the input solution. First, the algorithm terminates in a finite number of steps because $ |\C| = m$ and $|\succ_c|\leq|\S|=n$. Second, by the previous inductive argument, we have found the least valued comb for each school.} 
        
        \paragraph{Running-time analysis.}  %
        {The inner \textbf{for} loop (Step~\ref{step:sep_alg:for_students}) takes at most $n$ rounds and the most expensive computation within it is placing $s'$ in $\mathcal{T}$ to maintain the list $\mathcal{T}$ is descending order according to the values $v_{s,c}$. Note that this task  can be performed by comparing $s'$ with at most each element in $\mathcal{T}$, \ie, in a number of operations equal to $|\mathcal{T}|= q_c+t_c^\star$.  Hence, the number of elementary operations for finding the most violated comb of school $c$ is at most $
        \mathcal{O}(n\cdot(q_c+t_c^\star))$.
        Hence,  Algorithm~\ref{alg: separation method} can take at most $\mathcal{O}(m \cdot n\cdot \Bar{q})$ time, where $\Bar{q}= \max\limits_{c\in \text{block}(\mathbf{x}^\star)} \{ q_c +t^\star_c\}$ and $m=|\C|$.}
        \Halmos
        \endproof

    \subsection{Missing Examples in  Appendix~\ref{app:appendix_section_combs}}
    \subsubsection{Difference in the Sets of Fully-subscribed Schools.}\label{ex:fully_subscribed_difference}
        % \begin{example}
        Let $\mathcal{S} = \{ s_1, s_2, s_3 \}$ be the set of students and let $\mathcal{ C} =\{ c_1, c_2, c_3, c_4 \} $ be set of schools; every school has capacity 1.
        The preferences of the students are $s_1: c_1 \succ c_4 \succ c_3\succ c_2$; $s_2: c_1\succ c_2\succ c_3\succ c_4$; $s_3: c_4\succ c_1\succ c_2\succ c_3$, and the preferences of the schools are $c_1: s_2\succ s_1\succ s_3$; $c_2: s_2\succ s_1\succ s_3$; $c_3: s_1\succ s_2\succ s_3$; $c_4: s_3\succ s_1\succ s_2$.
        The optimal matching with no stability constraints (\ie, $\mathcal{J}=\emptyset$) and budget $B=0$,  is $x^\star_{s_1,c_1}=1$, $x^\star_{s_2,c_2}=1$, $x^\star_{s_3,c_4}=1$ and all other entries equal to zero.
        %$ \mathbf{x}$ = \{($c_1,s_1$), ($c_2,s_2 $), ($c_4,s_3 $)\},
        On the other side, the student-optimal stable matching is $\mu^\star$ = \{($s_2,c_1$), ($s_1,c_3 $), ($s_3,c_4 $)\}. Note that the set of fully-subscribed schools is mutually not-inclusive.
        %\label{example:counter-example}
    \subsubsection{Separation Algorithm: Example.}\label{ex: separation algorithm}
        % \begin{example}
%        \fede{adapted to the new algorithm}
        Let $\Gamma=\langle \S,\C,\succ, \mathbf{q} \rangle$ be the instance of the school choice problem with $\mathcal{S} = \{ s_1, s_2, s_3, s_4, s_5 \}$ as the set of students and $\mathcal{ C} =\{ c_1, c_2, c_3, c_4, c_5, c_6 \} $ as the set of schools; every school has capacity 1. The preferences of the students are
        $s_1: c_1 \succ c_4 \succ c_3\succ c_2\succ\ldots$; $s_2: c_1\succ c_2\succ c_3\succ c_4\succ\ldots$; $s_3: c_4\succ c_1\succ c_5\succ c_6 \succ\ldots$; $s_4: c_5\succ c_1\succ c_4\succ c_6 \succ\ldots$; $s_5: c_4\succ c_5\succ c_6\succ c_1 \succ\ldots$, and the preferences of the schools are $c_1: s_2\succ s_1\succ s_3\succ\ldots$; $c_2: s_2\succ s_1\succ s_3\succ\ldots$; $c_3: s_1\succ s_2\succ s_3\succ\ldots$; $c_4: s_3\succ s_4\succ s_5 \succ\ldots$; $c_5: s_4\succ s_3\succ s_5 \succ\ldots$; $c_6: s_5\succ \ldots$.
        The ``\ldots'' at the end of a preference list represent any possible strict ranking of the remaining agents on the other side of the bipartition.
        The optimal solution of the main program with no stability constraints (\ie, $\mathcal{J}=\emptyset$) and budget $B=1$,  is $(\mathbf{x}^\star, \mathbf{t}^\star)$ with $x^\star_{s_1,c_1}=1$, $x^\star_{s_2,c_2}=1$, $x^\star_{s_3,c_4}=1$, $x^\star_{s_5,c_4}=1$, $x^\star_{s_4,c_5}=1$, $t^\star_4 = 1$ and all other entries equal to zero.
        We take the instance $\Gamma_{\mathbf{t}^\star}$ (with the expanded capacities in accordance to $\mathbf{t}^\star$) and the matching $\mathbf{x}^\star$ as the input for Algorithm~\ref{alg: separation method}. At Step~\ref{step:sep_alg:initialization_const} we initialize the set of violated comb constraints $\mathcal{J}'$ to $\emptyset$. In order to proceed, we need to compute the set block($\mathbf{x}^\star$). First, note that the student-optimal stable matching of $\Gamma_{\mathbf{t}^\star}$ is $\mu$ = \{($s_2,c_1$), ($s_1,c_3 $), ($s_3,c_4 $), ($s_5,c_4 $), ($s_4,c_5 $)\}. Therefore, the set of schools that are fully subscribed in both $\mu$ and $\mathbf{x}^\star$ is $\{c_1, c_4, c_5 \}$. The only school that is fully subscribed in both $\mu$ and $\mathbf{x}^\star$ that has an exceeding student is $c_1$.
        Hence, at Step~\ref{step:sep_alg:for_schools}, we select $c_1$. We initialize the empty list $\mathcal{T}$, and we select the least preferred student enrolled in $c_1$, which is $\underline{s}=s_1$. Note that $\succ_{c_1}[s_1]=[s_2,s_1]$.
        The inner loop selects $s_2$ as the most preferred student. 
        At Step~\ref{step:sep_alg:tooth_value} we compute the value of $T^-(s_2,c_1)$, which is 0  (we set to 0 the value of an empty $T^-$).
        Then, at Step~\ref{step:sep_alg:completing_teeth}, we include $s_2$ in the teeth list, \ie, $\mathcal{T}=\{ s_2 \}$, which is composed of only one student because 1 is the capacity of school $c_1$ in $\Gamma_{\mathbf{t}^\star}$. Therefore, we can find the comb $C$ with base $(s_2,c_1)$ at Step~\ref{step:sep_alg:first_comb}.%, note that has value 0 in $\mathbf{x}^\star$.  
        At the next iteration, we have that $s'=s_1$ and we find that $s^\star$ is necessarily $s_2$. The initial comb $C$ built at the previous iteration is composed only by the tooth of $(s_2, c_1)$, \ie, $C=T(s_2, c_1)=\{(s_2, c_1) \}$. The value of $C$ in $\mathbf{x}^\star$ is 0. %The inner \textbf{for} loop at Step~\ref{step:comb-sel:7} can select only $s_1$. 
        Since the value of $T^-(s_1, c_1)$ is 0, the condition at Step~\ref{step:sep_alg:if_new_candidate} is false, meaning that it is not worth pursuing a comb built with basis $(s_1,c_1)$. Therefore, the algorithm jumps to Step~\ref{step:sep_alg:if_comb_value_small}, where it finds that the condition is satisfied since the value of $C$ is 0 and the capacity of $c_1$ is 1. Hence, at Step~\ref{step:sep_alg:add_comb_constraint}, we add $C$ to the set of cuts to be added to the main program. Note that at this point Algorithm~\ref{alg: separation method} terminates and returns to the main program the set of cuts $\mathcal{J}'$ containing only the comb based in $(s_2,c_1)$. Interestingly, the next optimal solution of the main program is the optimal solution of the problem.
        %$ \mathbf{x}$ = \{($c_1,s_1$), ($c_2,s_2 $), ($c_4,s_3 $)\},

\section{Appendix to Section~\ref{sec: implementation in chile}}

\subsection{Hybrid Approach}\label{app: hybrid details}

%\includegraphics[width=\textwidth]{OR/2nd_round/plots/rebuttal_performance_ratio_last_pref_1h.pdf}\caption{}\label{fig:Antofagasta_performance_profile_hybrid_pref+1}

%\includegraphics[width=\textwidth]{OR/2nd_round/plots/rebuttal_performance_ratio_C_pen_1h.pdf}\caption{}\label{fig:Antofagasta_performance_profile_hybrid_C+1}

    % \begin{figure}[htp!]
    %    \begin{subfigure}{0.49\textwidth}
    %   \caption{$|\succ|+1$}\label{fig:Antofagasta_performance_profile_hybrid_pref+1}
    %         \includegraphics[width=\textwidth]{OR/2nd_round/plots/rebuttal_performance_ratio_last_pref_1h.pdf}
    %     \end{subfigure}
    %     \begin{subfigure}{0.49\textwidth}
    %         \caption{$|\C|+1$}\label{fig:Antofagasta_performance_profile_hybrid_C+1}
    %         \includegraphics[width=\textwidth]{OR/2nd_round/plots/rebuttal_performance_ratio_C_pen_1h.pdf}
    %     \end{subfigure}
    %     \caption{Effect of Using Previous Preferences}\label{fig: effect prev pref}
    % \end{figure}

    Our hybrid approach combines in a mathematical program the compact formulation with L-constraints and the generalized comb formulation. The idea is that for each $(s,c) \in \E$, we must guarantee that it is not a blocking pair, and this can be done either through comb or L-constraints. To be more precise, in Algorithm~\ref{alg: cutting-plane}, we can enforce stability by introducing some comb constraints and L-constraints in the main program (Step~\ref{step:J}), and then by separating comb constraints (Steps~\ref{step: definition of C star} and~\ref{step:add new constraint}) at later iterations. We consider the following family of hybrid approaches parameterized by $\gamma\in[0,1]$, whereby for each school $c\in \C$: 
    %We now generalize the \emph{hybrid} formulation discussed in the main body considering the following family of hybrid formulations parameterized by $\gamma\in[0,1]$, whereby for each school $c\in \C$:
    \begin{enumerate}
        \item[(i)] we separate only the comb constraints~\eqref{constraint.binarization.stability} for those combs with base $(s,c)$ where $s$ has a priority in $c$ of at least $q_c+\lfloor \gamma \cdot B +0.5 \rfloor$; and
        \item[(ii)] we add the L-constraint~\eqref{eq:L_constraints} for $(s,c)$ for every student $s$ with a priority less than $q_c + \lfloor \gamma \cdot B +0.5 \rfloor$.
    \end{enumerate}     
    Note that when $q_c +  B < |\lrl{s\in \S \;:\; c\succ_s \emptyset}|$, \ie, the budget is not enough to cover for all students applying to $c$, then some pairs $(s,c)$ will have L-constraints even with $\gamma = 1$.

    In Figure~\ref{fig: sensitivity to gamma hybrid formulation}, we compare the computational time needed to solve the instance for the hybrid formulation considering $\gamma \in \lrl{0, 0.25, 0.5}$ for $B \in \lrl{10,25}$. On the one hand, we observe that $\gamma = 0.25$ leads to the best results, although the differences are relatively small when $B=10$. On the other hand, when $B=25$, we observe large differences when the penalty is low, with $\gamma = 0$ leading to the shortest computational time. Given these results, and to simplify the exposition and implementation, we decided to use $\gamma=0$ to obtain the results reported in Section~\ref{sec:  simulation results}.
    
    \begin{figure}[htp!]
       \begin{subfigure}{0.49\textwidth}
      \caption{$B=10$}\label{fig: sensitivity gamma b=10}
            \includegraphics[width=\textwidth]{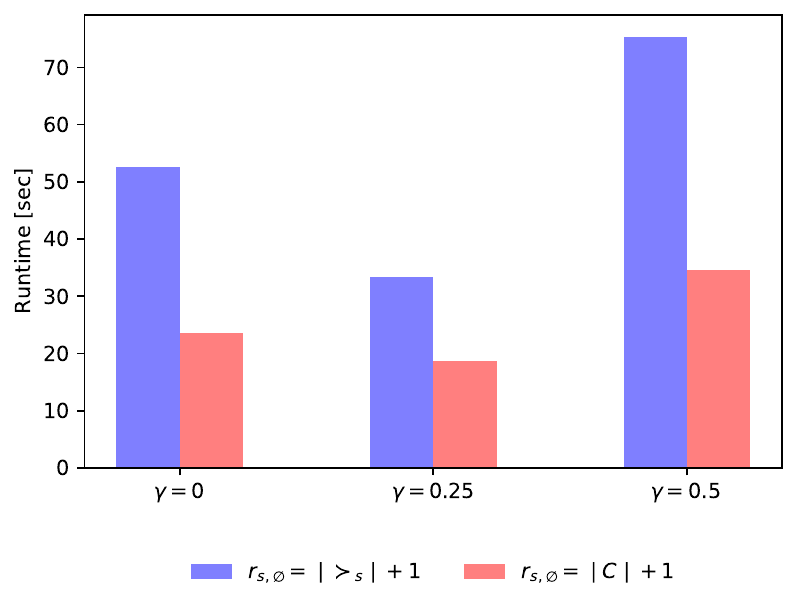}
        \end{subfigure}
        \begin{subfigure}{0.49\textwidth}
            \caption{$B=25$}\label{fig: sensitivity gamma b=25}
            \includegraphics[width=\textwidth]{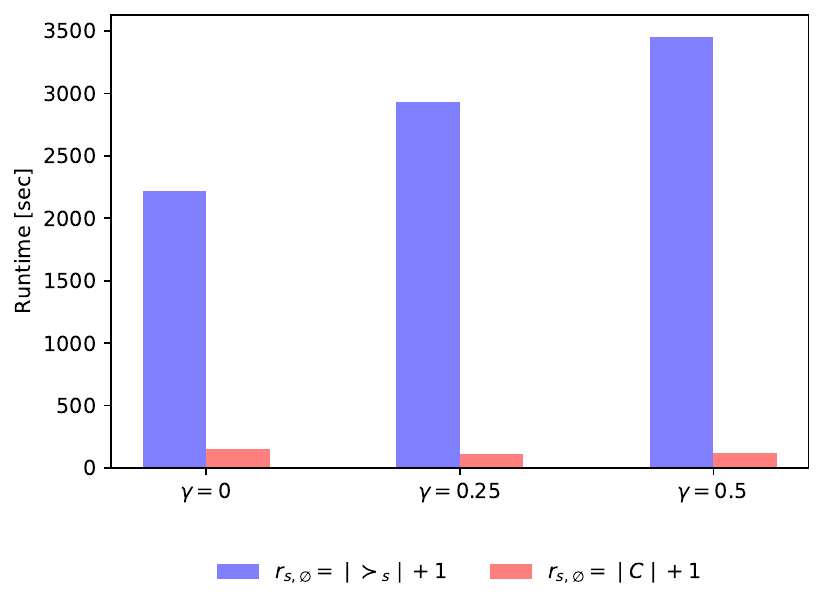}
        \end{subfigure}
        \caption{Sensitivity to $\gamma$}\label{fig: sensitivity to gamma hybrid formulation}
    \end{figure}

\subsection{Initialization Details and Acceleration of the Cutting-plane Method}\label{app: computational details of cutting-plane method}

    Many natural accelerations techniques can be added to the cutting plane method described in Algorithm~\ref{alg: cutting-plane} as well as its modified version, the hybrid approach. 
    %In our formulation, we improve the root node gap by adding the two following kinds of constraints:
    \begin{itemize}
        \item We implement the cutting-plane method as a branch-and-cut, incorporating the constraints generated by the separation algorithm as lazy constraints in Gurobi. %This avoids resetting the branch-and-bound tree for the main program whenever a set of comb constraints is added.
        \item We use the solution obtained from using \emph{LPH} as a warm-start. This allows the branch-and-cut to have immediately an upper bound, which may enable to prune earlier certain branches. 
        \item In Step~\ref{step:J}, we add no comb constraints, \ie, $\mathcal{J}=\emptyset$. However, we add several valid constraints to the initial main program (Step~\ref{step: solve MP}) that we derive from the student-optimal stable matching $\mu_{\mathbf{0}}$, obtained with no additional budget. 
        First, we know that each student $s$ must get an assignment they weakly prefer over $\mu_{\mathbf{0}}(s)$ when $B>0$. Hence, we introduce matching variables only among those pairs $(s,c)$ where $c$ is a school at least as preferred as the school to which student $s$ is matched in $\mu_{\mathbf{0}}$. 
        Second, 
        % we add (SOS1) constraints to ensure that one of the matching variables involving student $s$ obtains value 1 if $\mu_{\mathbf{0}}(s) \neq \emptyset$. 
        % \margarida{The SOS1 constraints that I used were for the first equality constraint in~\eqref{eq: feasible region generalized comb formulation}, \ie, instead of writing the equality constraint, I say the logic of the constraint to Gurobi: a student is assigned to at most one school (and this is called SOS1 constraint). Am I missing something? I am not, at the end of this bullet list and as new paragraph (because it is not a ``proven'' accelaration technique), we can just say: }
        the first equality constraints in~\eqref{eq: feasible region generalized comb formulation} were implemented as type 1 SOS constraints in Gurobi and the decision vector $\mathbf{x}$ forced to be binary. Although the latter is not necessary for the correctness of the method (in the worst-case, all comb constraints are added), we observed that Gurobi was able to leverage on this element and obtain faster execution times. 
        % For SOS 1, their impact on acceleration of solving times was not consistently advantageous compared to not using them.
        Third, we set $x_{s,c}=1$ for all $(s,c)\in \mu_{\mathbf{0}}$ such that $r_{s,c}=1$. Finally, we know that any school $c$ such that $\lra{\mu(c)} < q_c$ should receive no additional seats when $B> 0$. Hence, we set all $y^k_c = 0$ for all such schools. For the hybrid approach, additional constraints are added to the main program (see Appendix~\ref{app: hybrid details}).
        % \item \margarida{We do not do any of these constraints, unless I missed some active option in the code. Can we remove this point?} Related to the previous point, we can also add comb constraints that related to $\mu_{\mathbf{0}}$. Specifically, let \(\underline{\mu}_{\mathbf{0}}(c)\) be the lowest priority student admitted to school \(c\) in matching \(\mu_{\mathbf{0}}\) (according to \(\succ_c\)). Then, for each school \(c\) such that \(\lra{\mu_{\mathbf{0}}(c)} = q_c\), we add to \(\mathcal{J}\) the comb constraint involving \(C_c = S_{\underline{\mu}_{\mathbf{0}}(c), c} \cup \bigcup_{s: \mu_{\mathbf{0}}(s)=c} T_{s,c}^{-}\).
        % Moreover, for each fully subscribed school in $c$ in $\mu_{\mathbf{0}}$, we add the comb constraint with base $(s,c)$, where $s$ is the student ranked $q_c$-th by school $c$. Additionally, we also add a \emph{modified} comb constraint where the sum of all the teeth and the shaft starting from the last matched student in $\mu^0$ (i.e., $\underline{\mu}_{\mathbf{0}}(c)$) should be greater or equal to the length of the shaft starting from $\underline{\mu}_{\mathbf{0}}(c)$.           
    \end{itemize}

\section{Appendix to Section~\ref{subsec: properties of mechanism}}\label{app: properties mechanism}
\allowdisplaybreaks
    \subsection{Cardinality}\label{subsec:accessVsImpro}
        A crucial aspect of the solutions of Formulation~\ref{problem_def} is that they largely depend on the value of \(r_{s,\emptyset}\), i.e., the \emph{penalty} for having unassigned students. {We emphasize that $r_{s,\emptyset}$ does not indicate the position of $\emptyset$ in the ranking of $s$ over schools, but a penalty value for being unassigned.} One could expect that for greater penalty values, the optimal solution will prioritize \emph{access} by matching initially unassigned students and increasing the cardinality of the match. In contrast, for smaller penalty values, the focus will be on the \emph{improvement} of previously assigned students by prioritizing \emph{chains of improvement} that result in multiple students obtaining a better assignment than the initial matching with no extra capacities. In Theorems~\ref{thm: minimum cardinality} and~\ref{thm:card_maximized} we formalize this intuition.

        \begin{theorem}\label{thm: minimum cardinality}
            There is a sufficiently small and finite $r_{s, \emptyset}$ for all \(s\in \S\) such that the optimal solution of Formulation~\eqref{problem_def} returns a minimum cardinality student-optimal stable matching.
        \end{theorem}
        \proof{\it Proof.}
            Let \(\mu^0\) be the student-optimal assignment when there is no budget,  and let \(M(\mu) = \lrl{s\in \S: \mu(s)\in \C}\) be the set of students assigned in match \(\mu\).
            In addition, let \(r_{s,\emptyset} = r_{\emptyset} = \lra{\S}\cdot (1-\xi)\), where \(\xi = \max_{s\in \S} \lrl{\lra{\succ_s}}\) is the maximum length of a list of preferences among all students.

            Let \(\mu^*\) be the optimal allocation implied by the optimal solution of the problem \((\mathbf{x}^*, \mathbf{t}^*)\) considering a budget \(B\) and the aforementioned penalties.
            To find a contradiction, suppose that there exists an alternative match \(\mu'\) that uses the entire budget \(B\) and that satisfies
            \(\lra{M(\mu^*)} > \lra{M(\mu')}\). Without loss of generality, suppose that \(\lra{M(\mu^*)} = \lra{M(\mu')} + 1\).
            We know that \(M(\mu^0) \subseteq M(\mu^*)\) and \(M(\mu^0) \subseteq M(\mu')\), since no student who was initially assigned can result unassigned when capacities are expanded. Then, the difference in objective function between \(\mu^*\) and \(\mu'\) can be written as:
            \allowdisplaybreaks

            \begin{align}\label{eq: min cardinality starter}
                % \begin{split}
                    \Delta = \sum_{(s,c) \in \mu^*} r_{s,c} -  \sum_{(s,c) \in \mu'} r_{s,c}
                    &= \sum_{s\in \lrp{M(\mu^*)\cap M(\mu')} } r_{s,\mu^*(s)} - r_{s, \mu'(s)} \notag\\
                    &+ \sum_{s\in M(\mu^*)\setminus M(\mu') } r_{s,\mu^*(s)} - r_{s, \emptyset}
                    + \sum_{s\in M(\mu')\setminus M(\mu^*) } r_{s,\emptyset} - r_{s, \mu'(s)} \notag\\
                    &+ \sum_{s\in S\setminus \lrp{M(\mu^*)\cup M(\mu')}} r_{s,\emptyset} - r_{s, \emptyset} \notag\\
                    &= \sum_{s\in \lrp{M(\mu^*)\cap M(\mu')}} r_{s,\mu^*(s)} - r_{s, \mu'(s)} \notag\\
                    &+ \sum_{s\in M(\mu^*)\setminus M(\mu') } r_{s,\mu^*(s)} - r_{s, \emptyset}
                    + \sum_{s\in M(\mu')\setminus M(\mu^*) } r_{s,\emptyset} - r_{s, \mu'(s)} \notag\\
                    &\geq \sum_{s\in \lrp{M(\mu^*)\cap M(\mu')}} \lrp{1-\lra{\succ}_s} \notag\\
                    &+ \sum_{s\in M(\mu^*)\setminus M(\mu') } 1 - r_{s, \emptyset}
                    + \sum_{s\in M(\mu')\setminus M(\mu^*) } r_{s,\emptyset} - \lra{\succ_s} \notag\\
                    &\geq \lra{\lrp{M(\mu^*)\cap M(\mu')}}\cdot \lrp{1-\xi} \notag\\
                    &+ \lra{M(\mu^*)\setminus M(\mu')}\cdot (1-r_{\emptyset}) + \lra{M(\mu')\setminus M(\mu^*) }\cdot (r_\emptyset-\xi) \notag\\
                    &= \lra{\lrp{M(\mu^*)\cap M(\mu')}}\cdot \lrp{1-\xi}
                    + \lrp{\lra{M(\mu')\setminus M(\mu^*) } + 1}\cdot (1-r_{\emptyset})
                    \notag\\
                    &+ \lra{M(\mu')\setminus M(\mu^*) }\cdot (r_{\emptyset}-\xi) \notag\\
                    &\geq \lra{\lrp{M(\mu^*)\cap M(\mu')}}\cdot \lrp{1-\xi}
                    + \lrp{\lra{M(\mu')\setminus M(\mu^*) }}\cdot (1-\xi)
                    - r_{\emptyset}  \notag\\
                    &= \lrp{\lra{M(\mu^*)\cap M(\mu')} + \lra{M(\mu')\setminus M(\mu^*) }}\cdot \lrp{1-\xi}
                    - r_{\emptyset}  \notag\\
                    &= \lrp{\lra{M(\mu^*)\cap M(\mu')} + \lra{M(\mu')\setminus M(\mu^*) }}\cdot \lrp{1-\xi} - \lra{\S}\cdot (1-\xi) \notag\\
                    &= \lrp{\lra{M(\mu^*)\cap M(\mu')} + \lra{M(\mu')\setminus M(\mu^*) } - \lra{\S} }\cdot \lrp{1-\xi} \notag\\
                    &\geq 0, 
                % \end{split}
            \end{align}

            since both terms are negative as \(\lra{M(\mu^*)\cap M(\mu')} + \lra{M(\mu')\setminus M(\mu^*) } \leq \lra{S} \) and \(1\leq \xi \). Hence, we obtain that \(\Delta \geq 0\), which implies that the objective function evaluated at \(\mu'\) is strictly less than the objective function evaluated at \(\mu^*\), which contradicts the optimality of \(\mu^*\). Finally, note that this derivation holds for any set of penalties such that \(r_{s,\emptyset} \leq r_{\emptyset}\), and it also holds for the case when the clearinghouse can decide not to allocate all the budget (as long as this condition applies for both \(\mu^*\) and \(\mu'\)), so we conclude.
        \Halmos\endproof
        
Note that the penalty used in the proof is negative. A negative penalty can be interpreted as the policymaker willingness to allocate extra capacities to improve the assignment of students already in the system, for example, when merit scholarships are awarded to students who already have a secured position.  

By the stability constraints, every student that is initially assigned (i.e., when \(B=0\)) should be weakly better off when capacities are expanded. Hence, one may think that the minimum cardinality student-optimal stable matching is the one that is optimal for this subset of students. However, as we show in the example in Appendix~\ref{app: min cardinality is not optimal for initially assigned students} below, this is not the case.

        On the other hand, in Theorem~\ref{thm:card_maximized}, we show that if the penalty values are sufficiently high, then the optimal solution will prioritize \emph{access} by obtaining a student-optimal stable matching of maximum cardinality.
        \begin{theorem}\label{thm:card_maximized}
            There is a sufficiently high and finite \(r_{s, \emptyset}\) for all \(s\in \S\) such that the optimal solution of Formulation~\eqref{problem_def} returns a maximum cardinality student-optimal stable-matching.
        \end{theorem}
        \proof{\it Proof.}
            Let \(\mu^*\) be the stable-matching corresponding to the solution \((\mathbf{x}^*, \mathbf{t}^*)\). To find a contradiction, suppose there exists another stable matching \(\mu'\) that has a higher cardinality,  i.e., \(\lra{s\in \S \; : \; \mu'(s) = \emptyset} < \lra{s\in \S \; : \; \mu^*(s) = \emptyset}\); we also assume that \(r_{s,\emptyset} = \bar{r}\) for all \(s\in \S\), where \(\bar{r} > \sum_{s\in \S} \lra{\succ_s}\).
            By optimality of \((\mathbf{x}^*, \mathbf{t}^*)\), we know that
            \[
            \sum_{s\in \S} r_{s, \mu^*(s)} < \sum_{s\in \S} r_{s, \mu'(s)}.
            \]

            On the other hand, we know that
            \begin{equation}
                \begin{aligned}
                    \sum_{s\in \S} r_{s, \mu^*(s)} - \sum_{s\in \S} r_{s, \mu'(s)} &= \sum_{s\in \S: \mu^*(s) \in \C} r_{s, \mu^*(s)} - \sum_{s\in \S: \mu'(s) \in \C} r_{s, \mu'(s)} + \sum_{s\in \S: \mu^*(s) =\emptyset} r_{s,\emptyset} - \sum_{s\in \S: \mu'(s) =\emptyset} r_{s,\emptyset} \\
                    &= \sum_{s\in \S: \mu^*(s) \in \C} r_{s, \mu^*(s)} - \sum_{s\in \S: \mu'(s) \in \C} r_{s, \mu'(s)} \\
                    &+ \bar{r}\cdot \lrc{ \lra{s\in \S\;:\; \mu^*(s) = \emptyset} - \lra{s\in \S\;:\; \mu'(s) = \emptyset} } \\
                    &> -\sum_{s\in \S}\lra{\succ_s} + \bar{r}\cdot \lrc{  \lra{s\in \S\;:\; \mu^*(s) = \emptyset} - \lra{s\in \S\;:\; \mu'(s) = \emptyset} } \\
                    &> 0. \\
                \end{aligned}
            \end{equation}
            The first equality follows from \(r_{s,\emptyset} = \bar{r}\) for all \(s\in \S\). The first inequality follows from the fact that, given a student \(s\) that is assigned, the maximum improvement is to move from their last preference, \(\lra{\succ_s}\), to their top preference, and therefore \(r_{s, \mu^*(s)} - r_{s\mu'(s)} > 1-\lra{\succ_s} > -\lra{\succ_s}\). Then, we have that \(\sum_{s\in \S: \mu^*(s) \in \C} r_{s, \mu^*(s)} - \sum_{s\in \S: \mu'(s) \in \C} r_{s, \mu'(s)} \geq -\sum_{s\in\S} \lra{\succ_s}\). Finally, the last inequality follows from the fact that
            \[
            \lra{s\in \S\;:\; \mu^*(s) = \emptyset} - \lra{s\in \S\;:\; \mu'(s) = \emptyset} \geq 1
            \]
            and that \(\bar{r}\) is arbitrarily high. As a result, we obtain that
            \[
            \sum_{s\in \S} r_{s, \mu^*(s)} - \sum_{s\in \S} r_{s, \mu'(s)} > 0,
            \]
            which contradicts the optimality of \((\mathbf{x}^*, \mathbf{t}^*)\).
        \Halmos\endproof

\subsubsection{Example: Minimum Cardinality Suboptimality.}\label{app: min cardinality is not optimal for initially assigned students}

        %\nacho{
        Suppose there are \(n+2\) students and \(n\) schools, each with capacity \(q_c = 1\), and a total budget \(B=2\). For each student \(s_k\) with \(k\in \lrl{4, \ldots, n}\), we assume that their preferences are given by \(c_k \succ_{s_k} \emptyset \succ_{s_k} \ldots \), and we assume that the priorities at school \(c_k\) are \(s_k \succ_{c_k} \emptyset \succ_{c_k} \ldots\), where the $\succ \ldots$ represent an arbitrary ordering of the missing agents after $\emptyset$. In addition, we assume that the preferences and priorities of the other agents are:
          \begin{align*}
            s_1:& \;c_1 \succ \emptyset \succ \ldots && c_1: \; s_1 \succ s_2' \succ s_2 \succ \emptyset \succ \ldots \\
            s_2:& \;c_1 \succ c_3 \succ c_2 \succ \emptyset \succ \ldots && c_2: \; s_2 \succ s_2' \succ s_3' \succ \emptyset \succ \ldots \\
            s_3:& \;c_3 \succ c_2 \succ \emptyset \succ \ldots && c_3: \; s_3 \succ s_3' \succ s_2  \succ \emptyset \succ \ldots \\
            s_2':& \;c_n \succ \ldots \succ c_4 \succ c_1 \succ \emptyset \succ \ldots \\
            s_3':& \; c_3 \succ \emptyset \succ \ldots.
          \end{align*}
          In this case, the only option is to allocate the budget between schools \(c_1\) and \(c_3\), since allocating the budget to the other schools ($c_2$ or $c_k$ for $k\geq 4$) would have no effect. If $B=0$, then the student-optimal stable matching is $\mu^0 =\{ (s_1,c_1), (s_2,c_2), (s_3,c_3), (s_2',\emptyset), (s_3',\emptyset)  \}\cup \{ (s_k,c_k)\}_{k\geq 4}$. 
          Note that the optimal allocation for the set of students initially matched (when $B=0$) is to allocate the two extra seats to school \(c_1\).
          \begin{itemize}
            \item If \(t_{c_1} = 2\), then both students \(s_2\) and \(s_2'\) get assigned to \(c_1\), and thus the change among initially assigned students is \(-2\) (as student \(s_2\) moves from their third to their top preference).
            \item If \(t_{c_1} = t_{c_3} = 1\), then student \(s_2'\) and \(s_2\) get assigned to \(c_1\) and \(c_3\), respectively. Hence, the change among initially assigned students is \(-1\) (as student \(s_2\) moves from their third to their second preference).
            \item If \(t_{c_3} = 2\), then \(s_2\) and \(s_3'\) get assigned to \(c_3\). Hence, the change among initially assigned students is \(-1\) (as student \(s_2\) moves from their third to their second preference).
          \end{itemize}
          Finally, if $r_{s,\emptyset} = 0$ for every student $s$, it is easy to see that an optimal solution is to assign both additional seats to school \(c_3\), as the change in the objective function would be \(0\); in the other two cases analysed above (i.e., at least one extra capacity is allocated to $c_1$), the change in the objective function would be at least \(n-2\) as student \(s_2'\) would be assigned to $c_1$ ($c_1$ is the \(n-2\)-th preferred school of student $s_2'$). Hence, the optimal assignment is not the one that benefits the most those students initially assigned in $\mu^0$.
          Note that another optimal solution is to simply not assign any of the additional seats, but this case is not interesting as it would lead to no gains from the budget.
          % \textcolor{teal}{Margarida: For sake of completeness, I think that we should also indicate that another equivalent optimal solution is not to assign the budget:\\
          % The initial objective is $n-1+3=n+2$.
          % \begin{itemize}
          % \item If $t_{c_1} =2$ then, the objective$= n-1+1+n-2=2n-2$;
          % \item If $t_{c_1}=t_{c_3}=1$ then, $n-1+2+n-2=2n-1$;
          % \item If $t_{c_3}=2$ then, $n-1+2+1=n+2$.
          % \end{itemize}}
    \subsection{Incentives}\label{subsec: incentives}

        It is direct from~\citep{RothSotomayor1990} that our mechanism is not strategy-proof for schools, as we know that it is not in the base case with $B=0$. In Proposition~\ref{proposition_strategyproof_student}, we show that if the budget is positive and students know about it, then the mechanism that assigns students to schools and jointly allocates extra capacities is not strategy-proof for students.
        \begin{proposition}\label{proposition_strategyproof_student}
            The mechanism is not strategy-proof for students.
        \end{proposition}
        \proof{\it Proof.}
            Consider an instance with five students and five schools, each with capacity one. In addition, suppose that \(B=1\) and that the preferences and priorities are:
            \begin{subequations}
            \begin{alignat}{4}
             s_1&:  c_1\succ \ldots                     &\qquad & c_1:  s_1\succ  s_3\succ \ldots  \notag \\
             s_2&:  c_2\succ \ldots                     &\qquad &  c_2:  s_2\succ  s_3\succ \ldots  \notag \\
             s_3&:  c_1\succ c_2\succ  c_3\succ \ldots  &\qquad &  c_3:  s_3\succ \ldots  \notag \\
             s_1'&:  c_1'\succ \ldots                   &\qquad &  c_1':  s_1'\succ  s_2'\succ \ldots  \notag \\
             s_2'&:  c_1'\succ c_2'\succ \ldots         &\qquad &   c_2':  s_2'\succ \ldots   \notag
            \end{alignat}
            \end{subequations}
            where the \lq\lq$\ldots$\rq\rq represent an arbitrary completion of the preferences. If agents are truthful, the optimal allocation is to assign the extra seat to school $ c_1$, which will admit student $ s_3$; thus, the final matching would be $\{ (s_1, c_1), ( s_2,  c_2), ( s_3, c_1), (s_1', c_1'), (s_2',  c_2')  \}$. If student $ s_2'$ misreports her preferences by reporting
            $$s_2': c_1'\succ  c_1\succ  c_2\succ  c_2'\succ c_3,$$ the extra seat would be allocated to $ c_1'$ and $s_2'$ would get her favorite school.
        \Halmos\endproof

        Despite this negative result, in the next proposition we show that our mechanism is strategy-proof in the large.
        \begin{proposition}
            The mechanism is strategy-proof in the large.
        \end{proposition}
        \proof{\it Proof.}
            In an extension of their Theorem 1,~\cite{azevedo18} show that a sufficient condition for a semi-anonymous mechanism to be strategy-proof in the large is \emph{envy-freeness but for ties} (EF-TB), which requires that no student envies another student with a strictly worse lottery number. Hence, it is enough to show that our mechanism satisfies these two properties, i.e., semi-anonymity and EF-TB, for showing that our mechanism is strategy-proof in the large. Next, we provide definitions for semi-anonimity and for EF-TB, and prove for each property that our mechanism satisfies it.

            \paragraph{Semi-anonimity.} As defined in~\cite{azevedo18}, a mechanism is semi-anonymous if the set of students can be partitioned in a set of groups \(G\). Within each group \(g\in G\), each student belongs to a type \(t\); we denote by \(T_g\) the finite set of types that limits the actions of the students to the space \(A_g\). In our school choice setting, the groups are the set of students belonging to the same priority group (e.g., students with siblings, students with parents working at the school, etc.), the types are defined by the students' preferences \(\succ_s\), and the actions are the lists of preferences that students can submit. Then, two students \(s\) and \(s'\) that belong to the same group \(g\) and share the same type \(t\in T_g\) have exactly the same preferences and priorities and differ only their specific position in the schools' lists, which can be captured through their lottery numbers \(l_{s}, l_{s'} \in [0,1]\).\footnote{In other words, any ordering of students \(\succ_c\) in a given school \(c\) can be captured by \(s\succ_c s' \Leftrightarrow l_{s} > l_{s'}\) for any two students \(s,s'\) belonging to the same group \(g\in G\). Note that, for simplicity, we are implicitly assuming that ties within a group are broken using a single tie-breaker; this argument can be extended to multiple tie-breaking. } Note that \(G\) is finite because the number of priority groups is finite in most applications.\footnote{In the Chilean school choice setting, there are 5 priority groups: Students with siblings, students with working parents at the school, students who are former students, regular students and disadvantaged students.} Moreover, since the number of schools is finite, we know that the number of possible preference lists \(\succ_s\) is finite and, thus, the number of types within each group is finite. Hence, we conclude that our mechanism is semi-anonymous.

            \paragraph{EF-TB.} Given a market with \(n\) students, a direct mechanism is a function \(\Phi^n:\; T^n\rightarrow \Delta(\C \cup \lrl{\emptyset})^n \)
            that receives a vector of types \(T\) (the application list of each student) and returns a (potentially randomized) feasible allocation. In addition, let \(u_t(\tilde{c})\) be the utility that a student with type \(t\in T_g, g\in G\) gets from the lottery over assignments \(\tilde{c}\in \Delta(C\cup \lrl{\emptyset})\) (note that, by assumption, two students belonging to the same type have exactly the same preferences and, thus, get the same utility in each school \(c\in \C\cup \lrl{\emptyset}\)).
            Then, a semi-anonymous mechanism is envy-free but for tie-breaking if for each \(n\)
            there exists a function \(x^n: (T\times [0,1])^n \rightarrow \Delta(\C\cup\lrl{\emptyset})^n\) such that
            \[
            \Phi^n(\textbf{t}) = \int_{\textbf{l}\in [0,1]^n} x^n(\textbf{t},\textbf{l}), \mathrm{d}\textbf{l}
            \]
            and, for all \(i,j,n,\textbf{t}\) and \(\textbf{l}\) with \(l_i \geq l_j\), and if \(t_i\) and \(t_j\) belong to the same type, then
            \[
            u_{t_i}\lrc{x_i^n(\textbf{t},\textbf{l})} \geq u_{t_i}\lrc{x_j^n(\textbf{t},\textbf{l})}.
            \]
            In words, to show that our mechanism is EF-TB we need to show that whenever two students belong to the same group and one of them has a higher lottery, then the assignment of the latter cannot be worse that that of the former. This follows directly from the stability constraints, since for any budget allocation, we know that the resulting assignment must be stable. As a result, for any budget allocation, we know that given two students $s,s'$ that belong to the same type, the resulting assignment \(\mu\) satisfies \(\mu(s) \succ_s \mu(s')\) if \(s\succ_c s'\) for all \(c\in \C\). Then, it is direct that \(u_{t_s}\lrc{x_{s}^{n} (\textbf{t},\textbf{l})} \geq u_{t_s}\lrc{x_{s'}^{n} (\textbf{t},\textbf{l})}\), for whatever function \(x\) that captures our mechanism. Thus, we conclude that our mechanism is semi-anonymous and EF-TB, and therefore it is strategy-proof in the large.
        \Halmos\endproof

    \subsection{Monotonicity}\label{app: monotonicity}
        Suppose there are \(n\) students and \(n\) schools, each with capacity \(q_c = 1\), and a total budget \(B=1\). For each student \(s_k\) with \(k\in \lrl{5, \ldots, n}\), we assume that their preferences are given by \(c_{k-1} \succ_{s_k} c_k \succ_{s_k} \emptyset\), and we assume that the priorities at school \(c_k\) are \(s_4 \succ_{c_k} \ldots \succ_{c_k} s_n \succ_{c_k} \emptyset\) for all \(k \geq 4\). In addition, we assume that the preferences and priorities of the other agents are:
        \begin{align*}
          s_1:& \;c_1 \succ c_3 \succ \emptyset && c_1: \; s_1 \succ s_3 \succ \emptyset \\
          s_2:& \;c_2 \succ c_3 \succ \emptyset && c_2: \; s_2 \succ s_3 \succ \emptyset \\
          s_3:& \;c_1 \succ c_2 \succ \emptyset && c_3: \; s_1 \succ s_2 \succ \emptyset\\ 
          s_4:& \; c_4 \succ \emptyset. &&  
        \end{align*}
        It is easy to see that (when \(B=0\))
        \(\mu^0 = \lrl{(s_1, c_1), (s_2, c_2), (s_3, \emptyset), (s_4, c_4), \ldots, (s_n, c_n)}.\)
        In addition, if the penalty for having student \(s_3\) is sufficiently high (specifically, \(r_{s_3, \emptyset} > n-5\)), then the optimal budget allocation is to add one seat to school \(c_1\), so that the resulting match is \(\mu^1 = \lrl{(s_1, c_1), (s_2, c_2), (s_3, c_1), (s_4, c_4), \ldots, (s_n, c_n)}\). As a result, student \(s_3\) is now assigned to their top preference.

        Next, suppose that student \(s_3\) improves her lottery/score in school \(c_2\), so that now \(c_2\)'s priorities are
        \(c_2: s_3 \succ s_2 \succ \emptyset\). As a result, the initial assignment is \(\mu^{00} = \lrl{(s_1, c_1), (s_2, c_3), (s_3, c_2), (s_4, c_4), \ldots, (s_n, c_n)}\). In addition, note that (omitted allocations are clearly dominated):
        \begin{itemize}
          \item If \(t_{c_1} = 1\), then both \(s_2\) and \(s_3\) improve their assignment, and thus the change in the objective function is \(-2\) (both students move from their second to their top preference).
          \item If \(t_{c_2} = 1\), then only \(s_2\) improves their assignment, and thus the change in the objective function is \(-1\).
          \item If \(t_{c_4} = 1\), then a chain of improvements going from student \(s_5\) to \(s_n\) starts, with each of these students getting assigned to their top preference. As a result, the change in the objective function is \(-(n-4)\).
        \end{itemize}
        As a result, if \(n > 6\), it would be optimal to assign the extra seat to school \(c_4\), and the resulting assignment would be
        \(\mu^{11} = \lrl{(s_1, c_1), (s_2, c_3), (s_3, c_2), (s_4, c_4), (s_5, c_4), \ldots, (s_n, c_{n-1})}\). Note that student \(s_3\) is worse off in $\mu^{11}$ than in $\mu^{1}$, since she is now assigned to her second preference compared to the top one when her priority in school \(c_2\) was lower.

\subsection{Complexity}\label{subsec: complexity}
        In \cite{bobbio2022capacityvariation}, the authors analyze the complexity of Problem~\eqref{problem_def} and prove that it is NP-hard even when the preference lists of the students are complete, i.e., when students apply to all the schools. In this context, a possible approach is to design an approximation algorithm for this problem. Let us recall the definition of $f(\mathbf{t})$ for a fixed $\mathbf{t}\in\ZZ^{\C}_+$:
        \begin{equation}\label{def:function_matching}
        f(\mathbf{t}) := \min_{\mu}\left\{\sum_{(s,c)\in \mu} r_{s,c} : \ \mu \ \text{is a stable matching in instance} \ \Gamma_{\mathbf{t}} \right\}
        \end{equation}
        which can be evaluated in polynomial time by using the DA algorithm on instance $\Gamma_{\mathbf{t}}$. Therefore, one might be tempted to show that $f$ is a lattice submodular function due to the existence of known approximation algorithms (see, e.g., \citep{soma2015generalization}). As we show in the following result, $f$ is neither lattice submodular nor supermodular.
        \begin{proposition}\label{prop:submodularity}
        The function $f(\mathbf{t})$ defined in Expression \eqref{def:function_matching} is neither lattice submodular nor supermodular.
        \end{proposition}
        \proof{\it Proof.}
            Let us recall the definition of lattice submodularity. A function $f:\ZZ^{\C}_+\to\RR_+$ is said to be lattice submodular if $f(\mathbf{t}\vee \mathbf{t}')+f(\mathbf{t}\wedge \mathbf{t}')\leq f(\mathbf{t}) + f(\mathbf{t}')$ for any $\mathbf{t},\mathbf{t}'\in\ZZ^{\C}_+$, where $\mathbf{t}\vee \mathbf{t}:=\max\{\mathbf{t},\mathbf{t}'\}$ and $\mathbf{t}\wedge \mathbf{t}':=\min\{\mathbf{t},\mathbf{t}'\}$ component-wise. Function $f$ is lattice supermodular if, and only if, $-f$ is lattice submodular.
            %Now, let us recall Expression \eqref{def:function_matching}
            % \[
            % f(\mathbf{t}) = \min\left\{\sum_{(s,c)\in\mu}r_{s,c}: \ \mu \ \text{is a stable matching in instance} \ \Gamma_{\mathbf{t}} \right\}.
            % \]
            First, we show that \(f\) is not lattice submodular. Consider a set $\S=\{ s_1, s_2, s_3, s_4\}$ of students and a set $\C=\{ c_1, c_2, c_3, c_4, c_5 \}$ of schools. The preference lists are as follows
            \begin{subequations}
            \begin{alignat}{4}
            s_1&: c_1\succ c_2\succ \ldots &\qquad & c_1:  s_1\succ s_2\succ \ldots \notag \\
            s_2&: c_2\succ c_3\succ \ldots &\qquad & c_2:  s_2\succ s_3\succ \ldots  \notag \\
            s_3&: c_2\succ  c_5\succ c_4\succ \ldots &\qquad & c_3:  s_2\succ s_1\succ \ldots  \notag \\
            s_4&: c_5\succ \ldots &\qquad & c_4:  s_3\succ s_4\succ \ldots \notag \\
            &&\qquad &c_5:  s_4\succ s_3\succ \ldots   \notag
            \end{alignat}
            \end{subequations}

            School $c_1$ has capacity 0 and the other schools have capacity 1. We choose the following two allocations: $\mathbf{t}= (1,0,0,0,0)$ and $\mathbf{t'} = (0,1,0,0,0)$. Therefore, we obtain
            \[
            f(\mathbf{t}\vee\mathbf{t'}) + f(\mathbf{t}\wedge\mathbf{t'})  = f(1,1,0,0,0) + f(0,0,0,0,0) = 4   >  3 = 2+1 = f(\mathbf{t}) + f(\mathbf{t'}).
            \]

            Second, we show that \(f\) is not lattice supermodular. Consider a set $\S=\{ s_1, s_2, s_3\}$ of students and a set $\C=\{ c_1, c_2, c_3, c_4, c_5 \}$ of schools. The preference lists are as follows
            %\begin{align*}
            %s_1 : c_1, c_3, \ldots
            %s_2 : c_2, c_4, \ldots
            %s_3 : c_3,  c_4, c_5, \ldots
            %s_4 : c_5, \ldots
            %
            %\end{align*}
            \begin{subequations}
            \begin{alignat}{4}
            s_1&: c_1\succ c_3\succ \ldots \qquad \qquad  c_h:  s_1\succ s_2\succ \ldots   \qquad \text{ for all} \  h\in\{1,2,3\}. \notag\\
            s_2&: c_2\succ c_4\succ \ldots \notag \\
            s_3&: c_3\succ  c_4\succ c_5\succ \ldots  \notag \\
            s_4&: c_5\succ \ldots  \notag
            \end{alignat}
            \end{subequations}

            Schools $c_1$ and $c_2$ have capacity 0 and the other schools have capacity 1. We choose the following two allocations: $\mathbf{t}= (1,0,0,0,0)$ and $\mathbf{t'} = (0,1,0,0,0)$. Therefore, we obtain
            \[
            f(\mathbf{t}\vee\mathbf{t'}) + f(\mathbf{t}\wedge\mathbf{t'})  = f(1,1,0,0,0) + f(0,0,0,0,0) = 4   <  5 = 3+2 = f(\mathbf{t}) + f(\mathbf{t'}).\Halmos
            \]
        \endproof
\section{Model Extensions and Further Insights}\label{app: model extensions}

    \subsection{Extensions}

        Our model can be easily extended to capture several relevant variants of the problem. In what follows we name some direct extensions:
        \begin{itemize}
            \item Weighted budget constraints: If there is a unit-cost \(p_c\) of increasing the capacity of school \(c\), we can add an additional budget constraint of the form
            \[\sum_{c\in \C} t_c\cdot p_c \leq B',\]
            keeping all the other elements of the model unchanged. This extension could be used to allocate tuition waivers or other sort of scholarships that are school dependent.
            \item Different levels of granularity: Schools may not be free to expand their capacities by any value in \(\lrl{1,\ldots, B}\). This limitation can be easily incorporated into any of our formulations. Let us exemplify this for our Formulation~\eqref{Baiou_Balinski-binary-expansion}. Specifically, recall that \(y_c^k = 1\) if the capacity of school \(c\) is expanded by at least \(k\) seats, and \(y_c^k = 0\) otherwise. Then, we know that \(\sum_{k=1}^B  y_c^k = t_c\), so we must add the precedence constraints \( y^k_c\geq y^{k+1}_c\) for each \(c\in \C\) and $k=1,\ldots,B-1$.
            If the capacity of school \(c\) can only be expanded by values in a subset \(\mathcal{B}' \subseteq \lrl{1,\ldots, B}\), we can enforce this by adding the constraints \(y_c^k = y_c^{k+1}\) for all \(k \in \lrl{1,\ldots, B-1}\setminus \mathcal{B}'\) and $y_c^{B}=0$ if $B \notin \mathcal{B}'$. This could also be captured using knapsack constraints.
            \item Adding secured enrollment: The Chilean system guarantees that students that are currently enrolled and apply to switch will be assigned to their current school if they are not assigned to a more preferred one. This can be easily captured in our setting. Let $M$ be the set of students aiming to switch school. First, for each student $s\in M$, $\succ_s$ only contains schools that $s$ strictly prefers over the currently assigned school. Second, we add to the objective function the term
            $$\sum_{s \in M} \left(|\succ_s|+1\right) \left( 1 - \sum_{s: (s,c) \in \E} x_{s,c}\right),$$
            which corresponds to accounting in the objective function for the current matching of students in $M$ who are unable to switch to a more preferred school.
            % \item Adding secured enrollment: The Chilean system guarantees that students that are currently enrolled and apply to switch will be assigned to their current school if they are not assigned to a more preferred one. This can be easily captured in our setting by introducing a parameter \(m_{s,c}\), which is equal to 1 if student \(s\) is currently enrolled in school \(c\), and 0 otherwise, and defining \(M = \lrl{s\in \S: m_{s,c} = 1 \text{ for some } c \in \C}\). Then, we would only have to update a couple of constraints:
            % \begin{equation}
            %     \begin{split}
            %         \sum_{s \in \S} x_{s,c}\cdot (1-m_{s,c}) &\leq q_c + t_c, \quad \forall \ c \in \C, \\
            %         \sum_{c \in \C} x_{s,c} &= 1, \hspace{1cm} \forall \ s \in M.
            %     \end{split}
            % \end{equation}
            % The first constraint ensures that students currently enrolled do not count towards the capacity of the school they are currently enrolled. The second constraint ensures that all students that are currently enrolled are assigned to some schools (potentially, to the same school they are currently enrolled).
            % \item Room assignment: Schools report the number of vacancies they have for each level. This decision depends on the classrooms they have and their capacity. However, schools decide (before the assignment) what level goes in each classroom, and this determines the number of reported vacancies for that level. This may introduce some inefficiencies, since some levels may be more demanded, and thus assigning a larger classroom may benefit both students in that school but also in others.
            \item Quota assignment: Many school choice systems have different quotas to serve under-represented students or special groups. For instance, in Chile there are quotas for low-income students (15\% of total seats), for students with disabilities or special needs, and for students with high-academic performance. Moreover, some of these quotas may overlap, i.e., some students may be eligible for multiple quotas, and in most cases students count in only one of them.
            The number of seats available for each quota are pre-defined by each school, and schools have some freedom to define these quotas. Hence, our problem could be adapted to help schools define what is the best allocation of seats to quotas in order to improve students' welfare.
        \end{itemize}
        {Note that all these extensions can be easily accommodated in methods involving optimization, i.e., our exact methods and LPH, since these involve either (i) modifying the input (as in the case of incorporating quotas or secured enrollment) or (ii) adding constraints (as in the case of a monetary budget or different levels of granularity). However, note that Greedy may not be able to accommodate all of them, specially those requiring the addition of constraints. For instance, it is not direct how to adapt our greedy heuristic to incorporate monetary budget constraints since the school leading to the highest marginal improvement may not be cost-effective. Hence, this provides an additional argument in favor of LPH relative to Greedy.}

\subsection{Further Insights: An Application to the Chilean College Admission System}\label{sec: other applications of our framework}
        Since early 2023, we have collaborated with the Ministry of Education of Chile (MINEDUC) and the ``Sistema \'Unico de Admisi\'on'' (SUA; the Chilean college board) in several applications of our framework. Specifically, we have been working on two main projects: (i) evaluating which schools should be overcrowded in order to downsize under-demanded schools, and (ii) evaluating minimum requirements regarding seats to offer to under-represented groups. The former  problem can be addressed by adapting  what we discussed in the previous section, so we focus on the latter.

        The centralized part of the Chilean college admissions system includes an affirmative action policy (called ``Programa PACE''), which consists of reserved seats for under-represented students (over 10,000 students each year) and special funds for the institutions where these students enroll. To be eligible to participate in this affirmative action policy (and potentially receive these funds), education institutions must commit to reserving a specified number of seats following specific guidelines defined by MINEDUC, e.g., a minimum number of reserved seats per program and a minimum total number of reserved seats across all their programs. Conditional on satisfying these requirements, universities can independently decide how many seats to reserve. Many universities meet these requirements by fulfilling the minimum requirement (of one reserved seat) per program and then devoting the additional required seats to under-demanded programs. As a result, only 18\% of the reserved seats were used in the admissions process of 2022-2023.

        To tackle this issue, we have been collaborating with MINEDUC to evaluate the effects of potential changes to these requirements. Since we do not know what the objective function of each university is (and, thus, how they allocate these reserved seats), MINEDUC asked us to evaluate different combinations of requirements (e.g., a minimum number of seats for the most popular programs, changes to the way to define the total number of reserved seats to offer, etc.) and objective functions (e.g., minimize the preferences of assignment, maximize the utilization of the reserved seats, maximize the cutoffs, maximize the average scores in Math and Verbal of the admitted students, among others) to obtain a wide range of possible outcomes that could result from each set of requirements.

        For each of these combinations (of requirements, objectives, and other specific parameters), we adapted the framework described in this work and performed simulations considering the data from the admissions process of 2022-2023. Using current data to evaluate the implementation of these requirements in future years is without major loss, since students' preferences are relatively stable over time and, thus, the reported preferences of one year are the best predictor of those in the coming years. Finally, note that having a methodology to solve the problem relatively fast was instrumental in performing this analysis, as it required hundreds of simulations in a fairly large instance.

        This application showcases the flexibility of our methodology to answer different questions and stresses the need to solve these problems in a reasonable time.
\section{Additional Background}\label{app:background}

    \subsection{The Deferred Acceptance Algorithm}\label{app:DA}
        In this section, we recall the Deferred Acceptance algorithm in Algorithm~\ref{alg:DA} which was originally introduced by \citet{gale1962college}.
\begin{algorithm}[htp!]
            \caption{DA Algorithm}\label{alg:DA}
            \begin{algorithmic}[1]
            \Require  An instance $\Gamma=\langle \S,\C,\succ, \mathbf{q} \rangle$.
            \Ensure Student-optimal matching.
             \State Each student starts by applying to her most preferred school. Schools temporarily accept the most preferred applications and reject the less preferred applications which exceed their capacity.
             \State Each student $s$ who has been rejected, proposes to her most preferred school to which she has not applied yet; if she has proposed to all schools, then she does not apply. If the capacity of the school is not met, then her application is temporarily accepted. Otherwise, if the school prefers her application to one of a student $s'$ who was temporarily enrolled, $s$ is temporarily accepted and $s'$ is rejected. Vice-versa, if the school prefers all the students temporarily enrolled to $s$, then $s$ is rejected.
             \State If all students are enrolled or have applied to all the schools they rank, return the current matching. Otherwise, go to Step 2.
            \end{algorithmic}
        \end{algorithm}

\subsection{Separation Algorithm of \citep{baiou2000stable} and Counterexample}\label{ex: baiou separation algorithm counterexample}
        % \begin{example}
%\fede{new counter-example}
        
        In this counterexample, we show that Algorithm~\ref{alg: Baiou Balinski algo}, the separation algorithm provided in \cite{baiou2000stable}, does not find the most violated comb constraint as claimed in their Theorem 5. 

        \begin{algorithm}[htp!]
            \caption{Separation algorithm in \cite{baiou2000stable}}\label{alg: Baiou Balinski algo}
            \begin{algorithmic}[1]
            \Require  An instance $\Gamma = \langle \S,\C,\succ, \mathbf{q} \rangle$, and a matching $\mathbf{x}$. Schools are labeled as $c_1,\ldots, c_{|\C|}$.
            \Ensure A violated comb.
             \State $i \leftarrow \, 1$  \Comment{Index of the school}
             \State $\mathcal{Q} \leftarrow \, \emptyset$ \Comment{The students that will form the comb} \label{step: baiou bal: initialize comb set}
             \State Assign $q_i -1$ elements of $\S$ to $\mathcal{Q}$ such that $\sum_{(s,c_i)\in T^-(s',c_i)}x_{s,c_i}\leq \sum_{(s,c_i)\in T^-(s'',c_i)}x_{s,c_i}$, for every $s'\in \mathcal{Q}$, $s'' \in \S\setminus \mathcal{Q}$, where $T^-(\cdot)$ indicates the tooth without its base. \label{step: baiou bal: beginning}
             \State $U\leftarrow \S \setminus \mathcal{Q}$, and let $s$ be $c_i$'s least preferred applicant 
             \State If the rank of $s$ in $c_i$'s preference list is less than $q_i$, go to Step~\ref{step: baiou bal: end}  \label{step: baiou bal: if gamma plus}
             \If{$s\in \mathcal{Q}$}
             \State $\Bar{s} = \argmin\left\{ \sum_{(s,c_i)\in T^-(s',c_i)}x_{s,c_i}: \ s'\in U \right\}$
             \State $U\leftarrow U\setminus \{ \Bar{s}\} $
             \State $\mathcal{Q} \leftarrow (\mathcal{Q} \setminus \{s\}) \cup \{ \Bar{s}\}$
             \Else 
             \State $U\leftarrow U\setminus \{ s\} $
             \EndIf 
             \State $C \leftarrow S(s,c_i ) \cup T(s,c_i) \cup \bigcup_{s'\in \mathcal{Q}} T(s',c_i) $
             \If{$\sum_{(s',c')\in C}x_{s',c'}^\star < q_i $  }
             \State Return $C$
             \Else
             \State Replace $s$ by its immediate successor in $c_i$'s preference list and go to Step~\ref{step: baiou bal: if gamma plus}
             \EndIf 
             \State $i\leftarrow i+1 $ \label{step: baiou bal: end}
             \If{$i\leq |\C|$}
             \State Go to Step~\ref{step: baiou bal: initialize comb set}
             \Else 
             \State Return $\emptyset$
             \EndIf
            \end{algorithmic}
        \end{algorithm}
    
        Let $\Gamma=\langle \S,\C,\succ, \mathbf{q} \rangle$ be the instance of the school choice problem with $\mathcal{S} = \{ s_1, s_2, s_3, s_4, s_5, s_6, s_7, s_8 \}$ as the set of students and $\mathcal{ C} =\{c_1,c_2, c_3, c_4, c_5, c_6, \} $ as the set of schools; schools $c_1, c_6$ have capacity 2, while the others have capacity 1. 
        % The preferences of students
        %  $s_1, s_2$ are: $ c_6 \succ c_1\succ c_2 \succ c_3 \succ c_4 \succ c_5$.  
        %  The preferences of students
        %  $s_3, s_4$ are: $ c_6 \succ c_2 \succ c_3 \succ c_4 \succ c_5 \succ c_1 $. 
        %  The preference of student $s_5: c_2 \succ c_3 \succ c_4 \succ c_5 \succ c_1 \succ c_6$. 
        %  The preference of student $s_6: c_3 \succ c_2 \succ c_4 \succ c_5 \succ c_1 \succ c_6$. 
        %  The preference of student $s_7$ is: $c_4 \succ c_2 \succ c_3 \succ c_5 \succ c_1 \succ c_6$. 
        %  The preference of student $s_8$ is: $c_5 \succ c_2 \succ c_3 \succ c_4 \succ c_1 \succ c_6$.  
        %  Schools $c_6, c_1$ have the same preferences:  $ s_1\succ s_2\succ s_3\succ s_4 \succ s_5\succ s_6\succ s_7\succ s_8$. 
        %  The preference of school $c_2$ is:  $ s_5\succ s_6\succ s_7\succ s_8 \succ s_1\succ s_2\succ s_3\succ s_4$. 
        %  The preference of school $c_3$ is:  $ s_6\succ s_5\succ s_7\succ s_8 \succ s_1\succ s_2\succ s_3\succ s_4$. 
        %  The preference of school $c_4$ is:  $ s_7\succ s_6\succ s_5\succ s_8 \succ s_1\succ s_2\succ s_3\succ s_4$. 
        %  The preference of school $c_5$ is:  $ s_8\succ s_6\succ s_7\succ s_5 \succ s_1\succ s_2\succ s_3\succ s_4$. 

        The preferences and priorities are given by:
        \begin{subequations}
        \begin{alignat}{4}
         s_1, s_2 &:   c_6 \succ c_1\succ c_2 \succ c_3 \succ c_4 \succ c_5        &\qquad \qquad & c_1:  s_1\succ s_2\succ s_3\succ s_4 \succ s_5\succ s_6\succ s_7\succ s_8  \notag \\
         s_3, s_4&:  c_6 \succ c_2 \succ c_3 \succ c_4 \succ c_5 \succ c_1         &\qquad \qquad & c_2:  s_5\succ s_6\succ s_7\succ s_8 \succ s_1\succ s_2\succ s_3\succ s_4  \notag \\
         s_5&:  c_2 \succ c_3 \succ c_4 \succ c_5 \succ c_1 \succ c_6              &\qquad \qquad & c_3:  s_6\succ s_5\succ s_7\succ s_8 \succ s_1\succ s_2\succ s_3\succ s_4  \notag \\
         s_6&:  c_3 \succ c_2 \succ c_4 \succ c_5 \succ c_1 \succ c_6              &\qquad \qquad & c_4:  s_7\succ s_6\succ s_5\succ s_8 \succ s_1\succ s_2\succ s_3\succ s_4  \notag \\
         s_7&:  c_4 \succ c_2 \succ c_3 \succ c_5 \succ c_1 \succ c_6              &\qquad \qquad & c_5:  s_7\succ s_6\succ s_5\succ s_8 \succ s_1\succ s_2\succ s_3\succ s_4   \notag \\
         s_8&:  c_5 \succ c_2 \succ c_3 \succ c_4 \succ c_1 \succ c_6              &\qquad \qquad & c_6:  s_1\succ s_2\succ s_3\succ s_4 \succ s_5\succ s_6\succ s_7\succ s_8   \notag
        \end{alignat}
        \end{subequations}

        Let us consider the following optimal matching $\mathbf{x}^\star$  of the main program with no stability constraints (i.e., $\mathcal{J}=\emptyset$) and budget $B=0$:  $x^\star_{s_1,c_1}=1$, $x^\star_{s_2,c_1}=1$, $x^\star_{s_3,c_6}=1$, $x^\star_{s_4,c_6}=1$, $x^\star_{s_5,c_2}=1$, $x^\star_{s_6,c_3}=1$, $x^\star_{s_7,c_4}=1$, $x^\star_{s_8,c_5}=1$. %Note that $n=8$, $m=6$, $\Bar{q}= 2$. 

        If we use the separation algorithm provided by \citet{baiou2000stable}, we search for a violated comb from school $c_1$ to $c_5$ and we find none. Finally, we check for violated combs in school $c_6$, and we find the comb $C(s_3,c_6)$ with  teeth $T(s_3,c_6)$ and $T(s_2,c_6)$. Note that comb $C(s_3,c_6)$ has value 1 in $\mathbf{x}^\star$. %It took $\mathcal{O}(m \cdot n^2 )$ operations to find it.
        However, if we use our Algorithm~\ref{alg: separation method}, we have that the only school in block$(\mathbf{x}^\star)$ is $c_6$. We find the comb $C(s_2,c_6)$ with teeth $T(s_2,c_6)$ and  $T(s_1,c_6)$. Note that $C(s_2,c_6)$ has value 0 in $\mathbf{x}^\star$.% and it took $\mathcal{O}(m \cdot n)$ operations to find it, which is the number of operations needed to find block$(\mathbf{x}^\star)$.
        Therefore, if we use the separation algorithm of \citet{baiou2000stable}, we find a comb which is not the most violated one. 
% \input{OR/2nd_round/999_appendix_Z5_mincut_maxhrt}

% \clearpage
% \input{OR/2nd_round/999_old_appendix}
\end{APPENDICES}

%%%%%%%%%%%%%%%%%
\end{document}